\documentclass{article}

\usepackage{multirow}
\usepackage{algorithm}
\usepackage{algorithmic}
\usepackage{subcaption}
\usepackage{adjustbox}
\usepackage{amsmath}
\usepackage{amssymb}
\usepackage{mathtools}
\usepackage{amsthm}
\usepackage{diagbox}
\usepackage{wrapfig}
\usepackage{float}
\usepackage{caption}
\usepackage{enumitem}
\usepackage{booktabs}
\usepackage{tabularx}

\usepackage{comment}
\usepackage{placeins}

\newlength{\imggap}
\newlength{\colheight}
\newlength{\Hthree}
\newlength{\Hfour}

\usepackage{listings}
\lstdefinestyle{asciimap}{
  basicstyle=\ttfamily\small,
  columns=fullflexible,
  keepspaces=true,
  showstringspaces=false,
  frame=single,
  xleftmargin=0.5em, xrightmargin=0.5em,
}

\newtheorem{theorem}{Theorem}[section]

\newtheorem{lemma}[theorem]{Lemma}

\newtheorem{assumption}[theorem]{Assumption}

\def\tcb{\textcolor{blue}}	% Textcolor blue
\def\tcr{\textcolor{red}}	% Textcolor red

\usepackage[numbers]{natbib}
\usepackage[table]{xcolor}

\usepackage[main, final]{neurips_2026}

\usepackage[utf8]{inputenc} % allow utf-8 input
\usepackage[T1]{fontenc}    % use 8-bit T1 fonts
\usepackage{hyperref}       % hyperlinks
\usepackage{url}            % simple URL typesetting
\usepackage{booktabs}       % professional-quality tables
\usepackage{amsfonts}       % blackboard math symbols
\usepackage{nicefrac}       % compact symbols for 1/2, etc.
\usepackage{microtype}      % microtypography
\usepackage{xcolor}         % colors

\title{AGiR: Mitigating Gift Over-Reliance \\ in Mixed-Motive Games}

\author{%
  Woohyeon Byeon \quad Seongmin Kim \quad Jiwon Jeon \quad
  Woojun Kim\footnotemark[1] \quad Youngchul Sung\\[4pt]
  \textnormal{KAIST} \\
  \texttt{\{woohyeon.byeon, seongmin.kim, jiwon.jeon, woojun.kim, ycsung\}@kaist.ac.kr}
}

\begin{document}

\maketitle
{\renewcommand{\thefootnote}{\fnsymbol{footnote}}%
\footnotetext[1]{Correspondence to: Woojun Kim (woojun.kim@kaist.ac.kr).}}
% {\renewcommand{\thefootnote}{\fnsymbol{footnote}}%
% \footnotetext[1]{Co-advising authors. Correspondence to: Woojun Kim and Youngchul Sung.}}

\begin{abstract}
In this paper, we consider mixed-motive games (MMGs) which model various real-world multi-agent scenarios where self-interested agents must both cooperate and compete. 
We identify an issue with existing gifting methods, where agents become overly dependent on incoming gifts and fail to learn reward-yielding behaviors, a phenomenon we call \emph{gift over-reliance}. To address this, we propose a refusal-augmented gifting framework that balances cooperation and competition in MMGs. Our approach introduces an adaptive refusal mechanism that allows agents to selectively reject a portion of received gifts from their policy-update signals, thereby mitigating gift over-reliance and preventing the agents from being exploited by other agents. This mechanism is gifting-method-agnostic, serving as a plug-in for existing schemes while enabling decentralized learning. We demonstrate that across diverse MMG environments and gifting schemes, our mechanism consistently reduces gift over-reliance and prevents lock-in to sacrificing policies, improving $\alpha$-fairness as a result, without necessitating major changes to existing training pipelines.
\end{abstract}

\section{Introduction}
Multi-agent reinforcement learning (MARL) has made substantial progress in cooperative problems, where agents optimize a shared team goal~\cite{qmix, qtran, kim2023variational, kim2023adaptive}. However, many real-world settings are \emph{mixed-motive}: agents are neither strictly competitive nor their goals fully align with a social objective. Depending on how rewards are structured, cooperation may be beneficial in some situations, while self-interest dominates in others. For example, in Environmental Social Governance (ESG) investment~\cite{investesg}, a firm can either spend money to reduce emissions (cooperate) or cut costs by maintaining higher emissions (defect). Investors, in turn, may choose to favor ESG-friendly firms (cooperate) or focus solely on short-term profits (defect). When investors support responsible firms, their actions align private incentives with social value. Without such measures, responsible firms bear higher costs while others gain without contributing, leading to inefficient outcomes.

% \begin{figure}[t]
%   \centering
%     \includegraphics[width=\linewidth]{image/agir_intro.pdf}
%   \caption{Decentralized learning methods in mixed-motive games./fig will be modified.
% % AGiR jointly achieves social efficiency and fair learning opportunities in mixed-motive games.
% \textbf{Top: Independent learning} agents lead to socially inefficient outcomes, where no one is willing to bear the cost of cooperation.
% % \textbf{Middle: Gifting} improves social outcomes by incentivizing cooperation.
% % However, it can induce \emph{gift over-reliance}, creating \emph{unfair learning opportunities}: agents who rely on gifts fail to learn task-relevant behaviors (e.g., harvesting).
% % \textbf{Bottom: Gifting + Refusing (Ours)} mitigates gift over-reliance and restores fair learning opportunities, allowing agents to both contribute to society and learn how to earn rewards independently.
% \textbf{Middle: Gifting} improves social outcomes by incentivizing cooperation.
% However, agents who rely on gifts fail to learn task-relevant behaviors (e.g., harvesting).
% \textbf{Bottom: Gifting + Refusing (Ours)} restores fair learning opportunities, allowing agents to both contribute to society and learn how to earn rewards independently.
% }

%   \label{fig:intro}
%   \vspace{-15pt}
% \end{figure}

In such mixed-motive settings, one approach explored in the game-theoretic literature is to encourage cooperation through \emph{gifting}~\cite{akerlof1982labor}, where agents voluntarily transfer part of their reward to others to promote cooperative behavior. This mechanism helps to align individual incentives with collective welfare, as cooperative agents are compensated for their contribution. For example, investors may incentivize firms that engage in costly emission reduction. This gifting mechanism has also been studied in the context of MARL. For example, \citet{lupu2020gifting} proposed a peer-rewarding mechanism where agents share part of their rewards with others to encourage cooperation, while \citet{lase} extended this idea by incorporating empathy into the gifting process to adaptively balance altruism and self-interest in mixed-motive games.

% \tcr{real-world is asym or unfair. if we can say 'under agir 1) one who wants to leverage unfairness can maintain the adv and 2) one who wants to escape unfairness can escape the disadv' $\to$ then, write in this direction. (someone wants to resolve unfairness, but someone wants to keep unfairness in real world)}
While gifting can promote cooperation, it may also be interpreted as problematic from a perspective where such cooperation is sustained by gifts rather than task competence. 
In mixed-motive settings, this is undesirable, especially for self-interested agents: relying on incoming gifts can trap an agent in a suboptimal, earning-independent strategy, and discourage exploration of reward-yielding behaviors. For example, an agent with systematically lower earning opportunities may continue choosing cooperative actions mainly because gifts offset its short-term losses, reducing the incentive to learn how to earn direct environmental rewards on its own; meanwhile, other agents maintain their profit through task performance, supported by this agent's cooperation. Such outcomes may appear cooperative on the surface, but they distort both profit and learning opportunities and can be brittle when gift patterns change. We refer to the failure to develop self-reliant, reward-yielding strategies due to excessive dependence on gifts as \emph{gift over-reliance}, and our goal is to mitigate this.

% While gifting can effectively promote cooperation, it can also create unintended consequences when used excessively. When agents become overly dependent on gifts, a new form of imbalance emerges. Some agents may continue to act cooperatively mainly because of external gifts, placing less emphasis on their own profit. 
% As a result, certain agents rely primarily on gifts while others sustain their returns through task performance. 
% Although this outcome is cooperative, it remains unfair from a profit perspective.
% \tcr{In other words, agents that overly rely on gifts may fail to learn how to earn task-relevant rewards on their own, as they become complacent with incoming gifts before sufficient exploration.
% This results in an unfair learning dynamic, where certain agents face restricted opportunities to develop self-reliant task strategies.}
% We refer to this phenomenon as \emph{gift over-reliance}, which we aim to mitigate in this work.

To mitigate \emph{gift over-reliance} in mixed-motive games, we propose \textbf{A}daptive \textbf{Gi}ft \textbf{R}efusal (\textbf{AGiR}), a plug-in mechanism that augments existing gifting schemes by adaptively refusing incoming gifts. By regulating gift acceptance based on local reward context, AGiR reduces excessive dependence on external gifts while preserving cooperative behavior, often improving fairness measured on \emph{extrinsic} returns as a consequential benefit. In our formulation, agents observe only their own extrinsic rewards and received gifts during training, without access to others’ rewards, social-welfare signals, or a prescribed global objective.
We evaluate AGiR on mixed-motive multi-agent benchmarks, including Coin Game, Cleanup, and Harvest, representative of sequential social dilemmas~\cite{ssd_category}. We summarize the resulting outcomes with $\alpha$-fairness~\cite{mo2002fair}, a metric that jointly captures the trade-off between social welfare and inequity, with the emphasis controlled by $\alpha$.
In all three domains and with every base gifting method, AGiR lets agents with lower earning opportunities earn more on their own, improving $\alpha$-fairness in most settings as a result.
For example, consider the Harvest environment, which represents a commons dilemma.
In this environment, gifting can discourage some agents from learning to harvest by allowing others to monopolize the shared orchard. This may increase total return, but it restricts learning opportunities and worsens worst-agent outcomes.
AGiR prevents such gift over-reliance, alleviating monopolization and letting these agents learn to harvest on their own.

% We evaluate AGiR on mixed-motive multi-agent benchmarks, including CoinGame, Cleanup, and Harvest, which are representative environments of sequential social dilemmas (SSD)~\cite{ssd_category}. We assess performance in terms of $\alpha$-fairness~\cite{mo2002fair}, a metric that jointly captures task performance and fairness at levels controlled by $\alpha$. Across these domains, AGiR consistently improves $\alpha$-fairness and reduces gifting-induced lock-in. Importantly, it achieves competitive aggregate performance without relying on disproportionate individual sacrifices that artificially inflate team return, as observed in prior baselines.
% For instance, in Harvest environment, agents face a commons dilemma around a shared orchard. 
% Gifting can dissuade recipients from crowding the orchard, as they learn that they can be incentivized by gifts even without harvesting. 
% This can allow one or two agents to concentrate on harvesting, maintaining the sustainability of the shared orchard and increasing the total outcome. However, this apparent benefit masks a problematic unfairness: some agents fail to acquire the opportunity to learn task-relevant behaviors, such as reaching the orchard and harvesting, during training.
% Our AGiR prevents over-reliance on gifts, mitigating such monopolization and improving the worst-agent return.

Our contributions are threefold. (1) We identify and characterize 
\emph{gift over-reliance}, where some agents persistently sacrifice their extrinsic returns and fail to develop self-reliant, reward-yielding strategies due to dependence on gifts. (2) We introduce AGiR, a plug-in, method-agnostic mechanism that adaptively refuses gifts to mitigate gift over-reliance while preserving decentralized learning, requiring only each agent’s own rewards and received gifts. (3) Experiments on three sequential social dilemmas show that AGiR lets agents with lower earning opportunities earn more on their own, improving $\alpha$-fairness in most settings, and a two-agent study reports how AGiR behaves across scenarios that vary spawn position and competence.
Code is available at \url{https://github.com/whbyeon/AGiR}.

\section{Background and Related Works}\label{sec:background}
% MDP\&RL, SSD types(PD, public goods, commons) 

% $\alpha$-Fairness (here?)

% \subsection{SSD and MMG}
% SSD explicitly defines cooperative policies and defective policies where both are subsets of whole policy space and they do not have to be a partition \citep{ssd_seminal}. Then apply EGTA only on cooperative policy set and defective policy set, which aligns the concept of MG whose stage games are matrix games \tcb{(delete after ,?)}.\\

% MMG is broader. does not specify C or D. this is more align to RL.

\paragraph{Partially Observable Markov Game}
\label{subsec:pomg}
A Partially Observable Markov Game (POMG)~\cite{littman1994markov} models multi-agent decision-making under uncertainty. At each time step, each agent selects an action according to its own policy $\pi^i$ based on its local observation $o_t^i$ or its history, which jointly induces the next state $s_{t+1}$ and yields individual rewards $r^i_t$. Each agent aims to train a policy $\pi^i$ to maximize its expected discounted return, $V_\mu^i(\pi) = \mathbb{E}_{\mu,\pi}[\sum_{t=0}^\infty \gamma^t r^i(s_t, a_t)]$ where $\mu$ is an initial state distribution. Depending on how rewards are defined, a POMG can describe different types of multi-agent problems. In cooperative settings~\cite{jeon2022maser, kim2023adaptive}, all agents share the same reward ($r^{1} = \cdots = r^{n}$) and work toward a common goal. In adversarial settings~\cite{ia}, agents have directly conflicting objectives, often modeled as zero-sum games where $\sum_{i \in \mathcal{N}} r^i = 0$. In between, mixed-motive settings~\cite{lase, willis2025quantifying} capture situations where rewards are neither identical nor strictly opposed. Notably, mixed-motive settings give rise to incentives for both cooperation and competition at the same time.

\paragraph{Mixed-Motive Games}
\label{subsec:mmg}
In this paper, we consider mixed-motive games (MMGs), which correspond to the mixed-motive setting of POMG.
%introduced above.
Recent research on MMGs has advanced in a wide range of settings.
A large body of work has examined how agents can achieve cooperation under \textit{decentralized learning}, where each agent updates its policy independently. 
We treat the privacy of individual reward information and the absence of a central coordinator as essential conditions for modeling the independence of autonomous agents in real-world systems.
% \tcr{Think. comm, coordinator ins not allowed. but why gifting can be allowed? any valid example or scenario?}
Several studies adopt reward-shaping methods to promote cooperation. \citet{ia} penalizes advantageous and disadvantageous inequities, while \citet{svo} aligns rewards with social value orientations. Alternatively, opponent-shaping methods steer others' updates by modeling their learning dynamics~\citep{pola,cola,sos} or policies~\citep{loqa,aaa}. However, these approaches rely on access to others' reward information, which is unavailable in our decentralized setting. 
Separate lines of work explore inter-agent communication~\citep{si,ltos}, institutional mechanisms such as contracts or norms~\citep{moca,normative_system,norm_sanction}, or global social-welfare signals~\citep{sl,aga,woojun}. However, these methods are inapplicable to our setting, which excludes inter-agent communication, external institutional structures, and global welfare signals.
In contrast to these approaches, a complementary line of work explores \emph{gifting} mechanisms, where agents share part of their rewards to promote cooperation under decentralized learning.
We describe these methods in the next paragraph.

% %%%%%
% A separate line of work builds on inter-agent communication.
% \citet{si} designed intrinsic rewards from estimated cross-agent policy influence under explicit messaging, 
% while \citet{ltos} learned a networked MARL scheme via bi-level optimization, jointly training a high-level communication module and low-level policies.
% However, these methods depend on message exchange, which is more permissive than our setting.

% %%%%%
% Finally, other studies introduce institutional mechanisms embedded in the environment, such as formal contracts~\citep{moca} or social norms~\citep{normative_system,norm_sanction}, or assume access to a social-welfare signal~\citep{sl,aga,woojun}.
% Our approach does not assume access to social welfare or other agents’ rewards, and thus these methods cannot be directly applied in our setting. 

\paragraph{Gifting in Mixed-Motive Games}
In mixed-motive games, %where agents’ rewards are partly aligned and partly conflicting, \
\emph{gifting} provides a decentralized means of promoting cooperation.
By voluntarily sharing part of its own reward, an agent can encourage cooperative behaviors in others that ultimately improve the giver's own outcome. In doing so, gifting aligns individual and collective incentives in a self-organized manner.
The idea of \emph{gifting} originates from game theory, where it has been used to encourage cooperation in social dilemmas. In such situations, agents must choose between maximizing their own payoff and contributing to the group. Allowing agents to share part of their payoff can help balance this tension by rewarding cooperative behavior and supporting those who act prosocially.
Building on this foundation, recent work in MARL has investigated \emph{gifting} as a way to promote cooperation under decentralized learning.
\citet{lupu2020gifting} first introduced gifting in MARL, and \citet{minimal_reward_transfer} theoretically analyzed it in tabular social dilemmas, formulating a linear-programming-based approach that maximizes each agent’s retained self-reward while alleviating the dilemma.
\citet{lio} learned an additional incentive function, and \citet{lase} proposed an empathy-based gifting mechanism that allocates gifts according to inferred social relationships.
While these studies demonstrated the effectiveness of gifting, they overlooked a critical issue we address in this work: \emph{gift over-reliance}, where agents become dependent on external gifts and fail to develop self-reliant, extrinsic-reward-yielding behaviors. We now describe the general formulation of gifting below.

\noindent\textbf{General Formulation} 
% \begin{comment}
We illustrate a unified formulation of gifting in MMGs. Let $r_t^{i,\text{ext}}$ denote the extrinsic reward of agent $i$. Let $g_t^{i,j}$ denote the gift from agent $i$ to agent $j$ at time $t$, specified by agent $i$'s learnable or rule-based gifting function that maps local information to gifts for others.
Define $g_{t}^{i,-i}:=\sum_{j\neq i} g_t^{i,j}$ and $g_{t}^{-i,i}:=\sum_{j\neq i} g_t^{j,i}$. The shaped reward of agent $i$ is
$r_t^{i,\text{shaped}}=r_t^{i,\text{ext}}-g_t^{i,-i}+g_t^{-i,i}$.
Each self-interested agent updates its policy using this shaped reward, which includes extrinsic reward for self-interest and gift for cooperation. 
The gifting function of each agent is trained to improve the agent's own extrinsic return, as the ultimate goal of gifting is to enhance the giver's extrinsic return through cooperation.
We use this form as a conceptual template that covers prior gifting methods up to implementation differences; details are provided in Appendix~\ref{append:prior_gifting}.

\paragraph{Fairness in Multi-Agent Systems}

%% original ver.
\begin{comment}
In cooperative MARL, it is common to assume \emph{transferable utility}, where the team objective is paramount and the agents are largely indifferent to how the total return is distributed, provided that the aggregate performance is high. In mixed-motive settings, however, agents are self-interested and evaluate outcomes primarily through their own returns. 
In these contexts, it can be problematic if some agents consistently achieve lower extrinsic-reward-grounded performance, even when aggregate performance is high.
This raises fairness concerns specific to mixed-motive settings when agents learn in ways that limit some of them from acquiring reward-yielding skills.
While prior work such as \citet{lase} evaluates fairness using shaped rewards that include gifted rewards, our approach measures fairness solely on extrinsic rewards as in \citet{woojun}. 
We treat gifts as intrinsic incentives for cooperation rather than environmental outcomes, so fairness reflects extrinsic-reward-grounded performance rather than the redistribution of external incentives.
\end{comment}

In cooperative MARL, it is common to assume \emph{transferable utility}, where the team objective is paramount and the agents are largely indifferent to how the total return is distributed, provided that the aggregate performance is high. In mixed-motive settings, however, agents are self-interested and evaluate outcomes primarily through their own returns. A high aggregate return is then not sufficient on its own: it can be problematic if agents learn in ways that limit some of them from acquiring reward-yielding skills, a concern that lies at the level of each individual agent rather than of how returns are distributed. Accordingly, we do not aim to equalize returns across agents; an agent with more favorable circumstances may still earn more than others.
This concern also determines how we evaluate outcomes. While prior work such as \citet{lase} evaluates fairness using shaped rewards that include gifted rewards, we evaluate on extrinsic rewards as in \citet{woojun}, since what an agent earns by itself is what reflects its reward-yielding skills. We treat gifts as intrinsic incentives for cooperation rather than environmental outcomes.

\begin{comment}
In cooperative MARL, it is common to assume \emph{transferable utility}, where the team objective is paramount and the agents are largely indifferent to how the total return is distributed, provided that the aggregate performance is high. In mixed-motive settings, however, agents are self-interested and evaluate outcomes primarily through their own returns. 
% This can be problematic when 
In these contexts, it can be problematic if some agents consistently achieve lower extrinsic-reward-grounded performance, even when aggregate performance is high.
In particular, when agents learn in ways that limit some of them from acquiring reward-yielding skills, even in socially efficient outcomes, this raises fairness concerns that are specific to mixed-motive settings.
While prior work such as \citet{lase} evaluates fairness using shaped rewards that include gifted rewards, our approach measures fairness solely on extrinsic rewards as in \citet{woojun}. 
We treat gifts as intrinsic incentives for cooperation rather than environmental outcomes. Excluding them ensures that fairness reflects extrinsic-reward-grounded performance rather than the redistribution of external incentives.
\end{comment}

\section{Methodology}\label{sec:method}

\subsection{Motivation: Gift Over-Reliance}\label{sec:motivation}

\begin{figure*}[t]
% \begin{figure}[t]
  \centering
      \includegraphics[width=0.85\linewidth]      {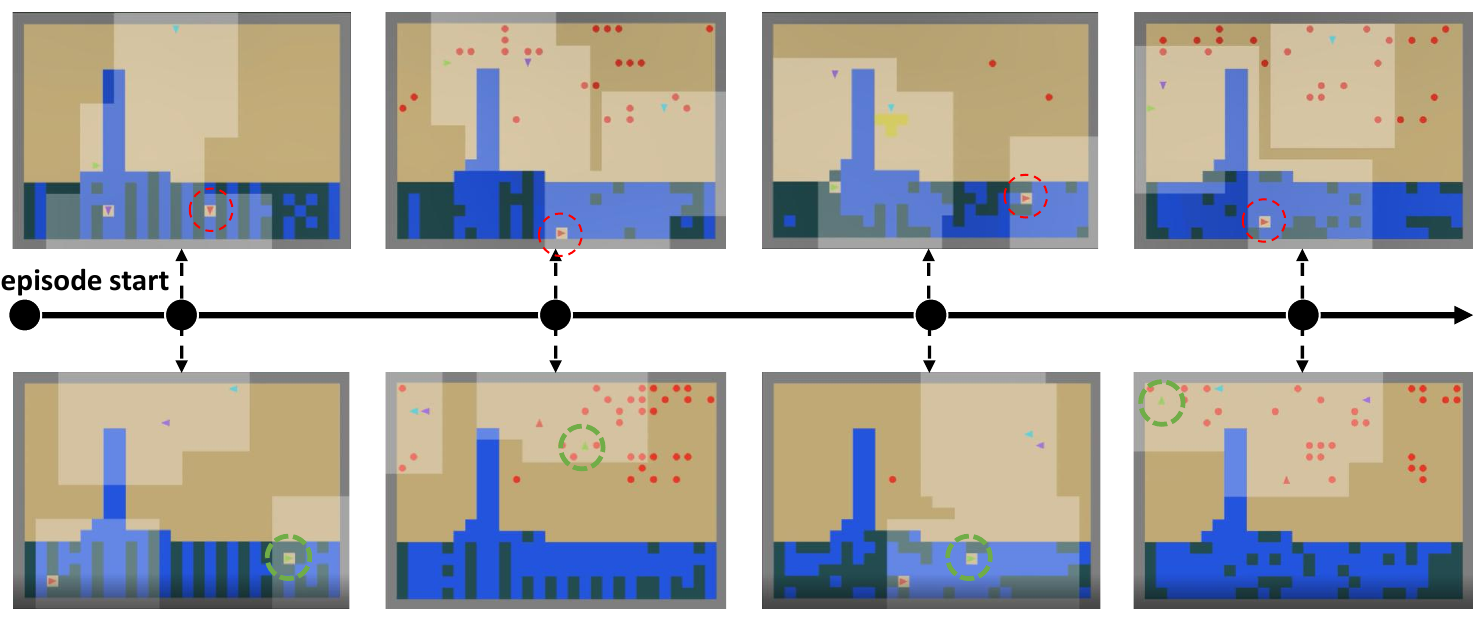}
      \caption{We track the behavior of a disadvantaged agent under different gifting mechanisms.
\textbf{Top}: LASE~\citep{lase}. A disadvantaged agent (red) is locked in cleanup (gift over-reliance).
\textbf{Bottom}: LASE with AGiR (ours). A disadvantaged agent (green) escapes gift over-reliance by alternating between cleanup and harvest, leading to more balanced and sustainable behavior.}
  \label{fig:motivation}
  %\Description{motivation: gift over-reliance}
\vspace{-0.5cm}

% \end{figure}
\end{figure*}

To investigate how existing gifting methods behave under inherent unfairness, we conducted an experiment with LASE~\citep{lase}, one of the recent state-of-the-art gifting methods, on an unfair variant of the original Cleanup environment~\citep{SSDOpenSource}, as shown in Fig.~\ref{fig:motivation}.

% use {figure*} to put image in two columns
% use {figure} to put image in one columns

% The goal of this environment is for each of the four agents (triangles in the figure) to harvest apples (red dots) as many as possible, but apples grow only when the pollution level of the river is below a threshold. Furthermore, apples regrow faster as the river pollution decreases. Initially, the river pollution is above the threshold with no apples in the field, and the pollution level continues to increase with time, so some of the agents should clean the river to reduce the pollution level in order to start and continue harvesting.
% Although the original environment spawns the four agents at four random positions, we modified the environment such that two agents spawn near the river and two others spawn away from the river.

The goal of this environment is for each agent (triangles in the figure) to harvest apples (red dots), which grow only when the river pollution is below a threshold and regrow faster as pollution decreases.
Initially, the pollution level is above the threshold and continues to increase over time, so some agents must clean the river to enable and sustain harvesting.
Although the original environment spawns agents at random positions, we modify it so that two disadvantaged agents spawn near the river and two advantaged agents spawn near the apple spawn sites.

The upper row of Fig.~\ref{fig:motivation} illustrates the behavior of agents trained with LASE.
The disadvantaged agent (the red agent marked with a red dotted circle), which spawns near the river, initiates cleaning due to exploration.
As a result, apples grow in the field and are harvested by other agents.
Those harvesting agents transfer gifts to the disadvantaged agent, allowing it to 
%continue receiving gifts and 
become locked into cleanup without ever harvesting apples throughout the episode.
In other words, during training, the gifts received for cleaning dominate its learning signal before it experiences extrinsic rewards from harvesting.

% This phenomenon, which we term \emph{gift over-reliance}, raises a broader concern for gifting methods:
% in multi-agent systems, \emph{fairness} is critical for enabling agents to defend themselves against exploitation arising from structural asymmetries.
% This motivates our goal of mitigating gift over-reliance to promote fair learning dynamics in mixed-motive settings.
% As a qualitative illustration, the lower row of Fig.~\ref{fig:motivation} shows that this issue can be alleviated.
% Under our approach, the disadvantaged agent is no longer locked into cleanup, but instead learns a more balanced behavior, alternating between socially beneficial cleaning and self-interested harvesting.

This phenomenon, which we term \emph{gift over-reliance}, is not specific to this asymmetric variant; we observe it in the standard environments as well (Appendix~\ref{append:symmetric}).
The issue is neither the cleaning role itself nor the gifts it earns: if the agent had tried harvesting and still chose to clean in return for gifts, this would be a legitimate division of labor.
What we observe instead is that the agent becomes fixed in the cleaning role before it ever experiences harvesting.
Our goal is therefore to enable each agent to escape such lock-in, not to equalize outcomes across agents.
As a qualitative illustration, the lower row of Fig.~\ref{fig:motivation} shows that this is achievable: under our approach, the disadvantaged agent no longer remains locked into cleanup and, having experienced harvesting, alternates between socially beneficial cleaning and self-interested harvesting.
We further examine this distinction under controlled positional and competence asymmetries in Section~\ref{sec:specialization}.

\subsection{Problem Statement}
%: Online Gift Refusal
% \tcb{replace G to another letter?}

We hypothesize that gift over-reliance stems from the agents’ unconditional acceptance of all incoming gifts in standard gifting methods. This lack of selectivity encourages an excessive dependence on gifted rewards at the expense of their own extrinsic performance. We therefore propose that agents selectively regulate gift acceptance based on their learning contexts. Such a mechanism is intended to prevent reward dependence while keeping the learning signal from the environment intact.

To this end, we introduce \emph{adaptive gift refusal}, where agents are given the ability to decide whether to accept or reject incoming gifts. %To address the issue of gift over-reliance, we formulate the problem of {\em online gift refusal}.
For this, we define the refusal function of agent $i$ as $w^i:\mathcal{H}\to [0,1]$, which maps the history of extrinsic rewards and received gifts to the acceptance ratio of the next gift.
Formally, the history space is given by
$\mathcal{H}=\cup_{t=1}^\infty \{(r_0,\ldots,r_{t-1},g_0,\ldots,g_{t-1})\mid r_k, g_k\in\mathbb{R}, 0\le k <t\}$,
where each element corresponds to a finite sequence of extrinsic rewards and received gifts.
At time $t$, we denote the refusal weight of agent $i$ at time $t$ as
$w_t^i := w^i(r_{0:t-1}^i,g_{0:t-1}^{-i,i})$,
where $r_{0:t-1}^i$ is the extrinsic reward history of agent $i$ and $g_{0:t-1}^{-i,i}$ is the sequence of gifts received from other agents.
Thus, $w_t^i$ specifies the fraction of the incoming gift $g_t^{-i,i}$ that agent $i$ accepts.
We augment the shaping term with refusal and  the resulting shaped reward is given by
% \begin{equation}\label{eq:refuse_reward}
% r_t^{i,\text{AGiR}}
% = r_t^{i,\text{ext}}
% - g_t^{i,-i}(\eta^i)
% + w_t^i(r_{0:t-1}^i,\,g_{0:t-1}^{-i,i})\cdot g_t^{-i,i}(\eta^{-i}).
% \end{equation}
% \begin{equation}\label{eq:refuse_reward}
% r_t^{i,\text{AGiR}}
% = r_t^{i,\text{ext}}
% - g_t^{i,-i}(\eta^i)
% + w_t^i\cdot g_t^{-i,i}(\eta^{-i}).
% \end{equation}
\begin{equation}\label{eq:refuse_reward}
r_t^{i,\text{AGiR}}
= r_t^{i,\text{ext}}
- g_t^{i,-i}
+ w_t^i\cdot g_t^{-i,i}.
\end{equation}
Given Eq.~\eqref{eq:refuse_reward}, \emph{our objective is to design an adaptive refusal weight computing strategy that operates in real time on local interaction history to mitigate gift over-reliance while preserving task performance}.

% \tcb{TODO. the problem is not stated yet.\\
% e.g. design real-time weight computing strategy}

% \tcb{about terminology: w is actually acceptance rate, equivalently refuse (1-w) when w is not 1 $\to$ refusal weight? acceptance weight}

\subsection{Adaptive Gift-Refusal Mechanism}\label{sec:main_method}
We now propose an \textbf{A}daptive \textbf{Gi}ft-\textbf{R}efusal (\textbf{AGiR}) mechanism that updates the refusal weight of each agent $i$. The key idea is that each agent computes a \emph{reward--gift orientation} (RGO) to assess its own level of gift reliance, and then adjusts weight $w_t^i$ to regulate gift acceptance and mitigate over-reliance. Concretely, comparing $\mathrm{RGO}$ to a baseline $B$ determines the \emph{direction} of change: if $\mathrm{RGO}<B$ (indicative of gift over-reliance), the agent \emph{decreases} $w_t^i$ to place more emphasis on extrinsic reward; if $\mathrm{RGO}>B$, it \emph{increases} $w_t^i$ to accept more gifts and thereby support cooperative behavior. In short, the signed magnitude of $\mathrm{RGO}-B$ determines the direction and size of the update, while a step-size modulator can further bound it for stability. The pseudo-code is provided in Appendix~\ref{append:pseudo_code}.

\noindent\textbf{Reward-Gift Orientation $\phi_t^i$ and Baseline $B_t^i$  }
To quantify the gift dependency of agent $i$ at time $t$, 
we define the RGO $\phi_t^i$ as the angle formed by the ratio between the cumulative reward and the cumulative gift until time $t$:
\vspace{-0.5em}
\begin{align}
    \phi_t^i := \arctan\!\left(\frac{R_t^i}{G_t^i}\right) \in (-\frac{\pi}{2},\frac{\pi}{2}),
\end{align}
% \begin{align}
%     \phi_t^i := \arctan\!\left(R_t^i/G_t^i\right) \in (-\pi/2,\pi/2),
% \end{align}
where $R_t^i := \sum_{k=t-H+1}^{t} r_k^{i,\text{ext}}$ denotes the cumulative extrinsic reward and $G_t^i := \sum_{k=t-H+1}^{t} g_k^{-i,i}$ denotes the cumulative gift received within a window size $H$.
This formulation permits any window size $H$, from the full history ($H=t+1$) to fixed windows over recent history.

$\phi_t^i$ reflects how strongly agent $i$ has relied on gifts relative to rewards for policy update. If $\phi_t^i \approx 0$, the reliance on gifts has overwhelmed rewards. In this case, until time $t$, agent $i$ has been discouraged from increasing its own extrinsic reward, but instead has engaged in prosocial or sacrificial behavior that benefits others and attracts gifts. In other words, the agent is in danger of gift over-reliance, and thus should reduce $w_t^i$ to down-weight the importance of gifts.
To implement this idea, we compare the orientation $\phi_t^i$ against a baseline $B_t^i$ and update the refusal weight $w_t^i$ accordingly. The baseline $B_t^i$ determines the \emph{sign} of the change in refusal weight: if $\phi_t^i - B_t^i > 0$, then $w_t^i$ increases, while if $\phi_t^i - B_t^i < 0$, then $w_t^i$ decreases. Moreover, the \emph{magnitude} of the change follows the signed deviation $\phi_t^i - B_t^i$ (with an independent step-size control for stability), so larger deviations induce proportionally larger adjustments while preserving the direction set by the baseline comparison. The baseline can be fixed or varying in time. We present our choice for the baseline in Appendix~\ref{append:practical_algo}.

% \tcb{add image like below?}\\
% \includegraphics[scale=0.3]{image/rgo_diagram_tmp.PNG}\\
% space issue

\noindent\textbf{Refusal Weight Update  }
We update the refusal weight according to
$w_{t+1}^i = w_t^i + \alpha\,\Delta_t^i$,
with step size $\alpha>0$, where $\Delta_t^i$ captures the change applied at time $t$.
% We denote $\Delta_t^i$ as the change in refusal weight, with $w_{t+1}^i = w_t^i + \alpha\,\Delta_t^i$ for step size $\alpha>0$.
Following the discussion above, we set the update to be proportional to the deviation of RGO from the baseline and scale it by the current weight and a reward-based step-size modulator:
% \begin{align}\label{eq:w_update}
% \Delta_t^i \;=\; w_t^i \cdot \underbrace{\big(\phi_t^i - B_t^i\big)}_{\text{gift reliance level}} \cdot \underbrace{S(R_t^i)}_{\substack{\text{step-size}\\ \text{modulator}}}.
% \end{align}
\begin{align}\label{eq:w_update}
\Delta_t^i ~=~ w_t^i ~\times \underbrace{\big(\phi_t^i - B_t^i\big)}_{\text{gift reliance level}} ~~~~\times \underbrace{S(R_t^i)}_{\text{step-size modulator}}.
\end{align}
Here, $S(R_t^i)$ dampens the update when the cumulative reward is high, reflecting the intuition that agents with sufficient returns have less need to rapidly change their refusal behavior:
% $S(R_t^i) \;:=\; 2\,\sigma\!\left(-\frac{R_t^i}{C}\right)$,
$S(R_t^i) := 2\sigma (-R_t^i/C )$,
where $\sigma(x)=\frac{1}{1+e^{-x}}$ and $C>0$ is a temperature parameter. Consequently, $S(0)=1$ and $S(R_t^i)$ decreases monotonically toward $0$ as $R_t^i$ grows, with larger $C$ yielding a slower decay.

% \subsection{Algorithmic Instantiation}
% \label{subsec:algo}
% \tcb{add PG equation?}\\

% \noindent\textbf{Refusal-aware policy.}
% We parameterize the policy as $\pi_\theta(a \mid o, w)$, allowing the refusal weight to enter as an additional input so that action selection is conditioned on the current observation $o$ and on the agent’s \emph{gift-acceptance stance} encoded by $w$. Intuitively, changing $w$ reshapes the agent’s objective. With $w=1$, incoming gifts enter the return with full weight, making actions that elicit or sustain others’ gifts more attractive. With $w=0$, on the other hand, gifts are ignored and the agent prioritizes actions that raise its own extrinsic reward. Hence, even under the same observation, the same agent will choose different actions at $w=1$ versus $w=0$.
% Conditioning the policy on $w$ therefore allows a single policy to adapt its behavior across refusal contexts. 
% In implementation, we treat $w$ as side information and concatenate it to the observation features before the policy network (equivalently, a contextual policy); this preserves decentralized learning while enabling behavior to vary with $w$.

\noindent\textbf{Refusal-Aware Policy  }
We parameterize the policy as $\pi_\theta(a \mid o, w)$, where the refusal weight $w$ is provided as an additional input alongside the observation $o$.
The scalar $w$ encodes the agent’s \emph{gift-acceptance stance}: when $w=1$, incoming gifts are fully accepted, encouraging actions that elicit others’ gifts; when $w=0$, gifts are ignored and the agent prioritizes extrinsic reward.
Conditioning on $w$ thus allows a single policy to adapt its behavior across refusal contexts.
\section{Theoretical Analysis}

In this section, we analyze AGiR along two axes. Along the rollout axis, with the joint policy profile and gifting functions held fixed, we show that the refusal-weight process $\{w^i_t\}_{t\geq 0}$ admits a stationary distribution and identify a sufficient condition for its uniqueness. Along the policy-update axis, we analyze how a policy update under the AGiR-updated refusal weight affects welfare in the Iterated Prisoner's Dilemma (IPD).
We first state standard assumptions for existence.
\begin{assumption}\label{main:assumption_1}
    The global state space $S\subset\mathbb{R}^d$ and
each agent's action space $A^i$ are finite.
\end{assumption}

\begin{assumption}[Continuity]\label{main:assumption_2}
    For each agent $i$, the baseline $B^i$ is continuous
in $w^i\in[0,1]$, and the policy
$\pi^i(a^i\mid s,w^i)$ is continuous in $w^i$
for every $s\in S$ and $a^i\in A^i$.
\end{assumption}

\begin{theorem}[Existence]\label{thm:existence}
For arbitrary fixed joint policy profile $\pi = (\pi^1, \dots, \pi^n)$ and gifting functions $\{G_{\eta^i}\}_{i=1}^n$, let $\{w_t^i\}_{t \ge 0}$ be the stochastic process of refusal weights for agent $i$ generated by AGiR with any fixed finite window size $H\ge1$. Under Assumptions~\ref{main:assumption_1} and~\ref{main:assumption_2}, for each agent $i$, this process admits a stationary marginal distribution.
\end{theorem}
\emph{Proof.} See Appendix~\ref{append:proof_existence} for the proof.

Intuitively, under a fixed joint policy, the refusal weights admit a stationary marginal distribution. We next identify conditions for its uniqueness and convergence.

\begin{assumption}\label{main:assumption_4}
For every agent $i$, the parameters satisfy $\alpha\in(0,\frac{1}{2\pi})$, $B^i,\phi^i(s,a)\in(-\frac{\pi}{2},\frac{\pi}{2})$, and $r^i(s,a)\in[-r_{\max},r_{\max}]$ for all $s,a$, where $B^i$ is independent of the refusal weights.
\end{assumption}

\begin{assumption}[Positive-Regime Stability]\label{main:assumption_5}
Under a fixed joint policy, let $Z=(s,\mathbf{w})$, $m^i(s,a)=1+\alpha S(r^i(s,a))(\phi^i(s,a)-B^i)$, and $Z_*=(s_*,\mathbf{1})$ for some $s_*\in S$. Assume:

(A) \textit{Low-weight recovery}: For some $\eta\in(0,1)$ and $\kappa>0$, $\mathbb{E}[\log m^i(s,a)\mid Z]\ge\kappa$ whenever $w^i\le\eta$.

(B) \textit{Uniform regeneration}: For every $c\in(0,1)$, there exist $N_c\ge1$ and $\beta_c>0$ such that $\Pr(Z_{t+N_c}=Z_*\mid Z_t=Z)\ge\beta_c$ uniformly over $w^i\ge c$ for all $i$, with $\Pr(Z_{t+1}=Z_*\mid Z_t=Z_*)>0$.

Formal conditions are given in Appendix~\ref{append:proof_uniqueness}.
\end{assumption}

\begin{theorem}[Stability and Uniqueness]\label{thm:uniqueness}
Suppose Assumptions~\ref{main:assumption_1}, \ref{main:assumption_2}, \ref{main:assumption_4}, and \ref{main:assumption_5} hold with $H=1$ and $w_0^i>0$ for all agents $i$. Then, the joint AGiR process admits a unique stationary distribution in the positive-weight regime and converges to it geometrically in total variation.
\end{theorem}
\emph{Proof.} See Appendix~\ref{append:proof_uniqueness} for the proof.

% Beyond existence, the uniqueness theorem characterizes when the long-run behavior settles into a single steady regime. Specifically, the signal $\phi$ governs this adjustment: $\phi > B$ drives $w$ upward, while $\phi \le B-\delta$ triggers a downward correction. Assumption~\ref{main:assumption_5} formalizes this push--pull balance by requiring the process to return to lower regions frequently enough to prevent runaway growth in gift reliance. This ensures that the dynamics converge to a unique stationary distribution.

Beyond existence, the uniqueness theorem characterizes conditions under which the refusal weights converge to a single steady regime. Assumption~\ref{main:assumption_5} captures two complementary mechanisms: low-weight recovery prevents the refusal weights from collapsing toward zero, while uniform regeneration allows the joint process to reach a common state, jointly ensuring uniqueness and convergence. 
% Under an additional condition permitting occasional downward updates, the stationary distribution assigns positive probability to both full and partial gift acceptance.

% Theorems~\ref{thm:existence} and~\ref{thm:uniqueness} characterize the long-run behavior of the refusal weights generated by AGiR under a fixed joint policy profile. On the policy-update axis, characterizing welfare under simultaneous policy learning in general environments and base gifting methods is challenging; we therefore analyze the IPD, a canonical model of social dilemmas. With Reward Exchange~\cite{minimal_reward_transfer} as the base gifting method and REINFORCE-style policy updates~\cite{reinforce}, a policy update under the AGiR-updated refusal weight improves the respective welfare measure, either the collective sum or the minimum return, against fixed cooperative, adversarial, and reciprocal opponent strategies. Theorem~\ref{thm:welfare_improvement_append} in Appendix~\ref{append:ipd} gives the formal statement and proof. Extending this analysis to the joint learning dynamics of general mixed-motive games is left for future work. Please refer to Appendix~\ref{append:theory_practice} for a discussion bridging theory and practice, including the empirical validity of our assumptions.

Theorems~\ref{thm:existence} and~\ref{thm:uniqueness} characterize the long-run behavior of the refusal weights generated by AGiR under a fixed joint policy profile. On the policy-update axis, characterizing welfare under simultaneous policy learning in general environments and base gifting methods is challenging; we therefore analyze the IPD, a canonical model of social dilemmas. With Reward Exchange~\cite{minimal_reward_transfer} as the base gifting method and REINFORCE-style policy updates~\cite{reinforce}, a policy update under the AGiR-updated refusal weight improves the respective welfare measure, either the collective sum or the minimum return, against fixed cooperative, adversarial, and reciprocal opponent strategies. Theorem~\ref{thm:welfare_improvement_append} in Appendix~\ref{append:ipd} gives the formal statement and proof. Extending this analysis to the joint learning dynamics of general mixed-motive games is left for future work. 
Please refer to Appendix~\ref{append:theory_practice} for a discussion of how our theoretical assumptions relate to practical settings.

\vspace{-0.5em}    %% 2 Thm in main body, 1 thm in append
\section{Experiments}\label{sec:experiments}

\vspace{-0.5em}

\subsection{Environments}

\begin{wrapfigure}[7]{r}{0.43\textwidth}
\vspace{-1.5em}
  \centering

  % \vspace{-14pt} 

  \includegraphics[width=\linewidth]{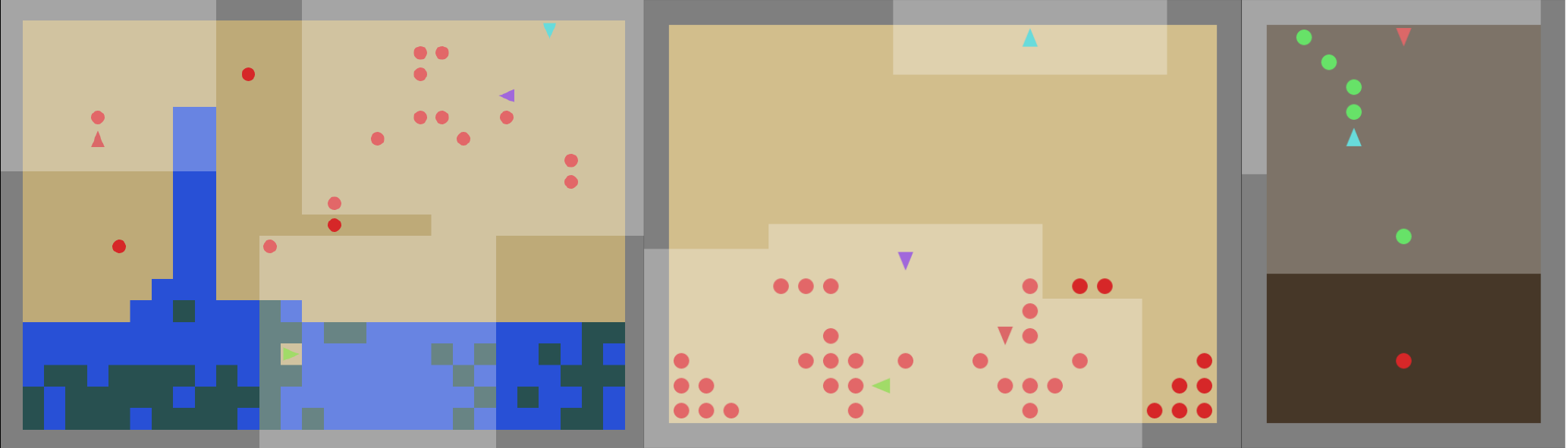}
  
  % \vspace{-1ex}
  \caption{Unfair Cleanup, Unfair Harvest, and Unfair Coin Game.}
  % \caption{Illustrations of the environments: Unfair Cleanup, Unfair Harvest, and Unfair Coin Game.}
  
  \label{fig:envs}
  
  % \vspace{-10pt}
\end{wrapfigure}
To evaluate our method, we adapt the SocialJax benchmark \citep{socialjax}, a JAX-based framework for MMGs. We use unfair variants of Coin Game, Cleanup, and Harvest by inducing structural asymmetry among agents, following \citet{woojun}. Such asymmetry leads to disadvantaged agents who may become dependent on gifts, and can enable advantaged agents to exploit this dependence to increase their own returns. 
We provide a brief summary of each environment below, with full details in Appendix~\ref{append:env_detail}.

\noindent\textbf{Unfair Cleanup } In the original Cleanup environment, agents collect apples for reward, but apple regrowth depends on maintaining low levels of environmental pollution through cleaning actions that provide no direct individual reward, giving rise to a social dilemma between self-interest and cooperation. We introduce structural inequality by placing two agents near the river (disadvantaged) and two near apple sites (advantaged). Since cleaning is required for apple regrowth, disadvantaged agents are biased toward a cleaning role. Under gifting, they may become locked into a sacrificial strategy, cleaning throughout the episode while relying on gifts instead of learning to harvest.

\begin{comment}
\begin{figure}[t]
  \centering
  \includegraphics[width=0.9\linewidth]{image/envs.pdf}
  \caption{Illustrations of the environments: Unfair Cleanup, Unfair Harvest, and Unfair Coin Game.}
  \vspace{-2ex}
  \label{fig:envs}
\end{figure}
\end{comment}

\noindent\textbf{Unfair Harvest } In the original Harvest environment, agents collect apples from orchard patches where regrowth is stochastic and depends on local resource density, creating a commons dilemma that requires implicit coordination to sustain long-term returns. We introduce structural inequality by placing two advantaged agents near the orchard and two disadvantaged agents farther away. Under gifting, disadvantaged agents may avoid competing for resources and instead rely on gifts from advantaged agents, failing to learn independent harvesting behaviors.

\noindent\textbf{Unfair Coin Game } The Coin Game extends the Iterated Prisoner’s Dilemma, where collecting a coin yields $+1$, while collecting the opponent’s coin incurs a $-2$ penalty to the opponent. We introduce structural inequality via asymmetric coin spawn probabilities, creating advantaged and disadvantaged agents. 
Under gifting, part of the $-2$ penalty is transferred as a negative gift to the advantaged agent, deterring it from taking the disadvantaged agent’s coins.

\subsection{Metrics}

In MMGs, prior work typically reports performance using separate metrics for efficiency (e.g., social welfare) and inequality (e.g., Gini), making their trade-off hard to assess at a glance. We therefore summarize the resulting outcomes with $\alpha$-fairness~\citep{mo2002fair}, a unified metric that jointly captures efficiency and inequality. A higher $\alpha$-fairness indicates both higher social welfare and less inequality, with $\alpha$ controlling their relative emphasis. We note that $\alpha$-fairness serves as a summary of outcomes rather than as the objective of AGiR. Full metric definitions are provided in Appendix~\ref{append:metrics}.

$\alpha$-fairness recovers several representative objectives, where $x_i$ denotes the return of agent $i$: as $\alpha\to\infty$, maximizing $\alpha$-fairness corresponds to maximizing the minimum return $\min_i x_i$ (\textbf{max-min fairness}, denoted \textbf{Min}); at $\alpha=1$, it corresponds to maximizing the geometric mean $\big(\prod_i x_i\big)^{1/n}$ (\textbf{proportional fairness}, denoted \textbf{GM}); and at $\alpha=0$, it corresponds to maximizing the arithmetic mean $\frac{1}{n}\sum_i x_i$ (\textbf{collective return}, denoted \textbf{AM}). Among these, Min, the lowest extrinsic return among agents, most directly reflects whether every agent learns to earn for itself. For completeness, we also report the \textbf{Gini} coefficient~\cite{david1968miscellanea}, while results for Jain's index~\cite{jain} are provided in Appendix~\ref{append:jain}.

Furthermore, we define the \textbf{Gift Reliance Ratio (GRR)} as $\mathrm{GRR}:=\frac{R^+}{E^+ + R^+}$, where $E$ and $R$ denote cumulative extrinsic rewards and received gifts, respectively, and $x^+=\max\{0,x\}$. 
While $\alpha$-fairness captures resulting outcomes, GRR captures reliance on gifts: a higher GRR indicates that a larger share of an agent's income comes from gifts. We use GRR as an auxiliary metric to quantify gift reliance and to assess its change under AGiR. We report the average GRR across all agents.

\subsection{Baselines}
We evaluate our method by plugging it into prior gifting methods: Reward exchange (RX)~\citep{minimal_reward_transfer}, LIO~\citep{lio}, and LASE~\citep{lase}.
For a broader context, we also report non-gifting methods for MMGs: independent PPO~\citep{ippo} with a common collective reward, denoted IPPO-C, and with individual rewards, denoted IPPO-I, as well as the reward-shaping approach IA~\citep{ia}. Note that IA requires access to other agents’ rewards during training, which is not allowed in our setting.

\subsection{Results}
For completeness, evaluation results on the standard environments are provided in Appendix~\ref{append:symmetric}.

% \begin{wrapfigure}[7]{r}{0.45\textwidth}
% \vspace{-1.4em}

% \centering
% \captionsetup{type=table,skip=0pt}
% \captionof{table}{$\alpha$-fairness, Gini coefficient, and GRR in Unfair Cleanup-Hard}
% \label{table_cleanup_hard}
% \tiny
% \setlength{\tabcolsep}{3pt}
% \resizebox{\linewidth}{!}{%
% \begin{tabular}{llrrrrr}
% \toprule
% \multicolumn{2}{l}{Algorithm} & AM $\uparrow$ & GM $\uparrow$ & Min $\uparrow$ & Gini $\downarrow$ & GRR $\downarrow$ \\
% \hline\hline
% \addlinespace[2pt]
% \multicolumn{2}{l}{IPPO-I} & 42 (84) & 42 (84) & 37 (74) & 0.01 (0.02) & -\\
% \multicolumn{2}{l}{IPPO-C} & 678 (24) & 0 (0) & 0 (0) & 0.52 (0.01) & - \\
% \multicolumn{2}{l}{IA}     & 226 (110) & 222 (108) & 175 (82) & 0.08 (0.03) & - \\
% \hline\hline
% \addlinespace[2pt]
% \multirow{2}{*}{LASE} & Vanilla & \textbf{314} (35) & 131 (161) & 112 (138) & 0.22 (0.12) & 0.25 \\
%                       & AGiR    & 288 (41) & \textbf{283} (42) & \textbf{221} (46) & \textbf{0.09} (0.01) & \textbf{0.07} \\
% \midrule
% \multirow{2}{*}{RX}   & Vanilla & \textbf{381} (61) & 273 (143) & 191 (113) & 0.19 (0.08) & 0.37 \\
%                       & AGiR    & 357 (33) & \textbf{352} (33) & \textbf{288} (29) & \textbf{0.09} (0.01) & \textbf{0.20} \\
% \bottomrule
% \end{tabular}%
% }
% \end{wrapfigure}

\noindent\textbf{Unfair Cleanup } 
We evaluate two difficulty levels, \emph{Easy} and \emph{Hard}, defined by the distance between the river and apple spawn sites. We focus on the \emph{Hard} setting, where distant spawns make exploration difficult and exacerbate gift over-reliance. Results for the \emph{Easy} setting are provided in Appendix~\ref{append:cleanup_easy}. Following prior studies~\citep{ltos}, LIO was omitted due to learning instabilities in Cleanup.

\begin{figure}[!t]
    \centering
    \begin{minipage}[t]{0.49\linewidth}
        \vspace{0pt}
        \centering
        \captionsetup{type=table,skip=2pt}
        \captionof{table}{$\alpha$-fairness, Gini coefficient, and GRR in Unfair Cleanup-Hard}
        \label{table_cleanup_hard}

        \tiny
        \setlength{\tabcolsep}{3pt}
        \resizebox{\linewidth}{!}{%
            \begin{tabular}{llrrrrr}
\toprule
\multicolumn{2}{l}{Algorithm} & AM $\uparrow$ & GM $\uparrow$ & Min $\uparrow$ & Gini $\downarrow$ & GRR $\downarrow$ \\
\hline\hline
\addlinespace[2pt]
\multicolumn{2}{l}{IPPO-I} & 42 (84) & 42 (84) & 37 (74) & 0.01 (0.02) & -\\
\multicolumn{2}{l}{IPPO-C} & 678 (24) & 0 (0) & 0 (0) & 0.52 (0.01) & - \\
\multicolumn{2}{l}{IA}     & 226 (110) & 222 (108) & 175 (82) & 0.08 (0.03) & - \\
\hline\hline
\addlinespace[2pt]
\multirow{2}{*}{LASE} & Vanilla & \textbf{314} (35) & 131 (161) & 112 (138) & 0.22 (0.12) & 0.25 \\
                      & AGiR    & 288 (41) & \textbf{283} (42) & \textbf{221} (46) & \textbf{0.09} (0.01) & \textbf{0.07} \\
\midrule
\multirow{2}{*}{RX}   & Vanilla & \textbf{381} (61) & 273 (143) & 191 (113) & 0.19 (0.08) & 0.37 \\
                      & AGiR    & 357 (33) & \textbf{352} (33) & \textbf{288} (29) & \textbf{0.09} (0.01) & \textbf{0.20} \\
\bottomrule
\end{tabular}%
        }
    \end{minipage}
    \hfill
    \begin{minipage}[t]{0.49\linewidth}
        \vspace{0pt}
        \centering
        \includegraphics[width=\linewidth]{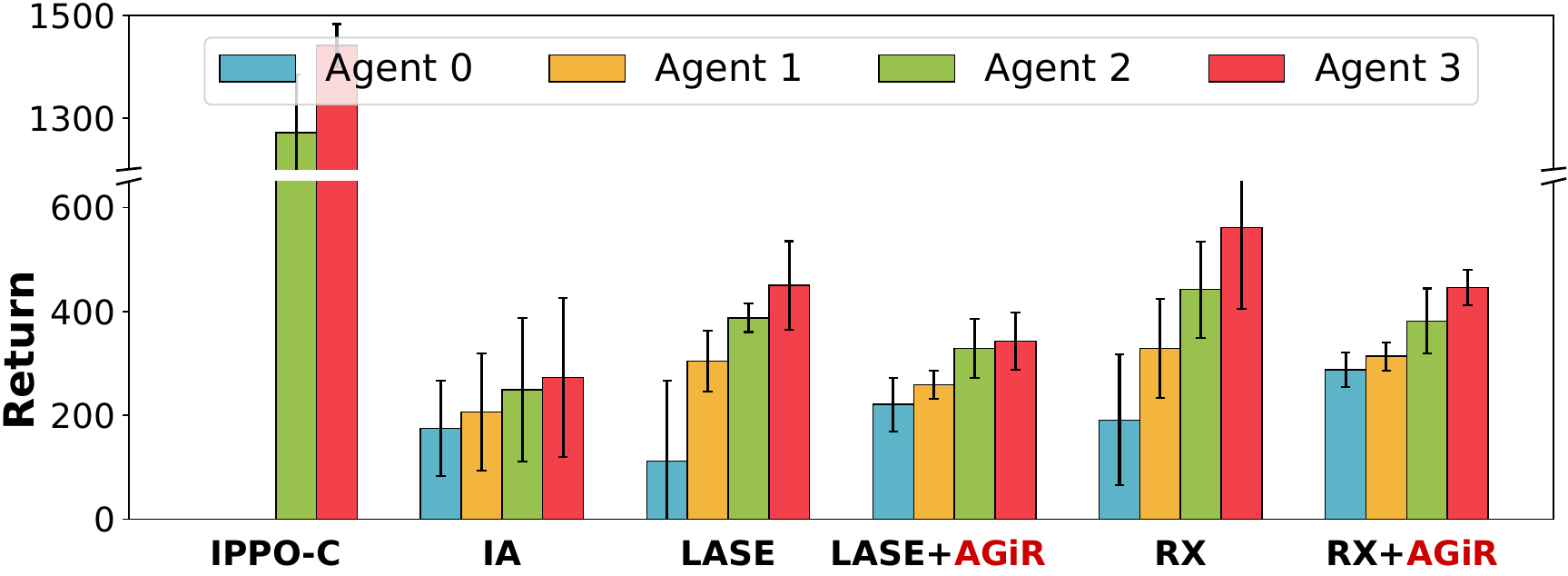}
        \captionsetup{type=figure,skip=2pt}
        % \captionsetup{skip=0pt}
        \captionof{figure}{Per-agent return distributions in Unfair Cleanup-Hard.}
        \label{fig:cleanup_agent_bar_hard}
    \end{minipage}
    % \vspace{-0.5em}
\end{figure}

Table~\ref{table_cleanup_hard} reports the results. Under vanilla gifting, disadvantaged agents near the river become locked into cleaning while relying on gifts from 
advantaged agents, leading to limited harvesting performance due to gift-dominated learning. 
In contrast, AGiR modulates incoming gifts
% \begin{wrapfigure}[8]{r}{0.45\textwidth}
% \vspace{-1em}
% \centering
% \includegraphics[width=\linewidth]{image/cleanup_hard_agents_bars_redrawn.pdf}%
% \captionsetup{type=table,skip=0pt}
% \captionof{figure}{Per-agent return distributions in Unfair Cleanup-Hard.}
% \label{fig:cleanup_agent_bar_hard}
% \end{wrapfigure}
to prevent such dominance, encouraging disadvantaged agents to explore away from the river and learn policies that balance cooperative cleaning with harvesting. Thus, AGiR raises Min and lowers GRR without a substantial loss in AM; GM and the Gini coefficient improve accordingly.
%% message: Min increased & while keeping AM

Fig.~\ref{fig:cleanup_agent_bar_hard} shows per-agent returns in Unfair Cleanup-Hard, where agents are ranked by return within each seed and averaged across five seeds. Across both LASE and RX, adding AGiR visibly raises the returns of the lower-ranked agents and narrows the gap to the best-ranked agents, consistent with the improvements in Min and GM reported in Table~\ref{table_cleanup_hard}. In contrast, IPPO-C achieves high AM but poor GM and Min, while IA achieves lower inequality at the cost of overall performance.

% \begin{wrapfigure}[16]{r}{0.45\textwidth}
% \vspace{-1.2em}
% \centering
% \captionsetup{type=table,skip=0pt}
% \captionof{table}{$\alpha$-fairness, Gini coefficient, and GRR in Unfair Harvest}
% \label{table_harvest}
% \tiny
% \setlength{\tabcolsep}{3pt}
% \resizebox{\linewidth}{!}{%
% \begin{tabular}{llrrrrr}
% \toprule
% \multicolumn{2}{l}{Algorithm} & AM $\uparrow$ & GM $\uparrow$ & Min $\uparrow$ & Gini $\downarrow$ & GRR $\downarrow$ \\
% \hline\hline
% \addlinespace[2pt]
% \multicolumn{2}{l}{IPPO-I} & 49 (0.1) & 43 (0.6) & 24 (1.4) & 0.24 (0.01) & - \\
% \multicolumn{2}{l}{IPPO-C} & 115 (8.7) & 11(17.1) & 2 (3.2) & 0.71 (0.06) & - \\
% \multicolumn{2}{l}{IA}     & 49 (0.1) & 43 (0.3) & 24 (0.7) & 0.25 (0.01) & - \\
% \hline\hline
% \addlinespace[2pt]
% \multirow{2}{*}{LASE} & Vanilla & \textbf{51} (1.1) & 29 (16.6) & 11 (9.1) & 0.35 (0.09) & 0.31 \\
%                       & AGiR    & 49 (0.3) & \textbf{41} (1.8) & \textbf{19} (4.5) & \textbf{0.27} (0.03) & \textbf{0.12} \\
% \midrule
% \multirow{2}{*}{RX}   & Vanilla & \textbf{61} (4.3) & \textbf{43} (3.7) & 15 (5.0) & 0.36 (0.03) & 0.43 \\
%                       & AGiR    & 49 (0.4) & 42 (0.7) & \textbf{21} (2.1) & \textbf{0.26} (0.01) & \textbf{0.25} \\
% \bottomrule
% \end{tabular}
% }

% \includegraphics[width=\linewidth]{image/harvest_agents_bars_redrawn.pdf}
% \captionof{figure}{Per-agent return distributions in Unfair Harvest.}
% \label{fig:harvest_agent_bar}

% \end{wrapfigure}

\begin{figure}[H]
    \centering

    % Left: Table 2
    \begin{minipage}[t]{0.49\linewidth}
        \vspace{0pt}
        \centering
        \captionsetup{type=table,skip=2pt}
        \captionof{table}{$\alpha$-fairness, Gini coefficient, and GRR in Unfair Harvest}
        \label{table_harvest}

        \tiny
        \setlength{\tabcolsep}{3pt}
        \resizebox{\linewidth}{!}{%
            \begin{tabular}{llrrrrr}
\toprule
\multicolumn{2}{l}{Algorithm} & AM $\uparrow$ & GM $\uparrow$ & Min $\uparrow$ & Gini $\downarrow$ & GRR $\downarrow$ \\
\hline\hline
\addlinespace[2pt]
\multicolumn{2}{l}{IPPO-I} & 49 (0.1) & 43 (0.6) & 24 (1.4) & 0.24 (0.01) & - \\
\multicolumn{2}{l}{IPPO-C} & 115 (8.7) & 11(17.1) & 2 (3.2) & 0.71 (0.06) & - \\
\multicolumn{2}{l}{IA}     & 49 (0.1) & 43 (0.3) & 24 (0.7) & 0.25 (0.01) & - \\
\hline\hline
\addlinespace[2pt]
\multirow{2}{*}{LASE} & Vanilla & \textbf{51} (1.1) & 29 (16.6) & 11 (9.1) & 0.35 (0.09) & 0.31 \\
                      & AGiR    & 49 (0.3) & \textbf{41} (1.8) & \textbf{19} (4.5) & \textbf{0.27} (0.03) & \textbf{0.12} \\
\midrule
\multirow{2}{*}{RX}   & Vanilla & \textbf{61} (4.3) & \textbf{43} (3.7) & 15 (5.0) & 0.36 (0.03) & 0.43 \\
                      & AGiR    & 49 (0.4) & 42 (0.7) & \textbf{21} (2.1) & \textbf{0.26} (0.01) & \textbf{0.25} \\
\bottomrule
\end{tabular}
        }
    \end{minipage}
    \hfill
    % Right: Figure 4
    \begin{minipage}[t]{0.49\linewidth}
        \vspace{0pt}
        \centering
        \includegraphics[width=\linewidth]{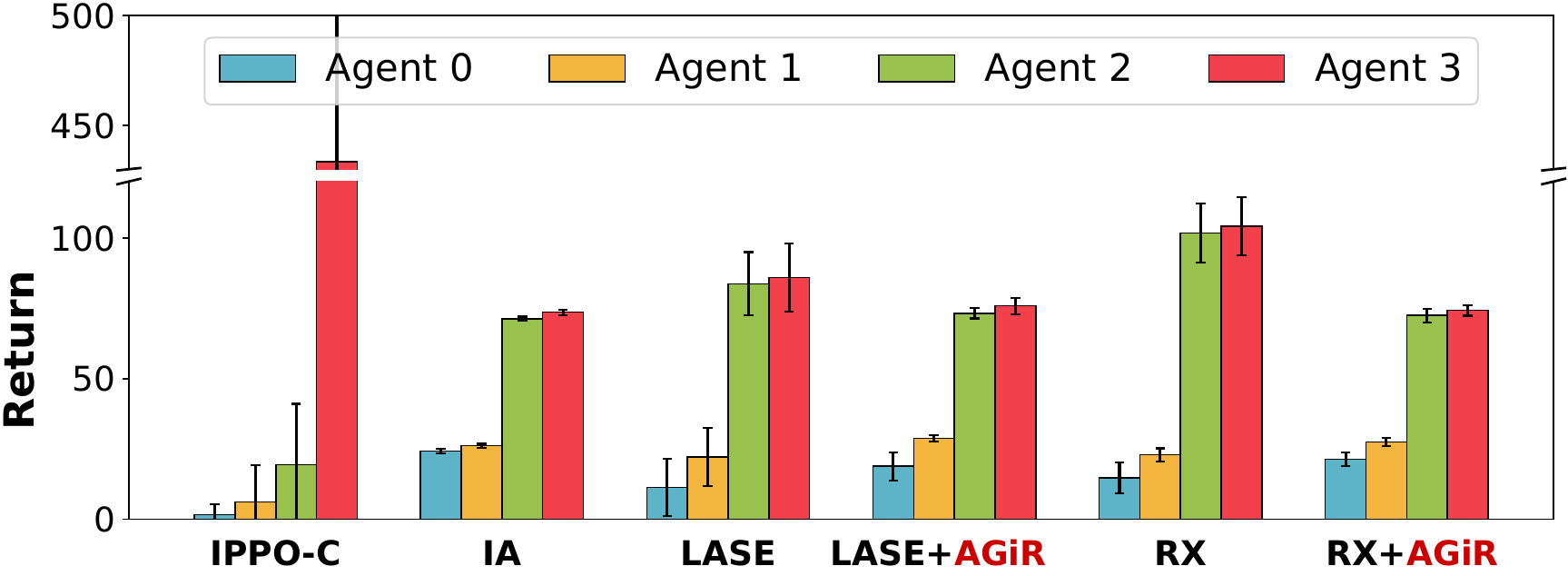}
        \captionsetup{type=figure,skip=2pt}
        \captionof{figure}{Per-agent return distributions in Unfair Harvest.}
        \label{fig:harvest_agent_bar}
    \end{minipage}
    \vspace{-0.5em}
\end{figure}

\noindent\textbf{Unfair Harvest } In the Unfair Harvest, two advantaged agents are initialized near the orchard and two disadvantaged agents farther away. Table~\ref{table_harvest} summarizes the results for this environment. 
As in Cleanup, LIO was excluded due to learning instabilities.

% As in the Unfair Cleanup environment, LIO failed to train in the Unfair Harvest environment and was therefore excluded.

Gift over-reliance allows advantaged agents to occupy the orchard early and send gifts that dampen disadvantaged agents' motivation to explore, leading to poor learning outcomes for harvesting. 
AGiR addresses this issue, raising Min while maintaining or improving GM.
Fig.~\ref{fig:harvest_agent_bar} shows the distribution of per-agent returns in Unfair Harvest, averaged over five seeds.
Compared to baseline gifting methods, AGiR raises the returns of the lower-ranked agents, reducing polarization across agents. In contrast, IPPO-C yields highly unequal outcomes. In each seed, one agent monopolizes the orchard, yielding a high AM but poor performance in terms of GM and Min return.

\begin{wrapfigure}[12]{r}{0.45\textwidth}
\vspace{-1.7em}
\centering
    \centering
    \captionsetup{type=table,skip=0pt}
    \caption{$\alpha$-fairness and GRR in Unfair Coin Game}
    \label{table_coingame_main}
    \tiny
    \setlength{\tabcolsep}{2pt} % 열 간격을 줄여서 폭 확보
    \resizebox{\linewidth}{!}{%
    \begin{tabular}{llrrrr}
    \toprule
    \multicolumn{2}{l}{Algorithm} & AM $\uparrow$ & GM $\uparrow$ & Min $\uparrow$ & GRR $\downarrow$ \\
    \hline\hline
    \addlinespace[2pt]
    % \multicolumn{2}{l}{SVO}    & 39.0 (0.5) & -329.3 (3.4) & -292.6 (3.5) & 468.9 (4.69) \\
    \multicolumn{2}{l}{IPPO-I} & 11 (0.9) & -66 (1.7) & -56 (1.8) & - \\
    \multicolumn{2}{l}{IPPO-C} & 97 (0.6) & 78 (1.7) & 40 (0.5) & - \\
    \multicolumn{2}{l}{IA}     & 12 (0.9) & -63 (0.4) & -53 (4.7) & - \\
    \hline\hline
    \addlinespace[2pt]
    \multirow{2}{*}{LASE} & Vanilla & 15 (33.1) & -61 (84.3) & -65 (107.3) & 0.56 \\
                          & AGiR  & \textbf{20} (1.3) & \textbf{-57} (3.4) & \textbf{-40} (3.9) & \textbf{0.54} \\
    \midrule
    \multirow{2}{*}{RX}   & Vanilla & \textbf{97} (1.0) & \textbf{78} (0.6) & 39 (0.9) & 0.39 \\
                          & AGiR   & 77 (4.7) & 74 (3.5) & \textbf{59} (1.5) & \textbf{0.28} \\
    \midrule
    \multirow{2}{*}{LIO}   & Vanilla & \textbf{-22} (11.2) & -128 (31.2) & -152 (43.8) & 0.5 \\
                          & AGiR   & -24 (17.7) & \textbf{-105} (73.3) & \textbf{-132} (93.1) & \textbf{0.30} \\
    \bottomrule
    \end{tabular}
}
\end{wrapfigure}

\noindent\textbf{Unfair Coin Game } 
In the Unfair Coin Game, the red agent is advantaged because red coins are rarer (about 20\% of spawns), reducing its risk of theft penalties and increasing opportunities to penalize the green agent. Under vanilla gifting, negative rewards may lead to negative gifts, which act as peer penalization. These punitive transfers are often beneficial in this environment as they discourage agents from taking opponent-colored coins and induce cooperative behavior. A natural concern is that a refusal mechanism might undermine this enforcement. However, our results show that AGiR avoids this risk, preserving the corrective power of gifting while improving the return of the disadvantaged agent.

% \begin{comment}
\begin{center}
\vspace{-0.4em}
% 1열: 3장
\begin{minipage}[t]{0.30\textwidth}\vspace{0pt}
  \includegraphics[width=0.9\linewidth,keepaspectratio]{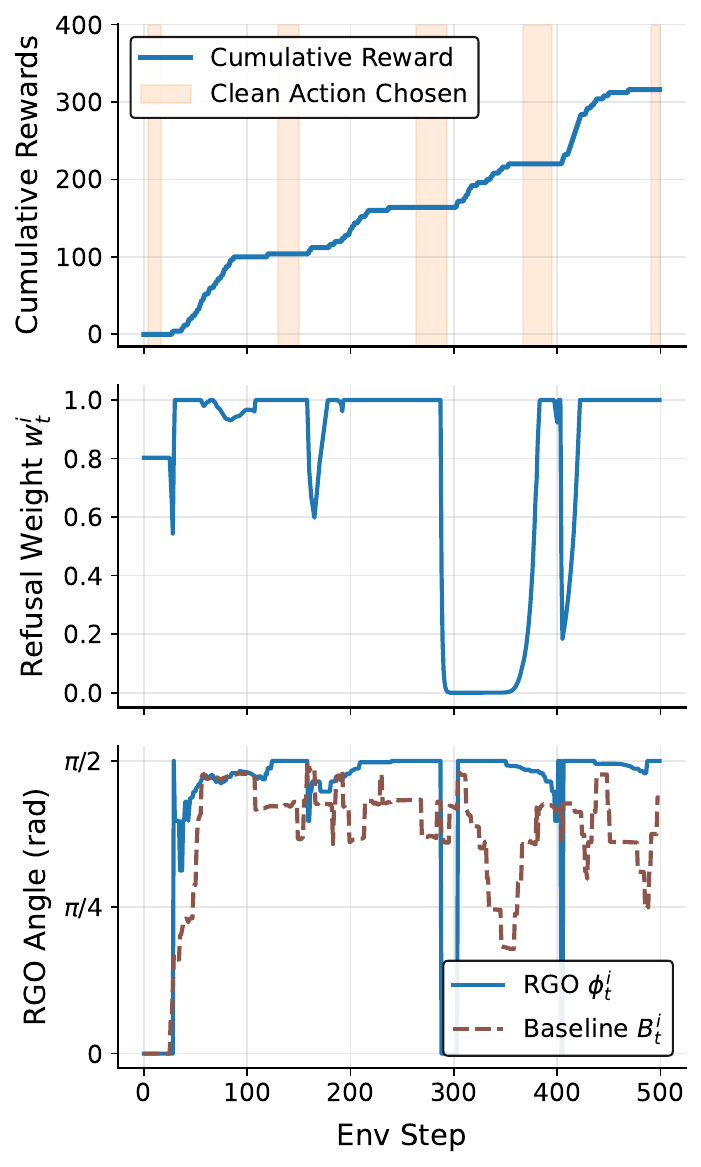}%
\end{minipage}\hfill%
\begin{minipage}[t]{0.17\textwidth}\vspace{0pt}
  \includegraphics[width=0.9\linewidth,height=\Hfour,keepaspectratio]{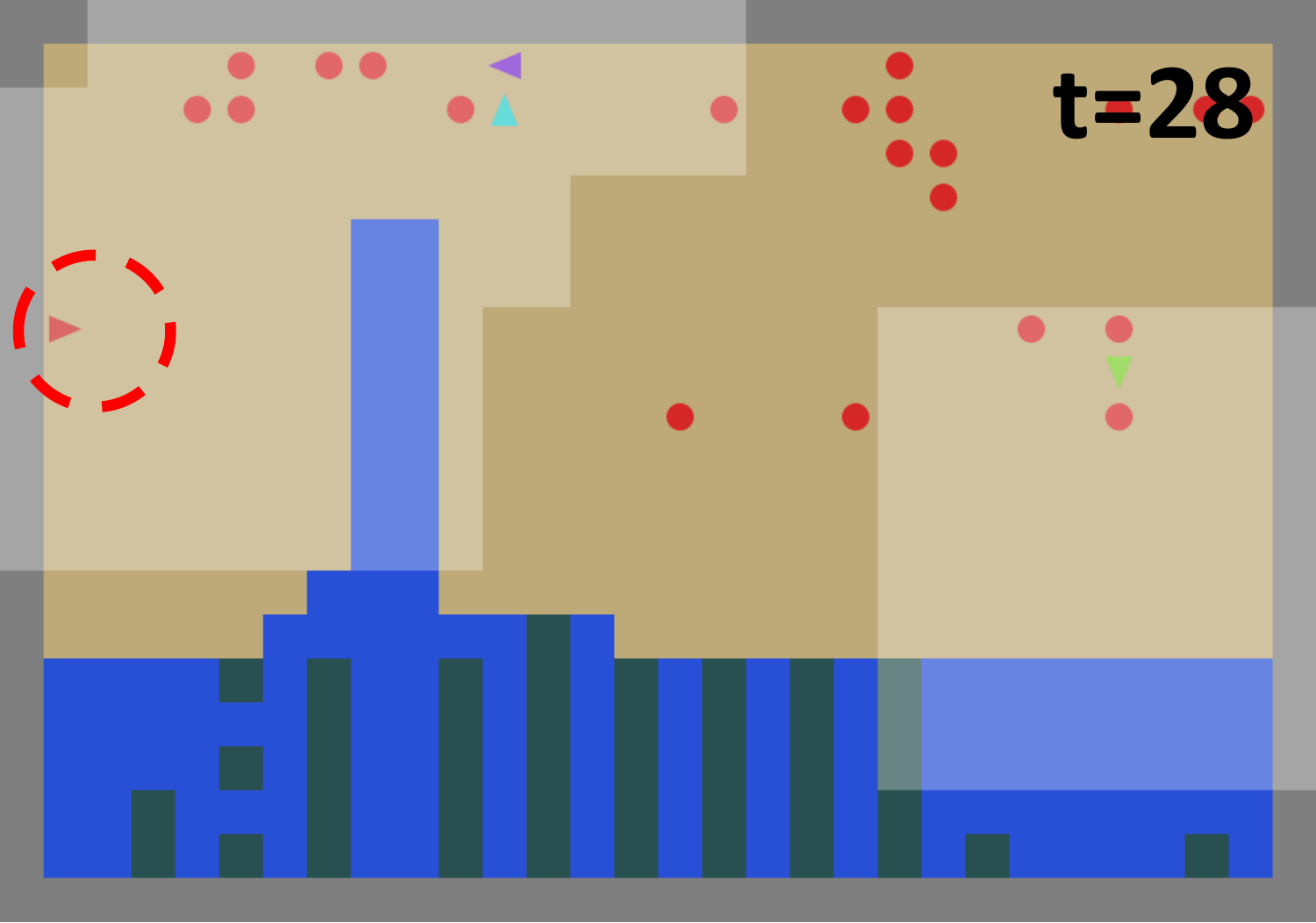}\vspace{\imggap}
  \includegraphics[width=0.9\linewidth,height=\Hfour,keepaspectratio]{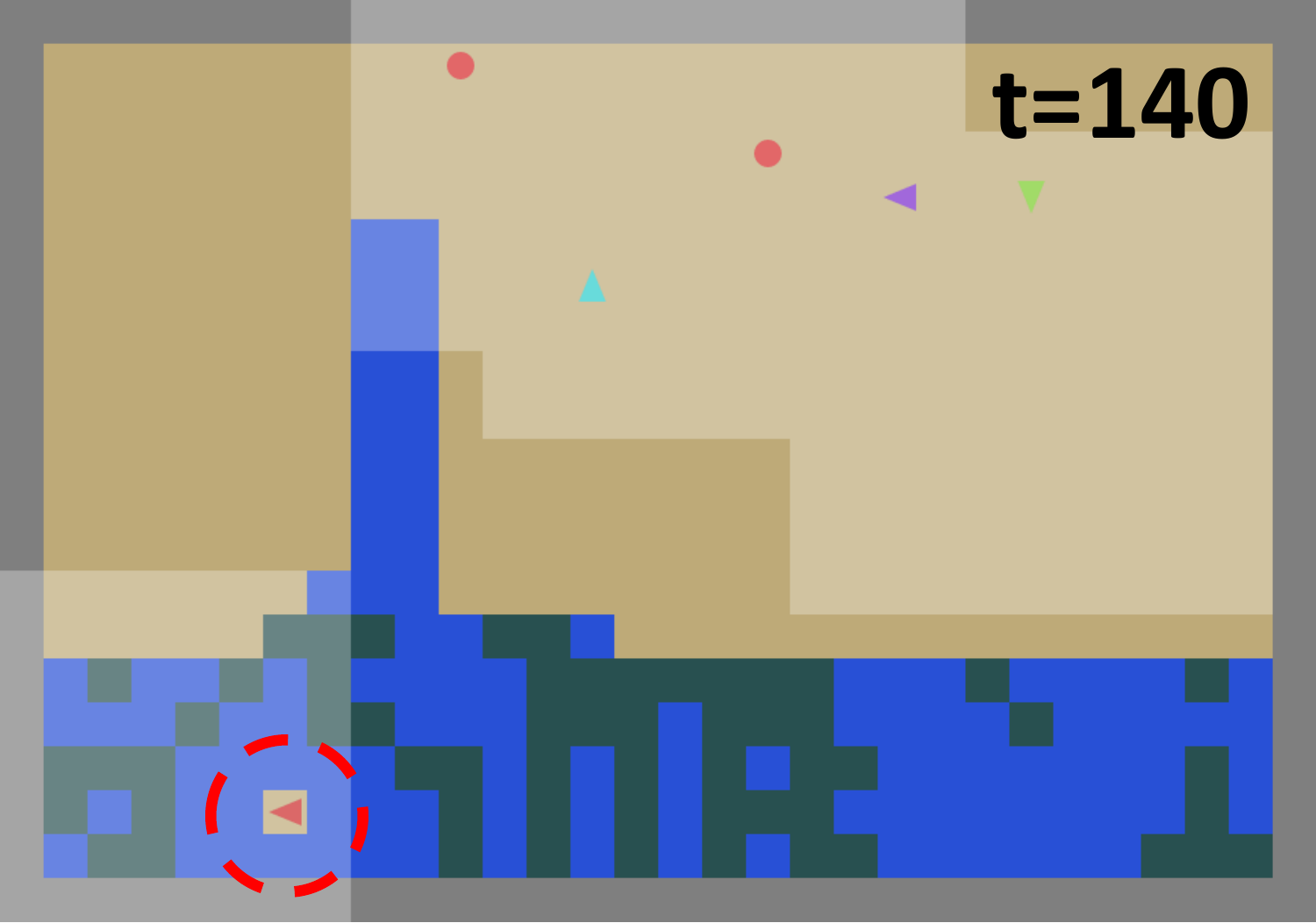}\vspace{\imggap}
  \includegraphics[width=0.9\linewidth,height=\Hfour,keepaspectratio]{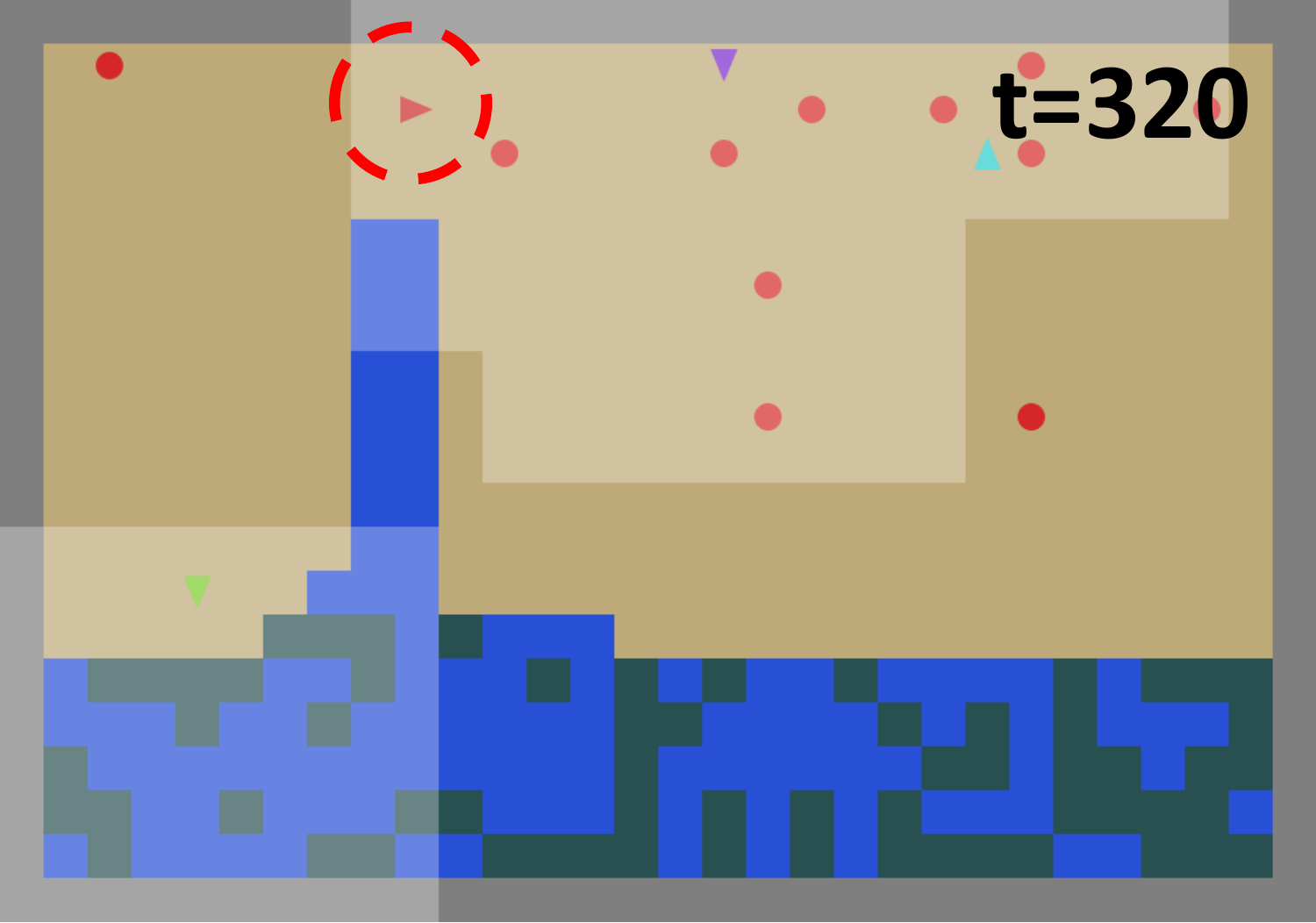}\vspace{\imggap}
  \includegraphics[width=0.9\linewidth,height=\Hfour,keepaspectratio]{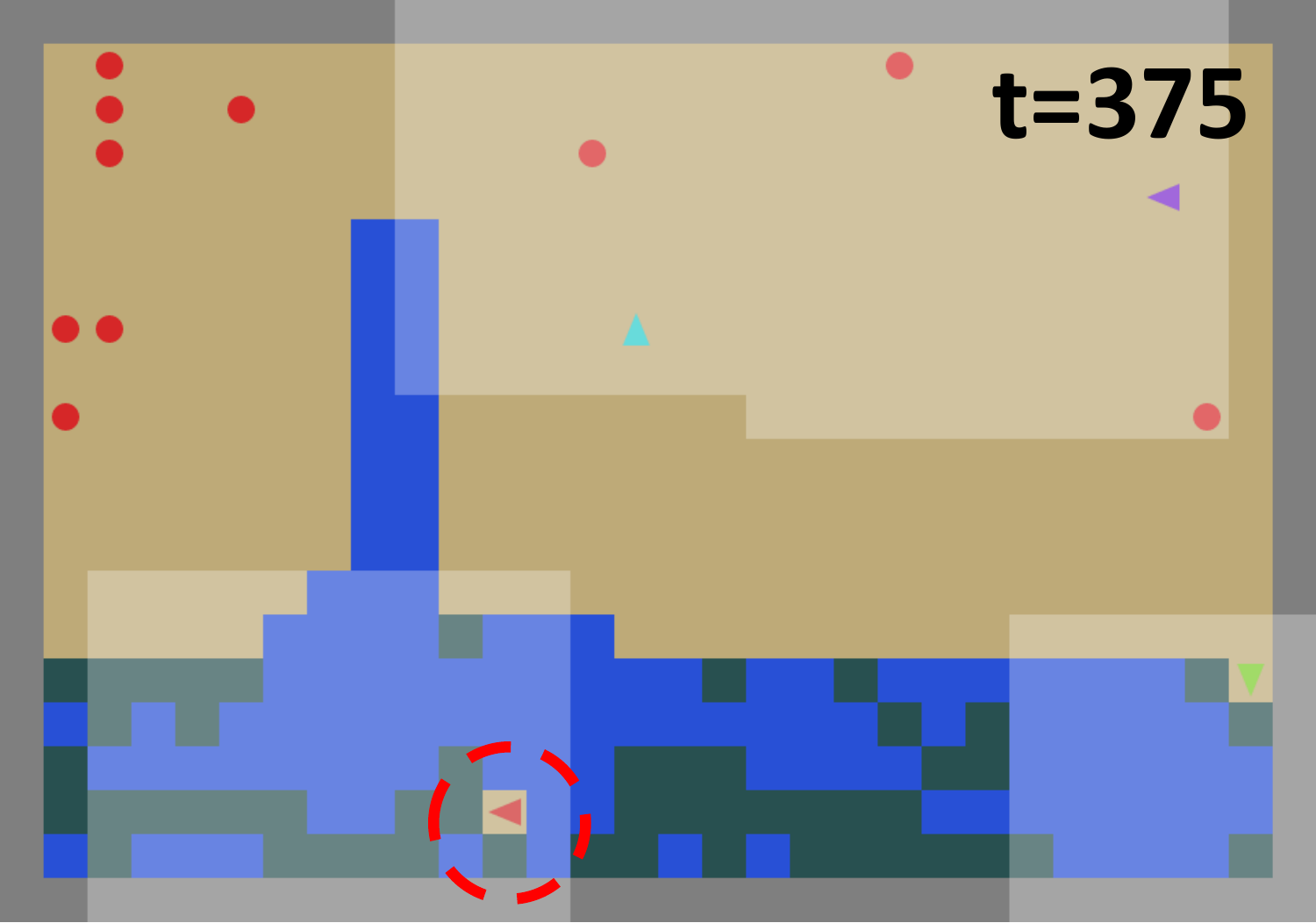}%
\end{minipage}\hfill%
\begin{minipage}[t]{0.30\textwidth}\vspace{0pt}
  \includegraphics[width=0.9\linewidth,keepaspectratio]{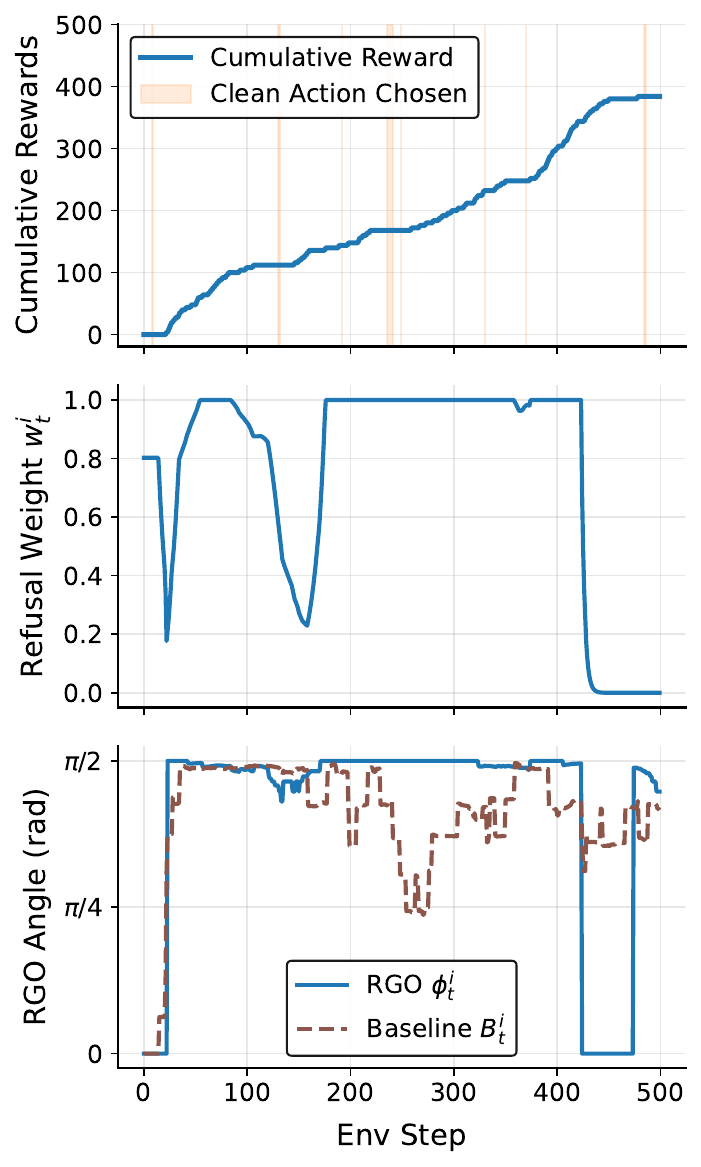}%
\end{minipage}\hfill%
\begin{minipage}[t]{0.17\textwidth}\vspace{0pt}
  \includegraphics[width=0.9\linewidth,height=\Hfour,keepaspectratio]{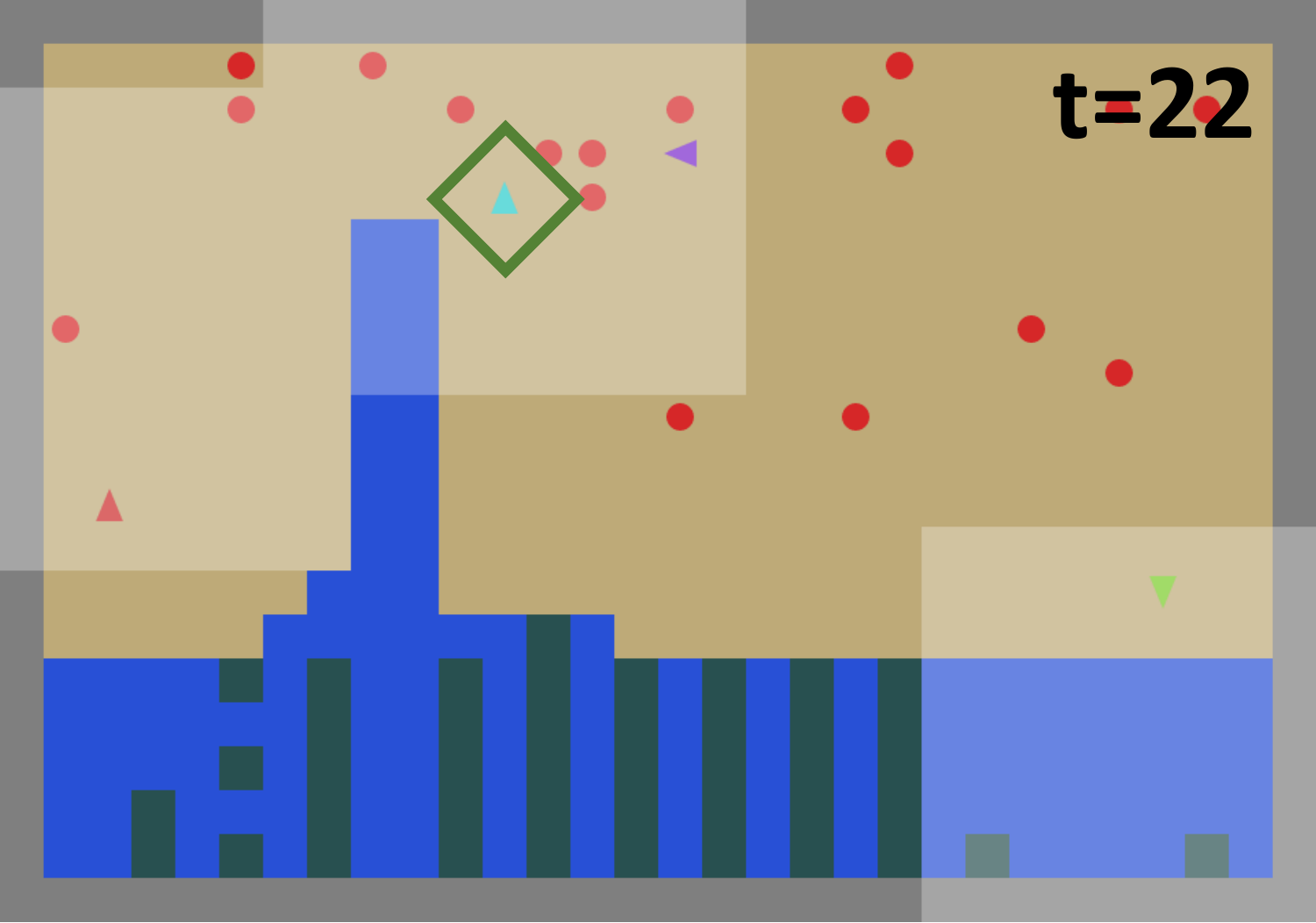}\vspace{\imggap}
  \includegraphics[width=0.9\linewidth,height=\Hfour,keepaspectratio]{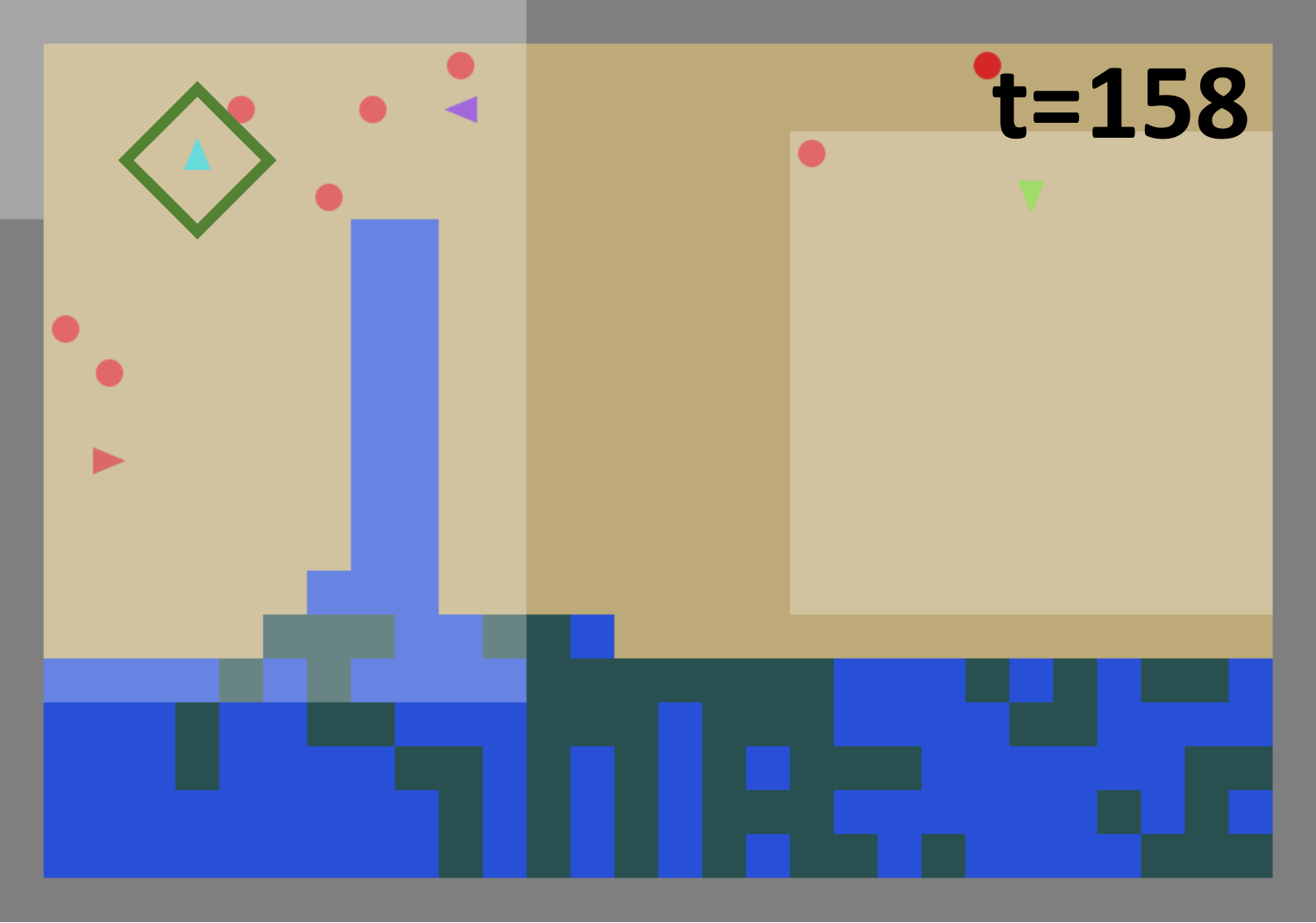}\vspace{\imggap}
  \includegraphics[width=0.9\linewidth,height=\Hfour,keepaspectratio]{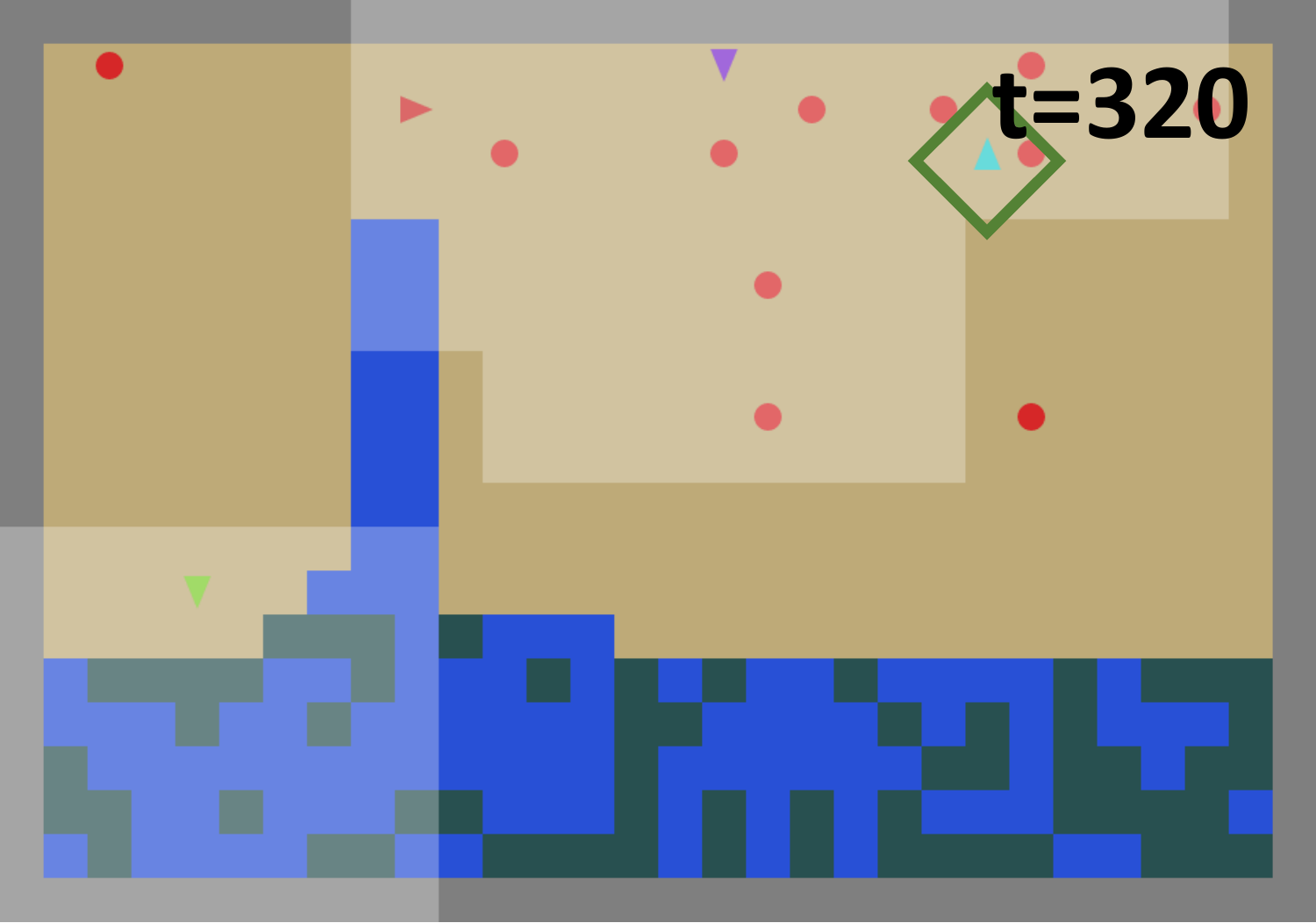}\vspace{\imggap}
  \includegraphics[width=0.9\linewidth,height=\Hfour,keepaspectratio]{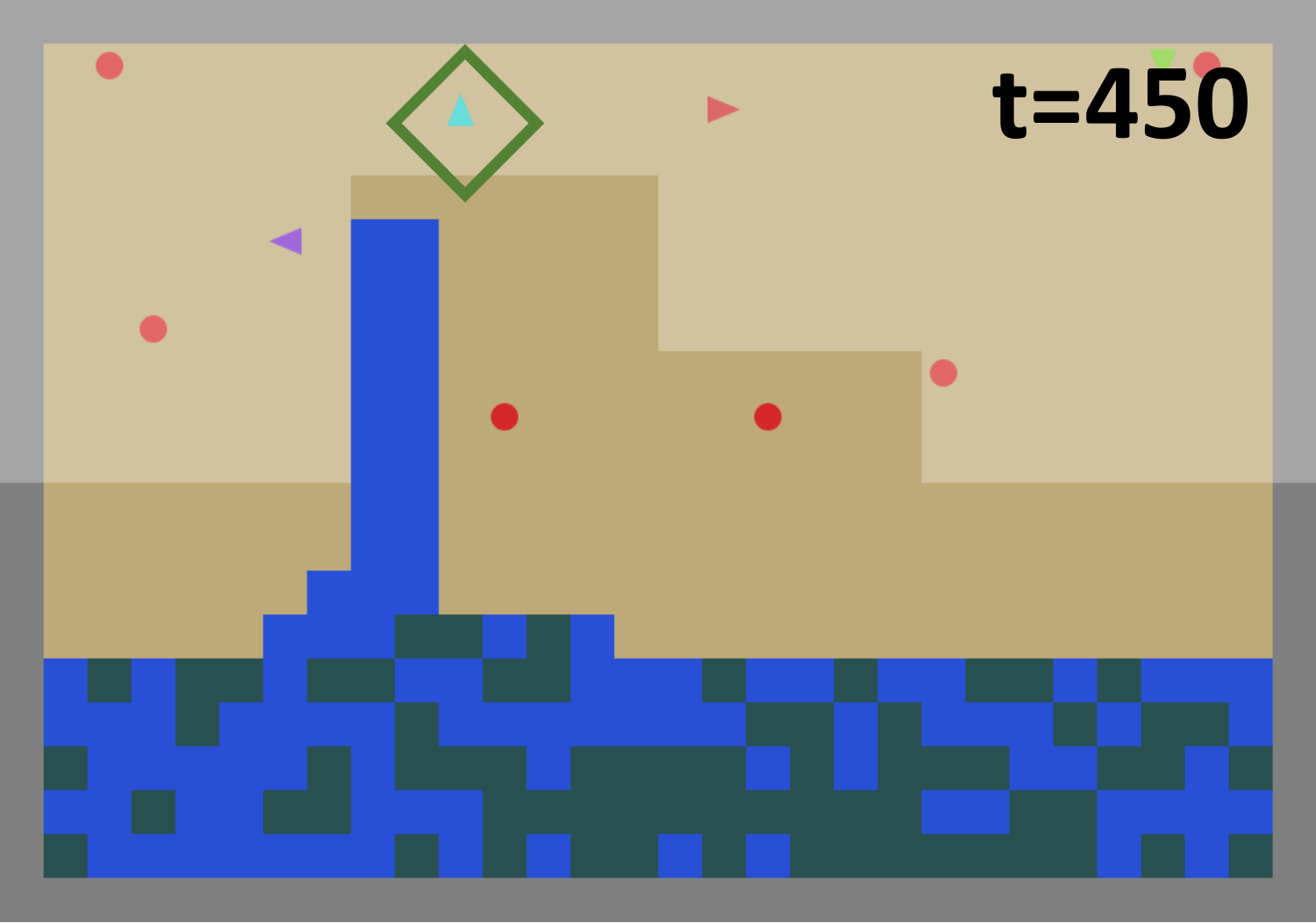}%
\end{minipage}

\captionsetup{type=figure,skip=0pt}
\captionof{figure}{Episode summary in Unfair Cleanup-Hard. Left: disadvantaged agent; Right: advantaged agent. For each agent, three time-series panels (bottom$\to$top): 
RGO with its baseline (dashed brown);
refusal weight $w_t^i$;
and cumulative extrinsic reward $\sum_{k\le t} r_k^{i,ext}$ with orange shading for clean actions. Snapshots show selected timesteps (timestamp at top-right). The focal agent is marked by a red dashed circle for the disadvantaged agent and a green solid diamond for the advantaged agent.}
\label{fig:in_epi_analysis}
\end{center}
% \end{comment}

Table~\ref{table_coingame_main} reports the results. Adding refusal improves Min, the return of the disadvantaged green agent, across all base gifting methods. Since a weakened deterrent would let the advantaged agent take more of the green agent's coins and thus lower Min, this rise indicates that refusal does not weaken the deterrent effect of negative gifts; the disadvantaged agent instead ends up better off.

\section{Analysis}\label{sec:analysis}

\subsection{Empirical Analysis on Within-Episode Dynamics}\label{sec:within_epi_dyn}

We analyze within-episode dynamics of refusal weights $w_t^i$, RGOs $\phi_t^i$, cleanup actions, and cumulative extrinsic reward in Unfair Cleanup-Hard. Fig.~\ref{fig:in_epi_analysis} shows a 500-step evaluation episode after training LASE + AGiR for $3\times10^8$ timesteps, comparing disadvantaged (left) and advantaged (right) agents.

We first examine the red disadvantaged agent (left column). At $t=28$, the refusal weight drops sharply; subsequently, the clean actions become sparse, snapshots show the agent leaving the river 
to harvest, and the cumulative return increases. Around $t=140$, the refusal weight remains near $1$,
 corresponding to frequent cleaning and flat cumulative return. Around $t=320$, the refusal weight is near $0$, snapshots show harvesting, and the cumulative return rises. At $t=375$, the refusal weight increases rapidly to $1$, accompanied by a return to cleaning behavior.
Overall, the disadvantaged agent exhibits prosocial behavior while still effectively pursuing its self-interest throughout the episode.

We next examine the cyan advantaged agent (right column). This agent’s behavior is largely insensitive to refusal weights. The cumulative extrinsic return rises steadily, as evidenced by snapshots across varying refusal regimes (e.g., sharp $w$-drops at $t=22, 158$; near-one $w$ at $t=320$; and near-zero $w$ at $t=450$). These snapshots consistently show the agent harvesting frequently with little cleaning. 
This agent maintains its payoff advantage throughout the episode, showing that AGiR does not pressure advantaged agents to sacrifice their returns to equalize outcomes.

\vspace{-0.1em}

\subsection{Behavior under Positional and Competence Asymmetries}
\label{sec:specialization}

%% P0

Section~\ref{sec:motivation} argued that the problem is not specialization itself but a role that becomes fixed before the agent experiences alternative behaviors. This raises a natural question: does AGiR dissolve only such lock-in, or legitimate specialization as well? We examine this with controlled experiments that separate the factors behind role assignment.

%% P1
\begin{wraptable}[7]{r}{0.38\textwidth}
% \vspace{-1.5em}
\centering
\captionsetup{type=table,skip=0pt}
\caption{Four scenarios in two-agent Cleanup.}
\label{tab:two_agent}
\scriptsize
\begin{tabularx}{\linewidth}{@{}l
  >{\raggedright\arraybackslash\hsize=0.8\hsize}X
  >{\raggedright\arraybackslash\hsize=1.2\hsize}X@{}}
\toprule
 & Equal competence & Asymmetric competence \\
\midrule
Random spawn & No gap & Competence gap \\
\addlinespace[3pt]
Fixed spawn & Spawn gap & Mismatch \\
\bottomrule
\end{tabularx}
\vspace{-1em}
\end{wraptable}

We consider two factors that can shape which agent cleans and which harvests: \emph{spawn position}, which affects which task an agent experiences first, and \emph{competence}, which affects how well it performs each task. Crossing the two in a two-agent Cleanup yields four scenarios (Table~\ref{tab:two_agent}). Agents spawn either at random or at fixed positions, one by the river and the other by the apples. Their competences are either equal or asymmetric: one agent is better at cleaning, with a wider cleaning beam, and the other is better at harvesting, with a larger reward per apple. We call them the \emph{talented cleaner} and the \emph{talented harvester}. 
We use LASE as the base gifting method. Further experimental details, quantitative results, and learning curves are provided in Appendix~\ref{append:two_agent}.

%% P2

\noindent\textbf{No gap.}
Without either asymmetry, LASE shows no lock-in in this setting, as both agents clean and harvest; accordingly, AGiR leaves its outcome essentially unchanged.

\noindent\textbf{Competence gap.} AGiR allows the talented cleaner to explore harvesting before settling on a role. When its harvesting reward is only $0.1\times$ that of the talented harvester, it eventually returns to cleaning; at $0.25\times$, it continues harvesting alongside cleaning. Thus, AGiR allows agents to retain specialization when it is beneficial rather than forcing them to abandon it.

% \noindent\textbf{Competence gap.}
% AGiR gives the talented cleaner the opportunity to try a new role, harvesting, and lets it decide for itself whether to keep it: the agent keeps harvesting when it pays off, and otherwise returns to cleaning, where it is compensated through gifts (Fig.~\ref{fig:competence_gap}). 
% We examine how this behavior 
% % \begin{wrapfigure}[8]{r}{0.5\textwidth}
% % % \vspace{-0.5em}
% % \centering
% % \includegraphics[height=0.36\linewidth,trim={0 0 12 0},clip]{image/competence_gap_eps0.1_cleaner_apples.pdf}%
% % \includegraphics[height=0.36\linewidth]{image/competence_gap_eps0.25_cleaner_apples.pdf}
% % \captionsetup{type=figure,skip=0pt}
% % \caption{Apples harvested per episode by the talented cleaner (left: $0.1\times$, right: $0.25\times$ reward).}
% % \label{fig:competence_gap}
% % % \vspace{-1em}
% % \end{wrapfigure}
% varies with the talented cleaner’s relative reward per apple, comparing 0.1 and 0.25 times that of the talented harvester.
% Under LASE, the talented cleaner 
% barely harvests at either reward level. Under AGiR, at 0.1 times, it learns to harvest early in training but soon returns to cleaning, ending with the same division of labor as under LASE; at 0.25 times, harvesting pays off, and it keeps harvesting alongside cleaning.

\noindent\textbf{Spawn gap.} With equal competence but asymmetric initial positions, LASE leaves the river-side agent locked into cleaning, a role that turns out to be largely uncompensated by gifts (246 in extrinsic return vs. 247 after net gift transfers). AGiR allows the agent to escape this role and explore harvesting, increasing its extrinsic return from 246 to 276 while also improving the apple-side agent's return from 291 to 302.

% \noindent\textbf{Spawn gap.}
% In this scenario, the division of labor under LASE is not a compensated specialization but an uncompensated burden on the river-side agent, and AGiR improves the outcome for both agents. Under LASE, the river-side agent earns considerably less extrinsic return than the apple-side agent (246 vs.\ 291), reflecting that it takes on more of the cleaning (49 vs.\ 36 cleaning actions per episode). Yet after accounting for net gift transfers, its return barely changes (from 246 to 247); the gifts do not compensate it for this sacrifice. Under AGiR, the river-side agent explores harvesting and its extrinsic return rises to 276, while that of the apple-side agent also rises to 302, indicating that the outcome under LASE was not efficient either.

\noindent\textbf{Mismatch.} When initial positions oppose the agents' specialties, LASE produces position-driven roles, leaving the talented harvester cleaning. With AGiR, the harvester shifts to harvesting while the talented cleaner takes over cleaning. Both agents' total returns also improve, from 128 to 176 for the talented harvester and from 8 to 81 for the talented cleaner.

% % \begin{wrapfigure}{r}{0.5\textwidth}
% % \vspace{-1em}
% % \centering
% % \includegraphics[height=0.36\linewidth,trim={0 0 12 0},clip]{image/mismatch_harvester_apples.pdf}%
% % \includegraphics[height=0.36\linewidth]{image/mismatch_harvester_cleaning.pdf}
% % \captionsetup{type=figure,skip=0pt}
% % \caption{Talented harvester in \emph{Mismatch} (left: apples harvested, right: cleaning actions per episode).}
% % \label{fig:mismatch}
% % \vspace{-1em}
% % \end{wrapfigure}
% \noindent\textbf{Mismatch.}
% When agents start on the side opposite to their specialty, LASE lets position shape the roles, whereas AGiR lets competence shape them (Fig.~\ref{fig:mismatch_additional}). 
% The cleaner’s reward per apple is 0.25$\times$ the harvester’s. Under LASE, the talented harvester, placed by the river, spends much of its effort on cleaning (119 cleaning actions per episode). It starts harvesting late in training, harvesting 68 apples per episode by the end.
% % With the cleaner’s apple reward fixed at 0.25$\times$ the harvester’s, under LASE the talented harvester, placed by the river, spends much of its effort on cleaning (119 cleaning actions per episode) and starts harvesting late in training, harvesting 68 apples per episode by the end. 
% Under AGiR, it begins harvesting early and nearly stops cleaning, harvesting 123 apples per episode, while cleaning is left to the talented cleaner. After accounting for net gift transfers, both agents’ returns increase, from 128 to 176 for the talented harvester and from 8 to 81 for the talented cleaner.

\subsection{Ablation Studies}
\noindent\textbf{Refunded Gifts  }
In our approach, a refused gift is simply \textit{not accounted for} in the recipient's subjective utility during policy updates. Refunding these transfers to the donor is a natural alternative, but it introduces strategic reasoning such as inferring intentions from rejection, which lies beyond the scope of this work. We therefore compare our mechanism with a naive refund scheme that adds returned gifts $b_t^{-i,i}$ to the shaped reward: $r_t^{i,\text{shaped}} = r_t^{i,\text{ext}} - g_t^{i,-i} + w_t^i g_t^{-i,i} + b_t^{-i,i}$. Our mechanism outperforms this naive refunding approach on average; detailed results are provided in Appendix~\ref{append:ablation_refund}.

%%% 원본 
% In our approach, a refused gift is simply \textit{not accounted for} in the recipient's subjective utility during policy updates. While refunding these transfers to the donor is a natural alternative, it introduces complex strategic reasoning such as inferring intentions from rejection, which lies beyond the scope of this work. Consequently, we evaluate our mechanism against a naive refund scheme where the shaped reward explicitly incorporates the returned gifts: $r_t^{i,\text{shaped}} = r_t^{i,\text{ext}} - g_t^{i,-i} + w_t^i g_t^{-i,i} + \text{(refunded gifts)}_t$. Across all metrics, our mechanism outperforms this naive refunding approach on average. Detailed results are provided in Appendix~\ref{append:ablation_refund}.

\begin{comment}
\noindent\textbf{Refunded gifts.}
Our method assumes refused gifts are \emph{thrown away}. While refunding them to the giver is a natural alternative, it introduces complex strategic reasoning such as inferring intentions from rejection, which is beyond our scope. Therefore, we evaluate a naive refund scheme where the shaped reward simply incorporates the returned gifts: $r_t^{i,\text{shaped}} = r_t^{i,\text{ext}} - g_t^{i,-i} + w_t^i g_t^{-i,i} + \text{(refunded gifts)}_t$. Across all metrics, our mechanism outperforms this naive refunding approach on average. Detailed results are provided in Appendix~\ref{append:ablation_refund}.
\end{comment}

\noindent\textbf{Refusal-Aware Policy  }
Removing $w$ from the policy input degrades performance, indicating that a refusal-aware policy is important for AGiR, as shown in Table~\ref{tab:ablation_no_w_pi} in Appendix~\ref{append:ablation_no_w_pi}.

\noindent\textbf{Hyperparameter Sensitivity Analysis  }
Performance is largely stable across most modulator coefficients $C$. Results remain robust to the step size $\alpha$ as long as it is not excessively large. Across a wide range of window sizes $H$, AGiR consistently achieves higher Min and GM than vanilla gifting. Finally, although a fixed baseline ($B_t = \pi/4$) already surpasses vanilla gifting, our adaptive baseline yields the highest Min and GM. Likewise, non-adaptive ways of limiting incoming gifts, such as a fixed acceptance weight, gift clipping, and linear annealing of the acceptance weight, yield lower GM and Min than AGiR in Unfair Cleanup-Hard. See Appendix~\ref{append:ablation_modulator} for full results.

\vspace{-0.4em}

\section{Conclusion}\label{sec:conclusion}
In this paper, we introduce the concept of gift over-reliance in MMGs, where agents prioritize gifted rewards over their own reward-yielding performance. To mitigate this, we propose AGiR, an adaptive gift-refusal method that regulates gift acceptance based on reward-gift orientation.
Experiments on unfair variants of Cleanup, Harvest, and Coin Game demonstrate that AGiR lets agents with lower earning opportunities earn more on their own. Our characterization of gift over-reliance is empirical; formalizing a criterion that distinguishes it from legitimate specialization in arbitrary environments is an important direction for future work. Limitations are discussed in Appendix~\ref{append:limitation_futurework}.
% Limitations and future work are discussed in Appendix~\ref{append:limitation_futurework}.

\section{Acknowledgments}
This work was supported by the National Research Foundation of Korea (NRF) grant
funded by the Korea government (MSIT) (No. RS-2025-00557589
, Generative Model Based Efficient Reinforcement Learning Algorithms for Multi-modal Expansion in
Generalized Environments),  
the National Research Foundation of Korea (NRF) grant funded by the Korea government (MSIT) (No. RS-2026-25510069), and the “Advanced GPU Utilization Support Program” funded by the Government of the Republic of Korea (Ministry of Science and ICT) (No. 02-26-01-0110).
The authors declare no competing interests.
We thank Fanqi Kong for sharing the original LASE implementation.

\bibliography{refer.bib}

%%%%%%%%%%%%%%%%%%%%%%%%%%%%%%%%%%%%%%%%%%%%%%%%%%%%%%%%%%%%

\appendix

%%%%%%%%%%%%%%%%%%%%%%%%%%%%%%%%%%%%%%%%%%%%%%%%%%%%%%%%%%%%

\newpage
\appendix
\onecolumn
\appendix
\section{Prior Gifting Methods in Formal Representation}\label{append:prior_gifting}

Each agent $i$ specifies its outgoing gifts using a function $G_{\eta^i}$ that maps to the gift vector $\{\,g_t^{i, j}\,\}_{j\in\mathcal{N}}$. 
Here, $\eta^i$ represents the learnable parameter associated with agent $i$'s gifting function.
The shaped reward of agent $i$ is
\begin{align}\label{eq:shape_reward}
    r_t^{i,\text{shaped}}=r_t^{i,\text{ext}}-g_t^{i,-i}(\eta^i)+g_t^{-i,i}(\eta^{-i}).
\end{align}
Note that when $\eta^i$ is not learned but fixed, the gifting function becomes rule-based. 
Each self-interested agent trains its own policy $\pi^i$ by optimizing the shaped return that includes the extrinsic reward for self-interest and the gift for cooperation. In addition, each agent trains its gifting function $G_{\eta^i}$ to maximize its own extrinsic return, under the rationale that the ultimate goal of gifting is to enhance the giver’s return through cooperation.
The notation $g_t^{i,j}(\eta^i)$ will be used if necessary to clarify the dependence of gift to the parameter $\eta^i$. 

\textbf{LIO.}
In LIO~\cite{lio}, each agent learns a separate gifting function in addition to its policy. The gifting function $G_{\eta^i}$ takes agent $i$'s observation and others' actions as inputs (i.e., $(o^i,a^{-i})$) and returns gifts for all others (i.e., $G_{\eta^i}(o^i,a^{-i}) = (g^{i,j}(\eta^i))_{j\neq i}$). LIO does not include an explicit cost for gifting in the shaped reward (i.e., $-g_t^{i,-i}(\eta^i)$ in Eq.~\eqref{eq:shape_reward}). However, the training loss for the gifting function contains a penalty term $-\sum_t\gamma^t \sum_{j\neq i} |g_t^{i,j}|$ that prevents excessive gifting, which acts as an implicit cost for gifting.

Formally, the shaped reward is defined by
\begin{align}\label{eq:shape_reward_lio}
    r_t^{i,\text{shaped}}=r_t^{i,\text{ext}}+\sum_{j\neq i} g_t^{j,i}(\eta^j),
\end{align}
and each agent updates its policy following the policy objective
\begin{align*}
    \max_{\theta^i} \mathbb{E}_{\pi_{\theta^i}}\left[\sum_t \gamma^t r_t^{i,\text{shaped}}\right].
\end{align*}

The parameter for each agent's gifting function, $\eta^i$, is updated by the following objective:
\begin{align}\label{eq:lio_gift_obj}
    \max_{\eta^i} \mathbb{E}_{\pi^{\eta}_{\theta^i}}\left[\sum_t \gamma^t r_t^{i,\text{ext}}\right]-\sum_t\gamma^t \sum_{j\neq i} |g_t^{i,j}(\eta^i)|
\end{align}
where $\pi^{\eta}_{\theta^i}$ denotes the updated policy by the policy objective above.

\textbf{LASE.}
In LASE~\cite{lase}, at each time $t$, each agent redistributes its extrinsic reward following the weight vector $w_t^i = (w_t^{i1}, \dots, w_t^{in})$, where $w_t^{ij} \ge 0$ and $\sum_{j\in\mathcal{N}} w_t^{ij} = 1$. Formally, the resulting shaped reward is as follows:

\begin{align}\label{eq:shape_reward_lase}
    r_t^{i,\text{shaped}}=\sum_{j\in\mathcal{N}} w_t^{ji}r_t^{j,\text{ext}}.
\end{align}
In this case, $r_t^{i,\text{shaped}}$ follows Eq.~\eqref{eq:shape_reward} with the gift $g_t^{i,j}({\eta^i})$ defined by $g_t^{i,j}({\eta^i})=w_t^{ij}(\eta^i)\cdot r_t^{i,\text{ext}}$. Each agent's policy is updated by the following objective:
\begin{align*}
    \max_{\theta^i} \mathbb{E}_{\pi_{\theta^i}}\left[\sum_t \gamma^t r_t^{i,\text{shaped}}\right].
\end{align*}

The gifting function $G_{\eta^i}$ consists of two modules: a Social Relation Inference (SRI) module and a Perspective-Taking (PT) module. 

The SRI module contains two networks, $Q^i_{SR}(o_t^i,a_t^{1:n};\eta^i_1)$ and $\pi_{SR}^i(a_t^i|\hat{o}_t^j;\eta^i_2)$, with model parameters $\eta^i_1$ and $\eta^i_2$, respectively. These networks are trained by a standard actor-critic method using extrinsic rewards. This means that $Q^i_{SR}$ and $\pi_{SR}^i$ are trained toward the optimal policy and value function with respect to their extrinsic rewards. This aligns with the philosophy of the training objective~\eqref{eq:lio_gift_obj} for the gifting function in LIO. Note that LASE has no penalty term in the gifting function's training objective; instead, it already includes the cost for gifting in its shaped reward.

The PT module contains an imagination network $Img^i(o_t^i, j;\eta^i_3)$ with model parameter $\eta^i_3$. The model $Img^i$ allows agent $i$ to predict agent $j$'s observation from agent $i$'s observation. The imagination network is trained by minimizing the cross-entropy loss between agent $j$'s actual action and the predicted action distribution.

The weight $w_t^{ij}$ is finally defined by 
\begin{align*}
    w_t^{ij}(\eta^i_1,\eta^i_2,\eta^i_3) = \frac{Q^i_{SR}(o_t^i,a_t^{1:n};\eta^i_1) - \sum_{a^j}\pi_{SR}^i(a| \hat{o}_t^j;\eta^i_2)Q^i_{SR}(o_t^i,(a_t^{-j}, a^j);\eta^i_1)}{(n-1)(\max_{a^j}Q^i_{SR}(o_t^i,(a_t^{-j}, a^j);\eta^i_1) - \min_{a^j}Q^i_{SR}(o_t^i,(a_t^{-j}, a^j);\eta^i_1))}
\end{align*}
where $\hat{o}_t^j = Img^i(o_t^i,j;\eta^i_3)$.
By letting $\eta^i = (\eta^i_1,\eta^i_2,\eta^i_3)$, the gifting function of LASE can be expressed as $G_{\eta^i}(o_t^i,a_t^{1:n},j) = (w_t^{ij}(\eta^i)\cdot r_t^{i,\text{ext}})_{j=1}^n = (g_t^{i,j}({\eta^i}))_{j=1}^n$. 

\textbf{Reward exchange.}
In Reward Exchange~\cite{minimal_reward_transfer}, each agent leaves its extrinsic reward with a ratio $s$ and equally distributes $(1-s)$ of its reward to others. Formally, the resulting shaped reward is as follows:
\begin{align}\label{eq:shape_reward_rx}
    r_t^{i,\text{shaped}} = s \cdot r_t^{i,\text{ext}} + \frac{1-s}{n-1} \sum_{j \neq i} r_t^{j,\text{ext}}.
\end{align}
In this case, $r_t^{i,\text{shaped}}$ follows Eq.~\eqref{eq:shape_reward} with the rule-based gift $g_t^{i,j}(\eta^i)$ determined by $g_t^{i,j}(\eta^i) = G_{\eta^i}(r_t^{i,\text{ext}}) = \frac{1-s}{n-1} r_t^{i,\text{ext}}$. 

Then, each agent updates its policy following the objective:
\begin{align*}
    \max_{\theta^i} \mathbb{E}_{\pi_{\theta^i}}\left[\sum_t \gamma^t r_t^{i,\text{shaped}}\right].
\end{align*}

\section{Pseudo Code}\label{append:pseudo_code}
% \tcb{pseudo code}\\
\begin{algorithm}[H]
\caption{Adaptive Gift Refusal (for each agent $i$)}
\label{alg:adaptive_refusal}
\begin{algorithmic}[1]
\REQUIRE initial observation $o_0^{i}$, initial refusal weight $w_0^{i}$, gifting function $G_{\eta^i}$, initial policy $\pi^{i}$
\FOR{$iter = 0,1,2,\ldots$}
\FOR{$t = 0,1,2,\ldots$}
    \STATE Sample action $a_t^{i} \sim \pi^{i}(\,\cdot \mid o_t^{i},\, w_t^{i}\,)$
    \STATE Execute $a_t^{i}$; receive extrinsic reward $r_t^{i,\text{ext}}$ and next observation $o_{t+1}^{i}$
    \STATE Compute gifts to others via $G_{\eta^i}$: \\
           % \hspace{1.5em}$\{\,g_t^{i,j}\,\}_{j\neq i} \gets G_{\eta^i}$
           \[
           \{\,g_t^{i,j}\,\}_{j\neq i} \gets G_{\eta^i}
           \]
    \STATE Send $\{\,g_t^{i,j}\,\}_{j\neq i}$ to recipients; receive incoming $\{\,g_t^{j,i}\,\}_{j\neq i}$
    \STATE Update refusal weight by Eq.~\eqref{eq:w_update}:
           \[
           w_{t+1}^{i} \;\gets\; clip\,\!\big(w_t^{i}+\alpha \Delta_t^i\,\big)
           \]
           % \[
           % w_{t+1}^{i} \;\gets\; \textsc{UpdateW(\tcb{TBD})}\!\big(w_t^{i},\, r_{t-H+1:t}^{i},\, \{g_{t-H+1:t}^{-i,i}\}\,\big)
           % \]
\ENDFOR
\STATE Compute shaped reward for each timestep $t$:
           \[
           r_t^{i,\text{AGiR}}
           \;=\;
           r_t^{i,\text{ext}}
           \;-\; \sum_{j\neq i} g_t^{i, j}
           \;+\; w_t^{i}~\!\!\sum_{j\neq i} g_t^{j, i},~\forall t
           \]
    \STATE Update policy $\pi^{i}$ using $r_t^{i,\text{AGiR}}$ (e.g., PPO step)
    \STATE If learnable, update $G_{\eta^i}$ according to the base gifting method.
\ENDFOR
\STATE \textbf{Output:} final policy $\pi^{i}$
\end{algorithmic}
\end{algorithm}

\section{Assumptions and Proofs for Theoretical Analysis}\label{append:proof}
\subsection{Assumptions and Proof for Theorem~\ref{thm:existence}}\label{append:proof_existence}
% \tcr{decide weak Feller / Feller} -> weak feller is correct
\begin{assumption}[Assumption~\ref{main:assumption_1}]\label{append:assumption_1}
    The global state space $S\subset\mathbb{R}^d$ and
each agent's action space $A^i$ are finite.
\end{assumption}

\begin{assumption}[Continuity, Assumption~\ref{main:assumption_2}]\label{append:assumption_2}
For each agent $i$, the baseline $B^i$ is continuous
in $w^i\in[0,1]$, and the policy
$\pi^i(a^i\mid s,w^i)$ is continuous in $w^i$
for every $s\in S$ and $a^i\in A^i$.
\end{assumption}

\begin{theorem} [Existence, Theorem~\ref{thm:existence}]
    For arbitrary fixed joint policy profile $\pi = (\pi^1, \dots, \pi^n)$ and gifting functions $\{G_{\eta^i}\}_{i=1}^n$, let $\{w_t^i\}_{t \ge 0}$ be the stochastic process of refusal weights for agent $i$ generated by AGiR with any fixed finite window size $H\ge1$. Under Assumptions~\ref{main:assumption_1} and~\ref{main:assumption_2}, for each agent $i$, this process admits a stationary marginal distribution.
\end{theorem}
\begin{proof}  
%%%%%%%%%%%%%% Part 1

For a fixed finite window size $H\geq 1$, define the
joint augmented state
\[
Z_t=(s_t,\mathbf{w}_t,\mathbf{r}_{t-H:t-1},
\mathbf{g}_{t-H:t-1})
\]
on the augmented state space
\[
\bar{S}_H
=S\times[0,1]^n\times\mathcal{R}^H\times\mathcal{G}^H,
\]
where $\mathbf{w}_t=(w_t^1,\ldots,w_t^n)$,
$\mathbf{r}_{t-H:t-1}=(\mathbf{r}_{t-H},\ldots,\mathbf{r}_{t-1})$,
and $\mathbf{g}_{t-H:t-1}=(\mathbf{g}_{t-H},\ldots,\mathbf{g}_{t-1})$
denote the joint refusal weights and the most recent $H$
joint extrinsic reward and received gift vectors, respectively.

Here, $s\in S$ and $a\in A$ denote the global state
and joint action, respectively.
The sets of possible joint reward and received gift vectors
are defined by
\begin{align*}
\mathcal{R}
&=\{\mathbf{r}(s,a):s\in S,a\in A\}\cup\{\mathbf{0}\}
\subseteq[-r_{\max},r_{\max}]^n,\\
\mathcal{G}
&=\{\mathbf{g}(s,a):s\in S,a\in A\}\cup\{\mathbf{0}\}
\subseteq[-g_{\max},g_{\max}]^n,
\end{align*}
where $\mathbf{r}(s,a)=(r^j(s,a))_{j=1}^n$ and
$\mathbf{g}(s,a)=(g^{-j,j}(s,a))_{j=1}^n$.
By Assumption~\ref{append:assumption_1}, both
$\mathcal{R}$ and $\mathcal{G}$ are finite.
For $t<H$, the unavailable entries in the reward
and gift histories are initialized to zero.

The refusal-aware policy of each agent $j$ depends only
on $(s,w^j)$ and can therefore be regarded as a policy
on $\bar{S}_H$ through
\[
\pi^j(da^j\mid Z)=\pi^j(da^j\mid s,w^j).
\]
Under the fixed joint policy profile
$\pi=(\pi^1,\ldots,\pi^n)$, the joint action is
sampled according to
\[
a_t\sim\pi(\cdot\mid Z_t),
\qquad
\pi(da\mid Z)
=\prod_{j=1}^n\pi^j(da^j\mid s,w^j).
\]

Given $Z_t$ and $a_t$, the next environment state
is generated as
\[
s_{t+1}\sim P(\cdot\mid s_t,a_t),
\]
and each agent $j$ receives
\[
r_t^j=r^j(s_t,a_t),
\qquad
g_t^j=g^{-j,j}(s_t,a_t).
\]

For each agent $j$, define
\[
R_t^j=\sum_{k=t-H+1}^{t}r_k^j,
\qquad
G_t^j=\sum_{k=t-H+1}^{t}g_k^j.
\]
The refusal weight is then updated by
\[
w_{t+1}^j
=\operatorname{clip}\left(
w_t^j+\alpha w_t^j S(R_t^j)
\left[
\arctan\left(\frac{R_t^j}{G_t^j+\epsilon}\right)
-B^j(w_t^j)
\right]\right).
\]
Choose sufficiently small $\epsilon>0$ such that
\[
0<\epsilon<
\min_{\substack{j\in\{1,\ldots,n\},\
\mathbf{x}_1,\ldots,\mathbf{x}_H\in\mathcal G\\
\sum_{k=1}^H x_k^j\neq 0}}
\left|\sum_{k=1}^H x_k^j\right|,
\]
where $x_k^j$ denotes the $j$-th component of
$\mathbf{x}_k$.

Let $\operatorname{shift}_H$ denote the history-shifting
operation defined by
\[
\operatorname{shift}_H((x_1,\ldots,x_H),y)
=(x_2,\ldots,x_H,y).
\]
Conditioned on $Z_t$ and $a_t$, the augmented
transition kernel is given by
\begin{align*}
&\bar{P}_H(dZ_{t+1}\mid Z_t,a_t)\\
&=P(ds_{t+1}\mid s_t,a_t)\\
&\quad\times\prod_{j=1}^n
\delta_{w_{t+1}^j}(dw_{t+1}^j)\\
&\quad\times\prod_{j=1}^n
\delta_{\operatorname{shift}_H(
r_{t-H:t-1}^j,r_t^j)}
(dr_{t-H+1:t}^j)\\
&\quad\times\prod_{j=1}^n
\delta_{\operatorname{shift}_H(
g_{t-H:t-1}^j,g_t^j)}
(dg_{t-H+1:t}^j),
\end{align*}
where $\delta_z(\cdot)$ denotes the Dirac measure at $z$.

The transition kernel under the fixed joint policy
profile $\pi$ is therefore
\[
\bar{P}_H^\pi(dZ'\mid Z)
=\int_A\bar{P}_H(dZ'\mid Z,a)\pi(da\mid Z).
\]
Since the conditional distribution of $a_t$ depends
only on $Z_t$, and all components of $Z_{t+1}$
are determined by $Z_t$, $a_t$, and the next
environment state, we have
\[
\Pr(Z_{t+1}\in E\mid Z_0,\ldots,Z_t)
=\bar{P}_H^\pi(E\mid Z_t)
\]
for every measurable $E\subseteq\bar{S}_H$.
Hence, $\{Z_t\}_{t\geq0}$ is a Markov chain with
transition kernel $\bar{P}_H^\pi$.

%%%%%%%%%%%%%% Part 2

We show that $\bar{P}_H^\pi$ is weak Feller.
The following proof is adapted from and expands upon the derivation of Proposition 6.1.2 in \citet{meyn2012markov}.

Let $Z'=F_H(Z,a,s')$ denote the joint augmented next state obtained from $Z$ and $a$, with the next environment state given by $s'$.
Since the state, action, and reward/gift history spaces are finite, their elements are isolated.
Furthermore, by Assumption~\ref{append:assumption_2} and the continuity of the clipping function $\operatorname{clip}(x)=\min\{\max\{0,x\},1\}$, the refusal-weight update is continuous in $\mathbf{w}$.
Thus, $F_H$ is continuous in $(Z,a,s')$.

Denote by $C_b(\bar{S}_H)$ the space of bounded continuous functions on $\bar{S}_H$.
For arbitrary $\psi\in C_b(\bar{S}_H)$, let
\begin{align*}
(\bar{P}_H^\pi\psi)(Z)
&:=\int_{\bar{S}_H}\psi(Z')\bar{P}_H^\pi(dZ'\mid Z)\\
&=\sum_{a\in A}\pi(a\mid Z)
\sum_{s'\in S}P(s'\mid s,a)
\psi(F_H(Z,a,s')).
\end{align*}
Since $\psi\in C_b(\bar{S}_H)$, there exists $C_\psi<\infty$ such that
$|\psi(Z')|\le C_\psi$ for all $Z'\in\bar{S}_H$.
Since $P(\cdot\mid s,a)$ and $\pi(\cdot\mid Z)$ are probability distributions,
\begin{align*}
|(\bar{P}_H^\pi\psi)(Z)|
&\le\sum_{a\in A}\pi(a\mid Z)
\sum_{s'\in S}P(s'\mid s,a)
|\psi(F_H(Z,a,s'))|\\
&\le C_\psi<\infty.
\end{align*}
Therefore, $\bar{P}_H^\pi\psi$ is bounded on $\bar{S}_H$.

We next show that $\bar{P}_H^\pi\psi$ is continuous.
Let $\{Z_m\}$ be an arbitrary sequence converging to $Z$ in $\bar{S}_H$.
Since the state and history spaces are finite, the corresponding components of $Z_m$ eventually coincide with those of $Z$, while $\mathbf{w}_m\to\mathbf{w}$.
By the continuity of each agent's policy in its refusal weight, the joint policy
\[
\pi(a\mid Z)=\prod_{j=1}^n
\pi^j(a^j\mid s,w^j)
\]
is continuous in $Z$ for every $a\in A$.
Furthermore, since $F_H$ and $\psi$ are continuous,
\[
\psi(F_H(Z_m,a,s'))
\to\psi(F_H(Z,a,s'))
\]
for every $a\in A$ and $s'\in S$.

Since $S$ and $A$ are finite, we can interchange the limit and the finite sums:
\begin{align*}
&\lim_{m\to\infty}(\bar{P}_H^\pi\psi)(Z_m)\\
&=\lim_{m\to\infty}
\sum_{a\in A}\pi(a\mid Z_m)
\sum_{s'\in S}P(s'\mid s_m,a)
\psi(F_H(Z_m,a,s'))\\
&=\sum_{a\in A}\pi(a\mid Z)
\sum_{s'\in S}P(s'\mid s,a)
\psi(F_H(Z,a,s'))\\
&=(\bar{P}_H^\pi\psi)(Z).
\end{align*}
Thus, $\bar{P}_H^\pi\psi$ is continuous on $\bar{S}_H$.

Combining this with the boundedness established above, we obtain
\[
\psi\in C_b(\bar{S}_H)
\implies
\bar{P}_H^\pi\psi\in C_b(\bar{S}_H).
\]
By Proposition 6.1.1 in \citet{meyn2012markov}, $\bar{P}_H^\pi$ is weak Feller.

We now show the existence of a stationary distribution for $\bar{P}_H^\pi$.
Let $\bar{\mu}\in\Delta(\bar{S}_H)$ be an arbitrary initial distribution of the joint augmented Markov chain with transition kernel $\bar{P}_H^\pi$.
Define
\[
\bar{d}_t
=\frac{1}{t}\sum_{k=0}^{t-1}
\bar{\mu}(\bar{P}_H^\pi)^k.
\]
Since $S$, $\mathcal R$, and $\mathcal G$ are finite, $[0,1]^n$ is compact, and $H$ is finite, the augmented state space
\[
\bar{S}_H
=S\times[0,1]^n\times\mathcal R^H\times\mathcal G^H
\]
is compact.
Therefore, the family of probability measures $\{\bar{d}_t\}$ on $\bar{S}_H$ is tight, implying that the joint augmented Markov chain is bounded in probability on average.

By Theorem 12.0.1 in \citet{meyn2012markov}, the weak Feller property and boundedness in probability on average imply the existence of at least one invariant probability measure $d^*$ on $\bar{S}_H$, satisfying
\[
d^*(E)
=\int_{\bar{S}_H}\bar{P}_H^\pi(E\mid Z)d^*(dZ)
\]
for every measurable $E\subseteq\bar{S}_H$.

For an arbitrary agent $i$, let $d_{w^i}^*$ denote the marginal distribution of $d^*$ on the refusal-weight coordinate $w^i\in[0,1]$, defined by
\[
d_{w^i}^*(B)
=d^*\bigl(\{Z\in\bar{S}_H:w^i\in B\}\bigr)
\]
for any measurable $B\subseteq[0,1]$.
Since $d^*$ is invariant under $\bar{P}_H^\pi$, the marginal distribution of $w_t^i$ remains $d_{w^i}^*$ when the joint augmented process is initialized according to $d^*$.
Thus, $d_{w^i}^*$ is a stationary marginal distribution of $\{w_t^i\}_{t\ge0}$ generated by AGiR.

\end{proof}

\subsection{Assumptions and Proof for Theorem~\ref{thm:uniqueness}}\label{append:proof_uniqueness}

% --------------------------------------------------
% Setting
% --------------------------------------------------

\begin{assumption}[Assumption~\ref{main:assumption_4}]\label{append:assumption_4}
The step size satisfies $\alpha\in(0,\frac{1}{2\pi})$.
For each agent $i$, $B^i\in(-\frac{\pi}{2},\frac{\pi}{2})$,
$\phi^i(s,a)\in(-\frac{\pi}{2},\frac{\pi}{2})$, and
$r^i(s,a)\in[-r_{\max},r_{\max}]$ for all $s,a$,
where $B^i$ is independent of the refusal weights.
\end{assumption}

For $H=1$, consider the joint process
\[
Z_t=(s_t,\mathbf{w}_t)
\]
on the state space
\[
\bar{S}=S\times(0,1]^n.
\]
% Here, $\mathbf{w}_t=(w_t^1,\ldots,w_t^n)$,
% and the initial refusal weights satisfy $w_0^i>0$
% for all agents $i$.

Under the fixed joint policy profile
$\pi=(\pi^1,\ldots,\pi^n)$, let
$\bar{P}^\pi$ denote the transition kernel of
$\{Z_t\}_{t\geq 0}$, where
\[
\pi(a\mid Z)=\prod_{j=1}^n
\pi^j(a^j\mid s,w^j).
\]

For each agent $i$, define
\[
m^i(s,a)
:=1+\alpha S(r^i(s,a))
\bigl(\phi^i(s,a)-B^i\bigr).
\]
Under Assumption~\ref{main:assumption_4},
$0<m^i(s,a)<2$, and the refusal-weight update
can be rewritten as
\[
w_{t+1}^i
=\min\{1,w_t^i m^i(s_t,a_t)\}.
\]
This follows because the lower clipping bound is
inactive when $w_t^i>0$, as $m^i(s,a)>0$.
Consequently, if $w_0^i>0$, then $w_t^i>0$
for all $t\geq0$.
Thus, $\bar{S}=S\times(0,1]^n$ is invariant
under the joint AGiR dynamics.

% --------------------------------------------------
% Assumption 4.5
% --------------------------------------------------

\begin{assumption}[Positive-Regime Stability, Assumption~\ref{main:assumption_5}]\label{append:assumption_5}
The following conditions hold under the fixed
joint policy profile $\pi$.

\textnormal{(A) Low-weight recovery.}
There exist $\eta\in(0,1)$ and $\kappa>0$
such that, for every agent $i$ and every
$Z=(s,\mathbf{w})\in\bar{S}$ satisfying
$w^i\le\eta$,
\[
\sum_{a\in A}\pi(a\mid Z)
\log m^i(s,a)\ge\kappa.
\]

\textnormal{(B) Uniform regeneration.}
There exists $s_*\in S$ such that, with
$Z_*=(s_*,\mathbf{1})$, the following holds.
For every $c\in(0,1)$, there exist
an integer $N_c\ge1$ and $\beta_c>0$ satisfying
\[
\inf_{\substack{
Z=(s,\mathbf{w})\in\bar{S}\\
w^i\ge c,\ \forall i
}}
(\bar{P}^\pi)^{N_c}
(\{Z_*\}\mid Z)
\ge\beta_c.
\]
Furthermore,
\[
\bar{P}^\pi(\{Z_*\}\mid Z_*)>0.
\]
\end{assumption}

% --------------------------------------------------
% Theorem 4.6
% --------------------------------------------------

\begin{theorem}[Stability and Uniqueness, Theorem~\ref{thm:uniqueness}]
Suppose Assumptions~\ref{main:assumption_1}, \ref{main:assumption_2}, \ref{main:assumption_4}, and \ref{main:assumption_5}
hold and $H=1$, with initial refusal weights
$w_0^i>0$ for all agents $i$.
Then, the joint AGiR process $\{Z_t\}_{t\ge0}$
on $\bar{S}$ admits a unique invariant
probability measure $\nu^*$.

Moreover, for every initial state $Z_0\in\bar{S}$,
\[
\left\|
(\bar{P}^\pi)^t(\cdot\mid Z_0)-\nu^*
\right\|_{\mathrm{TV}}
\longrightarrow 0
\]
geometrically fast as $t\to\infty$.

Furthermore, there exists $p>0$ such that
\[
\int_{\bar{S}}(w^i)^{-p}\nu^*(dZ)<\infty
\]
for every agent $i$.

Consequently, each agent's refusal-weight process
$\{w_t^i\}_{t\ge0}$ converges in distribution
to a unique stationary marginal distribution
$\nu_{w^i}^*$ satisfying
\[
\nu_{w^i}^*([0,u])
\le K_i u^p,
\qquad 0<u<1,
\]
for some finite constant $K_i>0$.
%%% non-degeneracy part
Furthermore, if for every agent $i$,
\[
\sum_{\substack{a\in A:\\
\phi^i(s_*,a)<B^i}}
\pi(a\mid Z_*)>0,
\]
where $Z_*=(s_*,\mathbf{1})$ is defined in
Assumption~\ref{main:assumption_5}(B), then the unique stationary
distribution $\nu^*$ is non-degenerate in each
refusal-weight coordinate, satisfying
\[
\nu^*(w^i=1)>0,
\qquad
\nu^*(0<w^i<1)>0.
\]

\end{theorem}

% --------------------------------------------------
% Proof
% --------------------------------------------------

\begin{proof}
We establish a geometric drift condition for
the joint AGiR process and then apply the
geometric ergodicity theorem of
\citet{meyn2012markov}.

\textbf{Step 1: Logarithmic transformation.}

For each agent $i$, define
\[
Y_t^i=-\log w_t^i.
\]
Since $w_0^i>0$ and $m^i(s,a)>0$,
we have $w_t^i>0$ for every finite $t$.
The refusal-weight update can therefore be
rewritten as
\[
Y_{t+1}^i
=
\max\left\{
0,\,
Y_t^i-\log m^i(s_t,a_t)
\right\}.
\]

This is a reflected stochastic recursion,
where the boundary $Y^i=0$ corresponds to
$w^i=1$.

\textbf{Step 2: Geometric drift.}

By Assumptions~\ref{main:assumption_1} and~\ref{main:assumption_4}, the state and
action spaces are finite, and
$0<m^i(s,a)<2$ for all $s,a,i$.
Thus, there exist constants
$0<m_{\min}\le m_{\max}<\infty$ such that
\[
m_{\min}\le m^i(s,a)\le m_{\max}
\]
for every $s,a,i$.
In particular, $\log m^i(s,a)$ is
uniformly bounded.

By Assumption~\ref{main:assumption_5}(A), whenever $w^i\le\eta$,
\[
\mathbb{E}_{a\sim\pi(\cdot\mid Z)}
[\log m^i(s,a)]\ge\kappa.
\]
Since $\log m^i(s,a)$ is uniformly bounded,
a uniform Taylor expansion around $p=0$ yields
\[
\begin{aligned}
\mathbb{E}_{a\sim\pi(\cdot\mid Z)}
[(m^i(s,a))^{-p}]
&=1-p\mathbb{E}[\log m^i(s,a)]+O(p^2)\\
&\le 1-p\kappa+O(p^2).
\end{aligned}
\]
Thus, for sufficiently small $p>0$, there exists
$\rho\in(0,1)$ such that
\[
\mathbb{E}_{a\sim\pi(\cdot\mid Z)}
[(m^i(s,a))^{-p}]\le\rho
\]
uniformly over all agents $i$ and states $Z$
satisfying $w^i\le\eta$.

Define the Lyapunov function
\[
V(Z)=1+\sum_{i=1}^n(w^i)^{-p}.
\]

For $w^i\le\eta$, since $x^{-p}$ is decreasing
for $p>0$, the refusal-weight update gives
\begin{align*}
(w_{t+1}^i)^{-p}
&=\max\left\{
1,\,
(w_t^i)^{-p}
(m^i(s_t,a_t))^{-p}
\right\}\\
&\le
1+(w_t^i)^{-p}
(m^i(s_t,a_t))^{-p}.
\end{align*}
Therefore,
\[
\mathbb{E}[
(w_{t+1}^i)^{-p}\mid Z_t=Z]
\le
1+\rho(w^i)^{-p}.
\]

On the other hand, if $w^i>\eta$,
then $m^i(s,a)\ge m_{\min}$ implies
\[
w_{t+1}^i
\ge\min\{1,\eta m_{\min}\}>0.
\]
Consequently,
\[
(w_{t+1}^i)^{-p}
\le
\left(\min\{1,\eta m_{\min}\}\right)^{-p}
=:D<\infty.
\]
Hence,
\[
\mathbb{E}[
(w_{t+1}^i)^{-p}\mid Z_t=Z]
\le D,
\]
where $D$ is independent of $Z$.

Combining the two cases, we have
\[
\mathbb{E}[(w_{t+1}^i)^{-p}\mid Z_t=Z]
\le \rho(w^i)^{-p}+D
\]
for every agent $i$ and every $Z\in\bar{S}$ since $D\ge1$ and $\rho(w^i)^{-p}>0$.
Summing over all agents yields
\begin{align*}
(\bar{P}^\pi V)(Z)
&:=\mathbb{E}[V(Z_{t+1})\mid Z_t=Z]\\
&\le 1+\rho\sum_{i=1}^n(w^i)^{-p}+nD\\
&=\rho V(Z)+b,
\end{align*}
where $b=1-\rho+nD<\infty$.

Choose
\[
L>\max\left\{n+1,\frac{b}{1-\rho}\right\},
\]
so that
\[
\lambda:=\rho+\frac{b}{L}<1.
\]
Define
\[
C_L=\{Z\in\bar{S}:V(Z)\le L\}.
\]
For $Z\notin C_L$, we have $V(Z)>L$,
and thus
\[
(\bar{P}^\pi V)(Z)
\le\lambda V(Z).
\]
Since $V$ is bounded on $C_L$, there exists
$b_L<\infty$ such that, for every $Z\in\bar{S}$,
\[
(\bar{P}^\pi V)(Z)
\le\lambda V(Z)+b_L\mathbf{1}_{C_L}(Z).
\]
This is a geometric Foster--Lyapunov
drift condition.

\textbf{Step 3: Regeneration and ergodicity.}

By the definition of $V$, every
$Z=(s,\mathbf{w})\in C_L$ satisfies
\[
w^i\ge L^{-1/p},
\qquad i=1,\ldots,n.
\]
Therefore, Assumption~\ref{main:assumption_5}(B) implies that
there exist $N\ge1$ and $\beta>0$ such that
\[
(\bar{P}^\pi)^N(E\mid Z)
\ge
\beta\,\delta_{Z_*}(E)
\]
for every $Z\in C_L$ and every measurable
$E\subseteq\bar{S}$.
Thus, $C_L$ is a small set.

Moreover, the geometric drift condition
implies that the process reaches $C_L$
almost surely from every $Z\in\bar{S}$.
Together with the regeneration condition,
this implies that $Z_*$ is accessible
from every initial state, and hence the
chain is $\delta_{Z_*}$-irreducible.

The condition
\[
\bar{P}^\pi(\{Z_*\}\mid Z_*)>0
\]
further implies aperiodicity.

Consequently, the geometric drift condition,
irreducibility, and aperiodicity satisfy
the conditions of the geometric ergodicity
theorem (Theorem 15.0.1) in \citet{meyn2012markov}.

Therefore, the joint Markov chain admits
a unique invariant probability measure
$\nu^*$ on $\bar{S}$.
Furthermore, there exist constants
$M<\infty$ and $q\in(0,1)$ such that
\[
\left\|
(\bar{P}^\pi)^t(\cdot\mid Z)-\nu^*
\right\|_{\mathrm{TV}}
\le M V(Z)q^t
\]
for every $Z\in\bar{S}$.

\textbf{Step 4: Stationary marginal distributions.}

The geometric drift condition also gives
\[
\int_{\bar{S}}V(Z)\nu^*(dZ)<\infty.
\]
By the definition of $V$,
\[
\int_{\bar{S}}
(w^i)^{-p}\nu^*(dZ)<\infty
\]
for every agent $i$.
For an arbitrary agent $i$, let
$\nu_{w^i}^*$ denote the marginal of
$\nu^*$ on the refusal-weight coordinate:
\[
\nu_{w^i}^*(B)
=
\nu^*(\{Z\in\bar{S}:w^i\in B\})
\]
for any measurable $B\subseteq[0,1]$.

Since the joint process converges to
$\nu^*$, each refusal-weight process
converges in distribution to
$\nu_{w^i}^*$.

Finally, by Markov's inequality,
for any $u\in(0,1)$,
\begin{align*}
\nu_{w^i}^*([0,u])
&=\Pr_{\nu^*}(w^i\le u)\\
&=\Pr_{\nu^*}((w^i)^{-p}\ge u^{-p})\\
&\le u^p\,
\mathbb{E}_{\nu^*}[(w^i)^{-p}]\\
&=:K_i u^p,
\end{align*}
where $K_i<\infty$, and $\Pr_{\nu^*}$ denotes probability under the invariant measure $\nu^*$.
Thus, the stationary marginal distribution
is unique within the positive refusal-weight
regime, and its probability mass near zero
is bounded as claimed.

Under the additional condition, the geometric drift condition
implies $\nu^*(C_L)>0$.
By invariance and Assumption~\ref{main:assumption_5}(B),
\begin{align*}
\nu^*(\{Z_*\})
&=\int_{\bar{S}}(\bar{P}^{\pi})^N
(\{Z_*\}\mid Z)\nu^*(dZ)\\
&\ge\beta\nu^*(C_L)>0.
\end{align*}
Hence, $\nu^*(w^i=1)>0$ for every agent $i$.
Moreover, since $\phi^i(s_*,a)<B^i$
implies $m^i(s_*,a)<1$, an agent with
$w^i=1$ has a positive probability of
transitioning to $0<w^i<1$ from $Z_*$.
By stationarity, $\nu^*(0<w^i<1)>0$.
This completes the proof.

\end{proof}

\subsection{One-step Analysis of AGiR in Iterated Prisoner's Dilemma}\label{append:ipd}
\textbf{Problem setting.} To investigate the theoretical properties of AGiR, we consider a standard two-player, two-action iterated normal form game with the actions Cooperation (C) and Defection (D). This setup is widely adopted in the theoretical analysis of sequential social dilemmas to characterize agent interactions. 
Our primary objective is to analyze the effect of AGiR’s one-step update on the value function.
The general payoff structure of the Prisoner’s Dilemma considered in this study is presented in Table~\ref{tab:pd}. We follow the standard assumptions for the PD payoff relationship, $T > R > P > S$ and $2R>T+S$, while assuming all payoff values are non-negative~\cite{press_dyson_zd_seminal}. 
In addition, for Case 3 (Tit-for-Tat Opponent), we assume that $2P > T + S$.

\begin{table}[h]
\centering
\caption{General payoff matrix for the two-player Prisoner's Dilemma.}\label{tab:pd}
\begin{tabular}{c|c|c}
\toprule
& C & D\\
\hline
C & (R,R) & (S,T) \\
\hline
D & (T,S) & (P,P)\\
\bottomrule
\end{tabular}
\end{table}

To analyze the behavior of AGiR, we introduce simplifying assumptions regarding its internal components. 
% First, to bypass the mathematical complexity arising from the non-linearity of the modulator function $S$, 
For simplicity, assume that $S=0$ and $T$ is sufficiently large so that we can treat $S(T)$ as 0. 
% we assume its values at the extreme payoffs to be $S(T) < \epsilon$ and $S(S) > 1 - \epsilon$, where $\epsilon$ is a sufficiently small positive constant. 
Furthermore, we define the baselines for both agents as $B^1 = B^2 = \frac{\pi}{4}$. This represents the midpoint of the interval $(0, \frac{\pi}{2})$.
% and is indeed a valid representation of the average across the four possible joint-action outcomes\tcr{CHECK!!}. 
% Consistent with Theorems~\ref{thm:existence} and~\ref{thm:uniqueness}, 
For simplicity, we assume $H=1$ in this analysis.
Finally, we assume a base gifting method follows a reward-exchange mechanism with a ratio $0 < \xi < 1$. Specifically, the gift $g^{i,-i}$ from agent $i$ to agent $-i$ is defined as $g^{i,-i} = \xi \cdot r^{i,ext}$.

There are four states $s_1=CC,s_2=CD,s_3=DC,s_4=DD$. 
We define the logistic policy for the player $i$ with policy parameter $\theta^i=(\theta^i_1,\theta^i_2,\theta^i_3,\theta^i_4)$ as follows: for $i=1,2$,
\begin{align*}
    \pi_{\theta^i}^i(a^i=C|CC) =& ~\frac{1}{1+e^{-\theta_1^i}} = ~p_1^i,\\
    \pi_{\theta^i}^i(a^i=C|CD) =& ~\frac{1}{1+e^{-\theta_2^i}} = ~p_2^i,\\
    \pi_{\theta^i}^i(a^i=C|DC) =& ~\frac{1}{1+e^{-\theta_3^i}} = ~p_3^i,\\
    \pi_{\theta^i}^i(a^i=C|DD) =& ~\frac{1}{1+e^{-\theta_4^i}} = ~p_4^i.
\end{align*}
Equivalently, the policy of player $i$ is commonly expressed as $p^i = (p^i_1, p^i_2, p^i_3, p^i_4)$.

For the policy update method,
we employed REINFORCE~\cite{reinforce} style policy update with sample-based policy gradient estimation: $r^i(s,a^i,a^{-i}) \nabla_{\theta^i}\log \pi_{\theta^i}^i(a^i|s)$.
% where $a^i\sim \pi_{\theta^i}^i(\cdot|s)$ and $a^{-i}\sim \pi_{\theta^{-i}}^{-i}(\cdot|s)$.
% Under the logistic policy, 
% REINFORCE with sample-based policy gradient estimation 
Under the logistic policy, the policy gradient update rule using this gradient estimator at state $s_j$ admits the following closed-form expressions:
\begin{align}\label{eq:reinforce_update}
    \tilde{\theta}_j^i = \begin{cases}
        \theta_j^i + \eta~ r^i(1-p_j^i) & \text{for }a^i=C~~\cdots{(a)}\\
        \theta_j^i - \eta~ r^i p_j^i & \text{for }a^i=D~~\cdots{(b)}
    \end{cases}
\end{align}
where $\Tilde{\theta}_j^i$ is the updated parameter and $\eta$ is a learning rate. 
\begin{theorem}[Value improvement under AGiR update]\label{thm:welfare_improvement_append}
Consider the IPD environment, the policy update algorithm, and the gifting method defined in the Problem setting in Section~\ref{append:ipd}. Let $w^i \in [0, 1]$ be the current refusal weight of agent $i$, and let $\tilde{w}^i$ denote the weight updated via AGiR. For any weight $w\in[0,1]$, let $\tilde{\pi}^i(w)$ denote the policy updated with respect to the shaped reward $r^{i,ext} - g^{i,-i} + w g^{-i,i}$. 
Let $v^i(\pi^1, \pi^2) = \mathbb{E}_{s \sim d^{\pi},a^i\sim\pi^i(\cdot|s)}[r^i(s, a)]$ denote the stationary expected reward for agent $i$ (corresponding to the expected average payoff in the game theory field), where $d^{\pi}$ is the stationary distribution under the joint policy $\pi = (\pi^1, \pi^2)$.
Then, the following properties hold:

\begin{enumerate}
    \item \textbf{(AllC Opponent)} If Player 2 adopts an All-Cooperate (AllC) strategy $\pi^2_{AllC}$, the update from $w^i$ to $\tilde{w}^i$ increases the social welfare:
    \[ v^1(\tilde{\pi}^1(\tilde{w}^i), \pi^2_{AllC}) + v^2(\tilde{\pi}^1(\tilde{w}^i), \pi^2_{AllC}) > v^1(\tilde{\pi}^1(w^i), \pi^2_{AllC}) + v^2(\tilde{\pi}^1(w^i), \pi^2_{AllC}). \]

    \item \textbf{(AllD Opponent)} If Player 2 adopts an All-Defect (AllD) strategy $\pi^2_{AllD}$, the update improves the maximin value:
    \[ \min_{j \in \{1,2\}} v^j(\tilde{\pi}^1(\tilde{w}^i), \pi^2_{AllD}) > \min_{j \in \{1,2\}} v^j(\tilde{\pi}^1(w^i), \pi^2_{AllD}). \]

    \item \textbf{(TFT Opponent)} If Player 2 adopts a Tit-for-Tat (TFT) strategy $\pi^2_{TFT}$, the update increases the social welfare:
    \[ v^1(\tilde{\pi}^1(\tilde{w}^i), \pi^2_{TFT}) + v^2(\tilde{\pi}^1(\tilde{w}^i), \pi^2_{TFT}) > v^1(\tilde{\pi}^1(w^i), \pi^2_{TFT}) + v^2(\tilde{\pi}^1(w^i), \pi^2_{TFT}). \]
\end{enumerate}
\end{theorem}

\begin{proof}
Without loss of generality, we show that for Player 1, the policy induced by the weight updated by AGiR is a better response to the opponent than the one induced by the current weight.

We denote by $\tilde{w}$ the refusal weight updated via AGiR from $w$. Likewise, $\tilde{\theta}(w)$ denotes the policy parameter obtained after a policy update using the AGiR-shaped reward (Eq.~\eqref{eq:refuse_reward}) with $w$. 
I.e., $\tilde{\theta}(w)$ is obtained by taking $r^i = r^{i,ext}-g^{i,-i}+wg^{-i,i}$ in Eq.~\eqref{eq:reinforce_update}, case (a) or (b). Similarly, we denote by $\tilde{p}_j^i(w)$ the updated probability that agent $i$ takes action $C$ in state $s_j$ for given $w$: $\tilde{p}_j^i(w) = \frac{1}{1+e^{-\tilde{\theta}_j^i(w)}}$.
To avoid confusion with the sucker payoff $S$, we denote the step-size modulator $S(\cdot)$ from Section~\ref{sec:main_method} by $\mathcal{S}(\cdot)$ throughout this subsection.

We now derive the change of $w$ and $\theta$ in each state.

1) $(a^1,a^2) = (C,C)$

% \begin{align*}
%     (\frac{r^{1,ext}}{g^{-1,1}},\frac{r^{2,ext}}{g^{-2,2}}) = (\frac{R}{kR},\frac{R}{kR}) = (\frac{1}{k},\frac{1}{k})
%     \implies (\phi^1,\phi^2) = (\arctan(\frac{1}{k}),\arctan(\frac{1}{k}))
% \end{align*}
% Here, $\arctan(\frac{1}{k})>\frac{\pi}{4}$ since $k<1$. Then, 
% \begin{align*}
%     &\Delta^1 = (\phi^1 - B^1) ~S(r^{1,ext}) = (\arctan(\frac{1}{k}) - \frac{\pi}{4}) ~S(R) >0,\\
%     &\Delta^2 = (\phi^2 - B^2) ~S(r^{2,ext}) = (\arctan(\frac{1}{k}) - \frac{\pi}{4}) ~S(R) >0.
% \end{align*}
% Then, $\tilde{w}^1>w^1$ and $\tilde{w}^1>w^1$. By (a), $\tilde{\theta}_1^1(\tilde{w})>\tilde{\theta}_1^1(w)$ and $\tilde{\theta}_1^2(\tilde{w})>\tilde{\theta}_1^2(w)$, and thus, $\tilde{p}_1^1(\tilde{w})>\tilde{p}_1^1(w)$ and $\tilde{p}_1^2(\tilde{w})>\tilde{p}_1^2(w)$.

\begin{align*}
    & \frac{r^{1,ext}}{g^{2,1}} = \frac{R}{\xi R} = \frac{1}{\xi }\\
    \implies & \phi^1 = \arctan(\frac{1}{\xi })\\
    \implies & \Delta^1 = (\phi^1 - B^1) ~\mathcal{S}(R) = (\arctan(\frac{1}{\xi }) - \frac{\pi}{4}) ~\mathcal{S}(R) >0
\end{align*}
Here, the last inequality is due to $\arctan(\frac{1}{\xi })>\frac{\pi}{4}$ since $\xi <1$. 
Then, $\tilde{w}^1 = w^1+\alpha\Delta^1>w^1$. 
By Eq.~\eqref{eq:reinforce_update}, case (a), $\tilde{\theta}_1^1(\tilde{w})>\tilde{\theta}_1^1(w)$ and thus, $\tilde{p}_1^1(\tilde{w})>\tilde{p}_1^1(w)$. More explicitly,
\begin{align*}
    \tilde{\theta}_1^1(\tilde{w}) 
= & \theta_1^1 + \eta~ (r^{1,ext}-g^{1,2}+\tilde{w}g^{2,1})(1-p_1^1)\\
= & \theta_1^1 + \eta~ (R-\xi R+\{w^1 +\alpha(\arctan(\frac{1}{\xi})-\frac{\pi}{4})\mathcal{S}(R)\}\cdot \xi R)(1-p_1^1)\\
= & \tilde{\theta}_1^1(w) + \eta\alpha\xi(\arctan(\frac{1}{\xi})-\frac{\pi}{4}) R\cdot \mathcal{S}(R) (1-p_1^1).
\end{align*}

2) $(a^1,a^2) = (C,D)$

% \begin{align*}
%     &(\frac{r^{1,ext}}{g^{-1,1}},\frac{r^{2,ext}}{g^{-2,2}}) = (\frac{S}{kT},\frac{T}{kS}) = (0,\infty)\\
%     \implies & (\phi^1,\phi^2) = (0,\frac{\pi}{2})\\
%     \implies & (\Delta^1,\Delta^2) = (-\frac{\pi}{4}S(S), \frac{\pi}{4}S(T)) = (-\frac{\pi}{4}, 0)
% \end{align*}

% Then, $\tilde{w}^1<w^1$ and $\tilde{w}^2=w^2$. By (a), 
% $\tilde{\theta}_2^1(\tilde{w})<\tilde{\theta}_2^1(w)$ and $\tilde{\theta}_2^2(\tilde{w})=\tilde{\theta}_2^2(w)$,
% and thus, $\tilde{p}_2^1(\tilde{w})<\tilde{p}_2^1(w)$ and $\tilde{p}_2^2(\tilde{w})=\tilde{p}_2^2(w)$.

% As a result, given $a_2 = D$, player 1 is less willing to take action $C$, as doing so would lead to exploitation; given $a_1 = C$, player 2 is already satisfied with the current action $D$, which allows exploitation and thus is inclined to maintain it.

\begin{align*}
    &\frac{r^{1,ext}}{g^{2,1}} = \frac{S}{\xi T} = 0\\
    \implies & \phi^1 = \arctan(0) = 0\\
    \implies & \Delta^1 = -\frac{\pi}{4}\mathcal{S}(S) = -\frac{\pi}{4}
\end{align*}

Then, $\tilde{w}^1 = w^1 - \frac{\pi}{4}\alpha<w^1$. By Eq.~\eqref{eq:reinforce_update}, case (a), $\tilde{\theta}_2^1(\tilde{w})<\tilde{\theta}_2^1(w)$ and thus, $\tilde{p}_2^1(\tilde{w})<\tilde{p}_2^1(w)$.
More explicitly,
\begin{align*}
    \tilde{\theta}_2^1(\tilde{w}) 
= & \theta_2^1 + \eta~ (r^{1,ext}-g^{1,2}+\tilde{w}g^{2,1})(1-p_2^1)\\
= & \theta_2^1 + \eta~ (S-\xi S+(w^1 - \frac{\pi}{4}\alpha)\xi T)(1-p_2^1)\\
= & \tilde{\theta}_2^1(w) - \frac{\pi}{4}\eta\alpha\xi T (1-p_2^1).
\end{align*}

As a result, given $a_2 = D$, Player 1 is less willing to take action $C$, as doing so would lead to exploitation.
% ; given $a_1 = C$, player 2 is already satisfied with the current action $D$, which allows exploitation and thus is inclined to maintain it.

3) $(a^1,a^2) = (D,C)$

% \begin{align*}
%     &(\frac{r^{1,ext}}{g^{-1,1}},\frac{r^{2,ext}}{g^{-2,2}}) = (\frac{T}{kS},\frac{S}{kT}) = (\infty,0)\\
%     \implies & (\phi^1,\phi^2) = (\frac{\pi}{2},0)\\
%     \implies & (\Delta^1,\Delta^2) = (\frac{\pi}{4}S(T), -\frac{\pi}{4}S(S)) = (0,-\frac{\pi}{4})
% \end{align*}
% Therefore, $\tilde{w}^1=w^1$ and $\tilde{w}^1>w^1$. Then,
% $\tilde{\theta}_3^1(\tilde{w})=\tilde{\theta}_3^1(w)$ and by (a), $\tilde{\theta}_3^2(\tilde{w})>\tilde{\theta}_3^2(w)$. As a result, $\tilde{p}_3^1(\tilde{w})=\tilde{p}_3^1(w)$ and $\tilde{p}_3^2(\tilde{w})<\tilde{p}_3^2(w)$.

% As a result, given $a_1 = D$, player 2 is less willing to take action $C$, as doing so would lead to exploitation; given $a_2 = C$, player 1 is already satisfied with the current action $D$, which allows exploitation and thus is inclined to maintain it.

\begin{align*}
    &\frac{r^{1,ext}}{g^{2,1}}= \frac{T}{\xi S} = \infty\\
    \implies & \phi^1 = \frac{\pi}{2}\\
    \implies & \Delta^1 = \frac{\pi}{4}\mathcal{S}(T) = 0
\end{align*}
Therefore by Eq.~\eqref{eq:reinforce_update}, case (b), $\tilde{w}^1=w^1$, $\tilde{\theta}_3^1(\tilde{w})=\tilde{\theta}_3^1(w)$ and thus, $\tilde{p}_3^1(\tilde{w})=\tilde{p}_3^1(w)$.

As a result, given $a_2 = C$, Player 1 is already satisfied with the current action $D$, which allows exploitation and thus is inclined to maintain it.

4) $(a^1,a^2) = (D,D)$

% \begin{align*}
%     &(\frac{r^{1,ext}}{g^{-1,1}},\frac{r^{2,ext}}{g^{-2,2}}) = (\frac{P}{kP},\frac{P}{kP}) = (\frac{1}{k},\frac{1}{k})\\
%     \implies & (\phi^1,\phi^2) = (\arctan(\frac{1}{k}),\arctan(\frac{1}{k}))\\
%     \implies & (\Delta^1,\Delta^2) = ((\arctan(\frac{1}{k})-\frac{\pi}{4})S(P), (\arctan(\frac{1}{k})-\frac{\pi}{4})S(P))
% \end{align*}
% Since $\arctan(\frac{1}{k})>0$, $\Delta^1=\Delta^2>0$ and thus, $w^1,w^2$ increase.
% By Eq. (b), $\tilde{p}_4^1(\tilde{w})<\tilde{p}_4^1(w)$ and $\tilde{p}_4^2(\tilde{w})<\tilde{p}_4^2(w)$. 

% As a result, when $a_2 = D$, player 1 is less willing to take action $C$, as it would lead to being exploited; similarly, when $a_1 = D$, player 2 is less willing to take action $C$ for the same reason.

\begin{align*}
    &\frac{r^{1,ext}}{g^{2,1}} = \frac{P}{\xi P} = \frac{1}{\xi}\\
    \implies & \phi^1 = \arctan(\frac{1}{\xi})\\
    \implies & \Delta^1 = (\arctan(\frac{1}{\xi})-\frac{\pi}{4})\mathcal{S}(P)
\end{align*}
Since $\arctan(\frac{1}{\xi})>\frac{\pi}{4}$, $\Delta^1>0$, then $\tilde{w}^1>w^1$.
By Eq.~\eqref{eq:reinforce_update}, case (b), 
$\tilde{\theta}_4^1(\tilde{w})<\tilde{\theta}_4^1(w)$ and thus, 
$\tilde{p}_4^1(\tilde{w})<\tilde{p}_4^1(w)$. 
More precisely, 
\begin{align*}
    \tilde{\theta}_4^1(\tilde{w}) 
= & \theta_4^1 - \eta~ (r^{1,ext}-g^{1,2}+\tilde{w}g^{2,1})p_4^1\\
= & \theta_4^1 - \eta~ (P-\xi P+\{w^1 +\alpha(\arctan(\frac{1}{\xi})-\frac{\pi}{4})\mathcal{S}(P)\}\cdot \xi P)p_4^1\\
= & \tilde{\theta}_4^1(w) - \eta\alpha\xi(\arctan(\frac{1}{\xi})-\frac{\pi}{4}) P\cdot \mathcal{S}(P) p_4^1.
\end{align*}

As a result, when $a_2 = D$, Player 1 is less willing to take action $C$, as it would lead to being exploited.

% Summarizing 1) to 4), 
% $\tilde{p}_1^1(\tilde{w})>\tilde{p}_1^1(w)$ and $\tilde{p}_1^2(\tilde{w})>\tilde{p}_1^2(w)$

% $\tilde{p}_2^1(\tilde{w})<\tilde{p}_2^1(w)$ and $\tilde{p}_2^2(\tilde{w})=\tilde{p}_2^2(w)$.

% $\tilde{p}_3^1(\tilde{w})=\tilde{p}_3^1(w)$ and $\tilde{p}_3^2(\tilde{w})<\tilde{p}_3^2(w)$.

% $\tilde{p}_4^1(\tilde{w})<\tilde{p}_4^1(w)$ and $\tilde{p}_4^2(\tilde{w})<\tilde{p}_4^2(w)$. 

We define the \emph{value} for each player's policy $p^i$ as the stationary expected reward under the joint stationary distribution $\mathbf{d} = [d_1, d_2, d_3, d_4]^\top$ which is induced by the policy profile $(p^1,p^2)$ (the \emph{expected payoff} in game-theoretic terms~\cite{press_dyson_zd_seminal}). The stationary distribution is defined over the four possible outcomes: $s \in \{(C,C), (C,D), (D,C), (D,D)\}$. The values for Player 1 ($v^1$) and Player 2 ($v^2$) are expressed as 
$v^1 = R\cdot d_1 + S\cdot d_2 + T\cdot d_3 + P\cdot d_4$ and 
$v^2 = R\cdot d_1 + T\cdot d_2 + S\cdot d_3 + P\cdot d_4$. 

% \tcr{SEE sec 3.2.2 in Malik. 1 only can set 2's score. so, use determinant expression directly!!}

% ----------------------------------\\
% \tcr{this proof do not use ZD strategy, but only use the result for the stationary expected payoff. Delete this later\\}

Let $F^1 = [R,S,T,P]^\top$ and $F^2 = [R,T,S,P]^\top$.
From Eq~(5) in \cite{press_dyson_zd_seminal}, the values of Player 1 and 2 induced by any policy profile $(p,q)$ (equivalently, the expected payoff with respect to the stationary distribution induced by strategy profile $(p,q)$) are expressed as follows:
\begin{align}\label{eq:value_det}
    v^1 = \frac{D(p,q,F^1)}{D(p,q,\mathbf{1})}, v^2 = \frac{D(p,q,F^2)}{D(p,q,\mathbf{1})}
\end{align}
where 
\begin{align}\label{eq:det}
    D(p,q,f) := det
    \begin{bmatrix}
        p_1q_1-1 & p_1-1 & q_1-1 & f_1\\
        p_2q_3 & p_2-1 & q_3 & f_2\\
        p_3q_2 & p_3 & q_2-1 & f_3\\
        p_4q_4 & p_4 & q_4 & f_4
    \end{bmatrix}
\end{align}
for any $f\in\mathbb{R}^4$. 
We analyze from this result against three representative strategies for IPD; \emph{AllC}: $q=[1,1,1,1]^\top$,\emph{AllD}: $q=[0,0,0,0]^\top$, and \emph{Tit-for-Tat}: $q=[1,1,0,0]^\top$.

\textbf{Case 1. \emph{AllC} Opponent.} Suppose that Player 2 takes the \emph{AllC} (All-Cooperate) strategy, i.e. $q=[1,1,1,1]^\top$. % Against AllC, AM increases.
Then, by straightforward calculation, we obtain
\begin{align*}
    D(p,AllC,f) = &det
    \begin{bmatrix}
        p_1-1 & p_1-1 & 0 & f_1\\
        p_2 & p_2-1 & 1 & f_2\\
        p_3 & p_3 & 0 & f_3\\
        p_4 & p_4 & 1 & f_4
    \end{bmatrix}\\
    = & (p_1-1)f_3 - p_3 f_1.
\end{align*}
Then, $v^1 = \frac{(1-p_1)T + p_3 R}{1-p_1 + p_3}$ and $v^2 = \frac{(1-p_1)S + p_3 R}{1-p_1 + p_3}$. Thus, $v^1+v^2 = \frac{(1-p_1)(S+T) + p_3 \cdot 2R}{1-p_1 + p_3}$.
From the results in 1) and 3), AGiR increases $p_1$ and does not change $p_3$.
Then, $\frac{1-p_1}{1-p_1 + p_3}$, the coefficient for $(S+T)$, decreases and $\frac{p_3}{1-p_1 + p_3}$, the coefficient for $2R$ increases. Due to the standard assumption for IPD, $2R>S+T$, we conclude that $v^1+v^2$ increases.

\textbf{Case 2. \emph{AllD} Opponent.} Suppose that Player 2 takes the \emph{AllD} (All Defect) strategy, i.e. $q=[0,0,0,0]^\top$. % Against AllD, Min increases.
Then, by straightforward calculation, we obtain
\begin{align*}
    D(p,AllD,f) =&  det
    \begin{bmatrix}
        -1 & p_1-1 & -1 & f_1\\
        0 & p_2-1 & 0 & f_2\\
        0 & p_3 & -1 & f_3\\
        0 & p_4 & 0 & f_4
    \end{bmatrix}\\
    = & (p_2-1) f_4 - p_4 f_2.
\end{align*}
% Then, $v^1 = \frac{(1-p_2)P + p_4 S}{1-p_2 + p_4}$ and $v^2 =  \frac{(1-p_2)P + p_4 T}{1-p_2 + p_4}$. Thus, $v^1+v^2 =  \frac{(1-p_2)\cdot 2P + p_4 (S+T)}{1-p_2 + p_4}$.
Then, $v^1 = \frac{(1-p_2)P + p_4 S}{1-p_2 + p_4}$ and $v^2 =  \frac{(1-p_2)P + p_4 T}{1-p_2 + p_4}$.
Since $S<T$, the minimum of the values is $\min\{v^1,v^2\}=v^1= \frac{(1-p_2)P + p_4 S}{1-p_2 + p_4}$.
Since $v^1$ depends only on $p_2$ and $p_4$, it suffices to consider policy updates only in states $s_2 = CD$ and $s_4 = DD$.

In the state $CD$, from the result in 2), $p_2$ is decreased by AGiR. Then, $\frac{1-p_2}{1-p_2 + p_4}$ increases and $\frac{p_4}{1-p_2 + p_4}$ decreases. Since $P>S$, $v^1$ increases.
In the state $DD$, from the result in 4), $p_4$ is decreased by AGiR. Then, $\frac{1-p_2}{1-p_2 + p_4}$ increases and $\frac{p_4}{1-p_2 + p_4}$ decreases. Since $P>S$, $v^1$ increases.
Therefore, AGiR increases $v^1$, which is the minimum value $\min\{v^1,v^2\}$.

\begin{comment}
From 2) and 4), $p_2$ and $p_4$ decrease by AGiR. 
Denote AGiR-updated strategy at state $CD,DD$ by $\tilde{p}_2 = p_2 - \Delta_2$ and $\tilde{p}_4 = p_4 - \Delta_4$ with $\Delta_2,\Delta_4>0$.
Then, since \tcr{[Here, write the reason]}, $\Delta_2>\Delta_4$.
\tcr{NOTE. update is one-by-one, i.e. p2 and p4 are not updated simultaneously. $\to$ for v1, both are good!\\}
We show the following inequality:
\begin{align*}
    \frac{1-\tilde{p}_2}{1-\tilde{p}_2 + \tilde{p}_4} = \frac{1-p_2 + \Delta_2}{1-p_2 + p_4 + (\Delta_2-\Delta_4)} > \frac{1-p_2}{1-p_2 + p_4}.
\end{align*}
By some calculations, the inequality becomes equivalent to $p_4 \Delta_2 > -(1-p_2)\Delta_4$, that obviously holds since $LHS>0>RHS$.
This means that, AGiR increases the weight of $P$ while decreases the weight of $S$ in $v^1$. Therefore, AGiR increases $v^1$, i.e. $\min\{v^1,v^2\}$ (Min).
\end{comment}

\textbf{Case 3. \emph{Tit-for-Tat} Opponent.} Suppose that Player 2 takes the \emph{Tit-for-Tat} (TFT) strategy, i.e. $q=[1,1,0,0]^\top$.
In this case, since $D(p,q,f) = D(p,q,\mathbf{1})=0$, we analyze with $q_\epsilon=[1-\epsilon,1-\epsilon,\epsilon,\epsilon]^\top$ for $\epsilon>0$ and take limit $\epsilon\to0+$.
Then, the third columns in Eq.~\eqref{eq:det} for both denominator and numerator become $\epsilon[-1,1,-1,1]^\top$.
Therefore, 
\begin{align*}
    &\frac{D(p,q_\epsilon,f)}{D(p,q_\epsilon,\mathbf{1})}\\
    = & \epsilon \cdot det
    \begin{bmatrix}
        p_1(1-\epsilon)-1 & p_1-1 & -1 & R\\
        p_2\epsilon & p_2-1 & 1 & S\\
        p_3(1-\epsilon) & p_3 & -1 & T\\
        p_4\epsilon & p_4 & 1 & P
    \end{bmatrix} 
    \left( \epsilon \cdot det
    \begin{bmatrix}
        p_1(1-\epsilon)-1 & p_1-1 & -1 & 1\\
        p_2\epsilon & p_2-1 & 1 & 1\\
        p_3(1-\epsilon) & p_3 & -1 & 1\\
        p_4\epsilon & p_4 & 1 & 1
    \end{bmatrix} \right)^{-1}\\
    = & det
    \begin{bmatrix}
        p_1(1-\epsilon)-1 & p_1-1 & -1 & R\\
        p_2\epsilon & p_2-1 & 1 & S\\
        p_3(1-\epsilon) & p_3 & -1 & T\\
        p_4\epsilon & p_4 & 1 & P
    \end{bmatrix} 
    \left( det
    \begin{bmatrix}
        p_1(1-\epsilon)-1 & p_1-1 & -1 & 1\\
        p_2\epsilon & p_2-1 & 1 & 1\\
        p_3(1-\epsilon) & p_3 & -1 & 1\\
        p_4\epsilon & p_4 & 1 & 1
    \end{bmatrix} \right)^{-1}.
\end{align*}
Notably, the singularity is eliminated due to the cancellation of $\epsilon$ in the third column, thus we now take $\epsilon\to0+$. 
After taking the limit $\epsilon\to0+$, a straightforward calculation yields
\begin{align*}
    &\frac{D(p,q,f)}{D(p,q,\mathbf{1})}\\
    = & det
    \begin{bmatrix}
        p_1-1 & p_1-1 & -1 & f_1\\
        0 & p_2-1 & 1 & f_2\\
        p_3 & p_3 & -1 & f_3\\
        0 & p_4 & 1 & f_4
    \end{bmatrix} 
    \left( det
    \begin{bmatrix}
        p_1-1 & p_1-1 & -1 & 1\\
        0 & p_2-1 & 1 & 1\\
        p_3 & p_3 & -1 & 1\\
        0 & p_4 & 1 & 1
    \end{bmatrix} \right)^{-1}\\
    = & \frac{1}{2}\frac{(1-p_1)f_3+p_3f_1}{1-p_1+p_3} + \frac{1}{2}\frac{(1-p_2)f_4+p_4f_2}{1-p_2+p_4}.
\end{align*}
Thus,
\begin{align*}
    v^1 = & \frac{1}{2}\frac{(1-p_1)T+p_3R}{1-p_1+p_3} + \frac{1}{2}\frac{(1-p_2)P+p_4S}{1-p_2+p_4},\\
    v^2 = & \frac{1}{2}\frac{(1-p_1)S+p_3R}{1-p_1+p_3} + \frac{1}{2}\frac{(1-p_2)P+p_4T}{1-p_2+p_4},~\text{   and}\\
    v^1+v^2 = & \frac{1}{2}\frac{(1-p_1)(T+S)+p_3\cdot 2R}{1-p_1+p_3} + \frac{1}{2}\frac{(1-p_2)\cdot 2P+p_4(T+S)}{1-p_2+p_4}.
\end{align*}

In the state $CC$, from the result in 1), $p_1$ is increased by AGiR. Then, $\frac{1-p_1}{1-p_1 + p_3}$ decreases and $\frac{p_3}{1-p_1 + p_3}$ increases. Since $2R>T+S$, $v^1+v^2$ increases.
In the state $CD$, from the result in 2), $p_2$ is decreased by AGiR. Then, $\frac{1-p_2}{1-p_2 + p_4}$ increases and $\frac{p_4}{1-p_2 + p_4}$ decreases. As we additionally assumed $2P>T+S$ for the Case 3, $v^1+v^2$ increases.
In the state $DC$, from the result in 3), $p_3$ is not changed by AGiR. Then, $v^1+v^2$ is also unchanged.
In the state $DD$, from the result in 4), $p_4$ is decreased by AGiR. Then, $\frac{1-p_2}{1-p_2 + p_4}$ increases and $\frac{p_4}{1-p_2 + p_4}$ decreases. As we additionally assumed $2P>T+S$ for the Case 3, $v^1+v^2$ increases.
In all states, the collective sum is unchanged or increased. Therefore, AGiR increases $v^1+v^2$. 
\end{proof}

\subsection{Bridging Theory and Practice}\label{append:theory_practice}

Several aspects of our theoretical assumptions align with our experimental setup.
In our experiments, the state and action spaces are finite, satisfying Assumption~\ref{main:assumption_1}; moreover, the neural-network policies are continuous in the refusal weights, consistent with Assumption~\ref{main:assumption_2}.
These assumptions concern the theoretical analysis and do not restrict the general formulation of AGiR.
While Theorem~\ref{thm:existence} establishes existence for any fixed finite $H\ge1$, Theorem~\ref{thm:uniqueness} and the IPD analysis assume $H=1$ for analytical simplicity.
Nevertheless, Fig.~\ref{fig:ablation_H} in Appendix~\ref{append:ablation_modulator} shows that AGiR maintains higher GM and Min than vanilla gifting across diverse window sizes.
The parameter bounds in Assumption~\ref{main:assumption_4} ensure positive multiplicative update factors.
Although the assumption treats $B$ as independent of the refusal weights, Fig.~\ref{fig:ablation_B} in Appendix~\ref{append:ablation_modulator} shows that even a fixed baseline yields higher GM and Min than vanilla gifting in practice.
Furthermore, Fig.~\ref{fig:in_epi_analysis} in Section~\ref{sec:within_epi_dyn} shows that RGO $\phi_t$ oscillates around the baseline $B_t$, leading to both upward and downward refusal-weight adjustments.
While this illustrates adaptive behavior, the low-weight recovery and uniform regeneration conditions in Assumption~\ref{main:assumption_5} remain sufficient theoretical conditions rather than empirically verified properties.
Finally, the Coin Game is a spatiotemporal extension of the IPD; while Theorem~\ref{thm:welfare_improvement_append} does not formally extend to it, it serves as an empirical testbed for the same dilemma structure in a richer multi-agent setting.

\section{Implementation Details}\label{append:Implementation_Details}

\subsection{Implementation Details for AGiR}\label{append:practical_algo}
We implement our gift refusal method on existing gifting methods on a PPO~\cite{schulman2017proximal} backbone. We set the baseline $B_t^i$ to the average orientation $\bar{\phi}_t^i$ across all parallel PPO environments and clip the updated refusal weight to $[0,1]$ at each time step. 
At every episode reset, each agent in each parallel environment independently samples its initial refusal weight uniformly at random in $[0,1]$. The initial weights therefore vary across agents and episodes and are not shared within an environment.
All experiments were conducted on a workstation equipped with 8 NVIDIA GeForce RTX 3090 GPUs (24GB VRAM each).
The pseudo-code for incorporating AGiR into general gifting is provided within Algorithm~\ref{alg:adaptive_refusal} of Appendix~\ref{append:pseudo_code}.

% \subsection{Hyperparameters for AGiR}
We used a step size of $\alpha=0.01$ and maintained a consistent window size of $H=50$ across all experimental environments. The specific values assigned to the step size modulator constant $C$ are summarized in Table~\ref{table_hyperparam_C}.

\begin{table}[h]
\caption{Values for $C$ across experimental environments}
\label{table_hyperparam_C}
\begin{center}
%\footnotesize
\begin{tabular}{lrrr}
\toprule
Env. & LASE & RX & LIO   \\
\midrule
Cleanup-Easy & 1 & 1 & -  \\
Cleanup-Hard & 100 & 100 & -  \\
Harvest & 100 & 100 & -  \\
Coin Game   & 50 & 100 & 10  \\

\bottomrule
\end{tabular}
\end{center}
\end{table}

\subsection{Environments}\label{append:env_detail}

\begin{figure}[t]
  \centering
  \includegraphics[width=\linewidth]{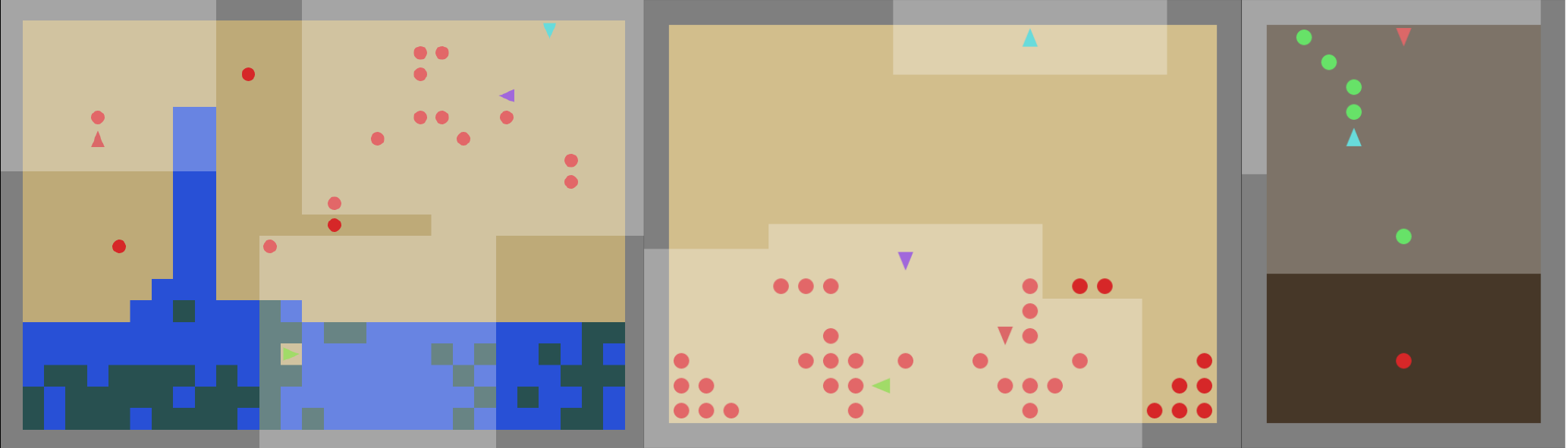}
  \caption{Illustrations of the environments: Unfair Cleanup, Unfair Harvest, and Unfair Coin Game.}
  \label{fig:envs_append}
\end{figure}

For all environments, we used an episode length of 500, 2 epochs, and a total of $3\times10^8$ training timesteps.
In the Unfair Cleanup environment, we set 100 rollout steps, 4 minibatches, and 256 parallel environments.
In the Unfair Harvest environment, we set 500 rollout steps, 4 minibatches, and 64 parallel environments.
In the Unfair Coin Game environment, we set 500 rollout steps, 500 minibatches, and 64 parallel environments. We provide the maps in Listing~\ref{fig:all_maps} and Fig.~\ref{fig:envs_append}, where each symbol follows the rules described in the SocialJax suite~\cite{socialjax}.
Advantaged agents are randomly spawned at positions labeled ‘Q’, whereas disadvantaged agents are randomly spawned at positions labeled ‘P’.

In the Unfair Cleanup and Unfair Harvest environments, the apple spawn probability rules follow the default configurations provided by SocialJax suite. For the Unfair Coin Game, the coin spawn probabilities are set to $0.001$ for green coins and $0.00025$ for red coins.

\begin{figure}[h]
\centering
\begin{lstlisting}[style=asciimap,caption={Cleanup (Easy and Hard) and Harvest maps},label={fig:all_maps}]
HFFFHFFHFHFHFHFHFHFHHFHFFFHF  HFFFHFFHFHFHFHFHFHFHHFHFFFHF  AAA    A      A    AAA
HFHFHFFHFHFHFHFHFHFHHFHFFFHF  HFHFHFFHFHFHFHFHFHFHHFHFFFHF  AA    AAA    AAA    AA
HFFHFFHHFHFHFHFHFHFHHFHFFFHF  HFFHFFHHFHFHFHFHFHFHHFHFFFHF  A    AAAAA  AAAAA    A
HFHFHFFHFHFHFHFHFHFHHFHFFFHF  HFHFHFFHFHFHFHFHFHFHHFHFFFHF        AAA    AAA      
HFFFFFFHFHFHFHFHFHFHHFHFFFHF  HFFFFFFHFHFHFHFHFHFHHFHFFFHF         A      A       
==============+~FHHHHHHf====  ==============+~FHHHHHHf====      AAAA      AAAA    
   P    P      ===+~SSf  P       P    P      ===+~SSf  P          Q        Q      
                   <~Sf                          <~Sf     
^T^T^T^T^T^T^ T^T <~S>^T^T^T                     <~S>     
    BBB    BBBB   <~S> B   B                    <~S>       
  BBBBBB  BBBB   B<~S>  BBB   ^T^T^T^T^T^T^T^ ^T<~S> T^T^T
  BBB BB   BBBB   <~S>  BBB    BBB BB           <~S>   BBB
  BBB       BB   B<~S>         BBBBBB    BB     <~S>   BBB
   Q  BBBBBB      <~S>  BB              BBBB    <~S>          PPPPPPPPPPPPPPPPPP  
    BBBBBBQ       ;~S, QBB                      ;~S,         PPPPPPPPPPPPPPPPPPPP 
    BBBBQBB    QBBBssBBB            BBBBB   BBBBssBBB       PPPPPPPPPPPPPPPPPPPPPP
    BBBBBBBB   BBBBBBBBBB        Q BBBBBBBB  BBBBBBBBBBBB 
  BBBBBBBBBBBBBBBBBBBBBB      BBBBBBBBBBQBBBBBBBBBQBBBBBB 
                              BBBBBBBBQBBBBBBQBBBBBBBBBBBB
\end{lstlisting}
\end{figure}

\subsection{Baselines}
For common settings, following the SocialJax suite, we use actor-critic policies incorporating a convolutional neural network (CNN). The CNN module consists of $5\times5$, $3\times3$, and $3\times3$ kernels with 32 channels each, followed by a dense layer with 64 units. We use ReLU as the activation function. Both the actor and critic networks share the aforementioned CNN architecture, followed by two dense layers with 64 units and ReLU activations. We employed a learning rate of $0.0005$ with annealing, a discount factor of $\gamma=0.99$, and PPO-related hyperparameters (including a GAE coefficient of $\lambda=0.95$, a clip coefficient of $0.2$, an entropy coefficient of $0.01$, a value-function coefficient of $0.5$, and a maximum gradient norm of $0.5$), following the default configurations in~\citet{socialjax}.

% For SVO~\cite{svo}, we used an ideal angle of $45^\circ$ and an SVO weight of $0.5$. For IA~\cite{ia}, we set $\alpha=5$ and $\beta=0.05$. 
For LASE~\cite{lase}, to design the social-relationship modules $\pi_{SR}^i$ and $Q_{SR}^i$, we used a lightweight CNN for training stability: a $3\times3$ kernel with 16 channels, followed by two dense layers with 64 units and the action dimension, respectively, with ReLU activations. We excluded the imagination module to isolate the effect of the gifting mechanism from potential training instability. Additionally, we set the learning rates for $\pi_{SR}^i$ and $Q_{SR}^i$ to $0.0005$ and the number of gradient steps to 50. 
For Reward Exchange~\cite{minimal_reward_transfer}, we set the transfer ratio to $0.5$. For LIO~\cite{lio}, to design the separate gifting function, we used a lightweight convolutional module: a $3\times3$ kernel with 16 channels, followed by two dense layers with 64 units and the number of agents, respectively, with ReLU activations. 
To leverage the provided baselines in the SocialJax suite, we reimplemented all baseline methods using a PPO backbone.

% \newpage
\section{Additional Metric Definitions and Results} 

\subsection{Metric Definitions} \label{append:metrics}

We provide the formal definitions of the metrics below, where $x_i$ denotes the return of agent $i$. Since returns can be negative in the Coin Game, we use a signed geometric mean for this environment.

\begin{itemize}
    \item \textbf{$\alpha$-fairness}~\citep{mo2002fair}:
    \begin{equation}
    U_\alpha(x_1,\ldots,x_n) =
    \begin{cases}
    \sum_{i=1}^n \frac{x_i^{1-\alpha}}{1-\alpha} & \alpha \neq 1 \\
    \sum_{i=1}^n \log x_i & \alpha = 1
    \end{cases}
    \end{equation}

    \item \textbf{Gini coefficient}~\cite{david1968miscellanea}:
    \begin{equation}
    \mathrm{G}(x_1,\ldots,x_n) = \frac{\sum_{i=1}^n \sum_{j=1}^n |x_i - x_j|}{2n \sum_{i=1}^n x_i}
    \end{equation}

    \item \textbf{Jain's index}~\cite{jain}:
    \begin{equation}
    J(x_1,\ldots,x_n)= \frac{(\sum_i x_i)^2}{n\sum_i x_i^2}
    \end{equation}
\end{itemize}

\subsection{Comparison in Jain's Index}\label{append:jain}
We also consider a fairness metric not compared in the main text: Jain’s index, $J(x_1,\ldots,x_n)= \frac{(\sum_i x_i)^2}{n\sum_i x_i^2}$~\cite{jain}. It equals 1 under best fairness and $\frac{1}{n}$ under worst fairness; thus, larger is fairer.
Note that while this metric captures fairness, it does not capture efficiency because it is normalized metric. For this reason, we used 
$\alpha$-fairness in the main text. For completeness, however, we also report Jain’s index here. For the Unfair Coin Game, we report Jain’s index only for positive values.
Table~\ref{tab:jain} shows that AGiR improves Jain’s index across all environments and all base gifting methods. Entries for which Jain’s index is undefined due to zero or negative values are denoted as NaN.

\begin{table}[h]
\caption{Jain's index in Unfair Cleanup, Unfair Harvest, and Unfair Coin Game environments.}
\label{tab:jain}
\begin{center}
\footnotesize
\begin{tabular}{llrrrr}
\toprule
\multicolumn{2}{l}{Algorithm} & Cleanup-Easy $\uparrow$ & Cleanup-Hard $\uparrow$ & Harvest $\uparrow$ & Coin Game $\uparrow$ \\
\hline\hline
\addlinespace[2pt]
% \multicolumn{2}{l}{SVO}    & NaN     & NaN     & 0.6391 & 0.0140 \\
\multicolumn{2}{l}{IPPO-I} & NaN     & NaN     & 0.8135 & 0.0267 \\
\multicolumn{2}{l}{IPPO-C} & 0.5743 & 0.4970 & 0.2894 & 0.7150 \\
\multicolumn{2}{l}{IA}     & 0.9810 & 0.9741 & 0.8101 & 0.0335 \\
\hline\hline
\addlinespace[2pt]
\multirow{2}{*}{LASE} & Vanilla & 0.9748 & 0.8260 & 0.6917 & NaN \\
                      & AGiR    & \textbf{0.9838} & \textbf{0.9683} & \textbf{0.7859} & NaN \\
\midrule
\multirow{2}{*}{RX}   & Vanilla & 0.9034 & 0.8781 & 0.6770 & 0.7386 \\
                      & AGiR    & \textbf{0.9785} & \textbf{0.9687} & \textbf{0.7984} & \textbf{0.9505} \\
\bottomrule
\end{tabular}
\end{center}
\end{table}

\newpage

\section{Experimental Results on the Standard Environments} \label{append:symmetric}

The empirical results in Table~\ref{tab:fairenv} demonstrate that gift over-reliance is not limited to our asymmetric variants and can also arise in standard MMG settings. Across the standard environments, adding AGiR generally increases GM and Min relative to vanilla gifting. Notably, in Cleanup-Hard, AGiR significantly boosts GM and Min metrics while effectively suppressing the Gini coefficient. Even in Harvest, where baseline performance is already relatively balanced, AGiR maintains stability and further minimizes inequality. Overall, these results indicate that AGiR can mitigate gift over-reliance even without the structural asymmetries introduced in our main experiments.

\begin{table}[h]
\centering
\caption{$\alpha$-fairness and Gini coefficient in standard environments}
\label{tab:fairenv}
\small
\begin{tabular}{llcccc}
\toprule
Env. & Algorithm & AM $\uparrow$ & GM $\uparrow$ & Min $\uparrow$ & Gini $\downarrow$ \\
\midrule
\multirow{4}{*}{Cleanup-Hard} & LASE & 291 (61) & 104 (148) & 82 (116) & 0.25 (0.12) \\
 & LASE + AGiR & 252 (66) & 177 (119) & 150 (104) & 0.12 (0.09) \\
 \cmidrule{2-6}
 & RX & 620 (136) & 73 (147) & 60 (120) & 0.24 (0.07) \\
 & RX + AGiR & 527 (206) & 216 (176) & 187 (154) & 0.14 (0.11) \\
\midrule
\multirow{4}{*}{Harvest} & LASE & 72 (1) & 71 (0.8) & 58 (2.7) & 0.08 (0.01) \\
 & LASE + AGiR & 72 (0.7) & 71 (0.6) & 61 (0.9) & 0.08 (0.01) \\
 \cmidrule{2-6}
 & RX & 73 (0.6) & 71 (0.7) & 55 (2) & 0.12 (0.02) \\
 & RX + AGiR & 71 (0.6) & 71 (0.8) & 61 (1.8) & 0.07 (0.01) \\
\bottomrule
\end{tabular}
\end{table}

% \newpage

\section{Additional Experimental Results in Unfair Cleanup-Easy}\label{append:cleanup_easy}
% \subsection{Experimental results in Unfair Cleanup-Easy}\label{append:cleanup_easy}

\begin{figure*}[h]
  \centering
  
  % --- 왼쪽: 표 (Table - Unfair Harvest) ---
  \begin{minipage}[t]{0.48\textwidth}
    \vspace{0pt} % 상단 기준점 고정
    \centering
    \captionof{table}{$\alpha$-fairness, Gini coefficient, and GRR in Unfair Cleanup-Easy}
    \label{table_cleanup_easy}
    \tiny
    \setlength{\tabcolsep}{3pt}
    \begin{tabular}{llrrrrr}
        \toprule
    \multicolumn{2}{l}{Algorithm} & AM $\uparrow$ & GM $\uparrow$ & Min $\uparrow$ & Gini $\downarrow$ & GRR $\downarrow$ \\
    % \cmidrule(lr){1-2}
    % Method & Variant &  &  &  &  \\
    \hline\hline
    \addlinespace[2pt]
    % \multicolumn{2}{l}{SVO}     & 0.115 (0.2) & 0.093 (0.2) & 0.047 (0.1) & 0.15 (0.19) \\
    \multicolumn{2}{l}{IPPO-I} & 206 (169) & 204 (167) & 170 (142) & 0.13 (0.18) & - \\
    \multicolumn{2}{l}{IPPO-C} & 310 (226) & 58 (82)   & 6 (6)     & 0.46 (0.13) & - \\
    \multicolumn{2}{l}{IA}     & 196 (135) & 194 (133) & 159 (113) & 0.07 (0.03) & - \\
    \hline\hline
    \addlinespace[2pt]
    \multirow{2}{*}{LASE} & Vanilla & 311 (78)  & 307 (77)  & 253 (71)  & 0.08 (0.02) & 0.14 \\
                          & AGiR  & \textbf{349} (19)  & \textbf{346} (19)  & \textbf{290} (17)  & \textbf{0.07} (0.01) & \textbf{0.06} \\
    \midrule
    \multirow{2}{*}{RX}   & Vanilla &\textbf{ 467} (68)  & 335 (168) & 242 (137) & 0.16 (0.1) & 0.37 \\
                          & AGiR  & 415 (14)  & \textbf{410} (13)  & \textbf{339} (25)  & \textbf{0.08} (0.01) & \textbf{0.19} \\
    \bottomrule
    \end{tabular}
  \end{minipage}
  \hfill % 두 요소 사이 간격
  % --- 오른쪽: 그림 (Figure - Unfair Harvest) ---
  \begin{minipage}[t]{0.48\textwidth}
    \vspace{0pt} % 표의 캡션 높이와 이미지 윗부분을 맞춤
    \centering
    \includegraphics[width=\linewidth]{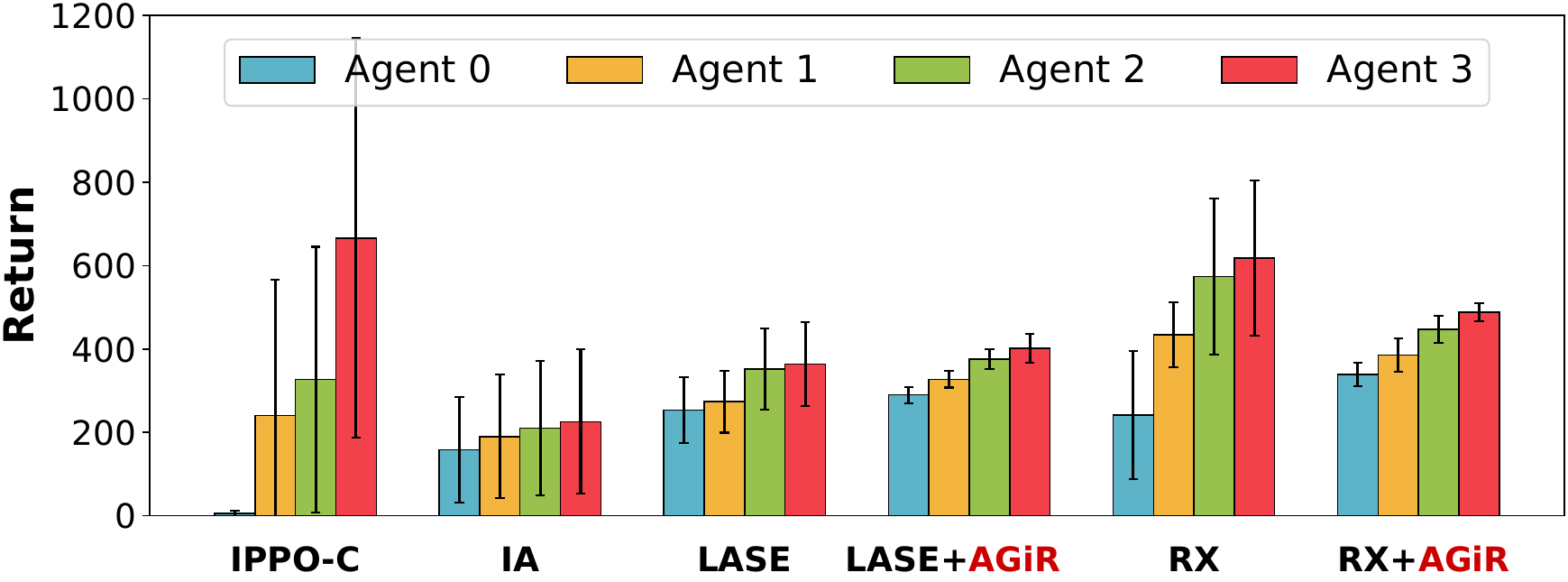}
    \captionof{figure}{Per-agent return distributions averaged over five seeds in Unfair Cleanup-Easy.}
    \label{fig:cleanup_easy_agent_bar}
  \end{minipage}

\end{figure*}

In Unfair Cleanup, we determine difficulty by the accessibility of resources. In the \emph{Easy} level, apples spawn near the river, allowing disadvantaged agents to receive gifts for cleaning while still easily discovering harvesting through exploration. In contrast, the \emph{Hard} level places apples far from the river. This makes reaching apple sites through exploration alone substantially more difficult, often leading agents to become locked into cleaning roles as they rely on gift signals instead of discovering independent harvesting behaviors.
The results for the \emph{Easy} level, specifically the $\alpha$-fairness metrics and Gini coefficient, are summarized in Table~\ref{table_cleanup_easy}, while the per-agent return distributions are illustrated in Fig.~\ref{fig:cleanup_easy_agent_bar}.

\section{Details of Two-Agent Cleanup Experiments}\label{append:two_agent}

\subsection{Experimental Setup}
\label{append:two_agent_cleanup_setup}

To further examine the behavioral effects of positional and competence asymmetries discussed in Section~\ref{sec:specialization}, we construct a controlled two-agent variant of Cleanup. The environment uses the same map as Unfair Cleanup-\emph{Easy}, but contains two agents. Unless otherwise specified below, the remaining environment parameters follow those used in the main Cleanup experiments.

We control agent competence along two task-specific dimensions. First, each agent $i$ is assigned a harvesting efficiency $\epsilon_i$, such that harvesting one apple yields a reward of $2\epsilon_i$ in the two-agent environment. Second, each agent is assigned a cleaning-beam length $L_i$. The cleaning beam extends $L_i$ cells forward and $L_i$ cells to each side, covering $3L_i+1$ cells including the agent's current cell; $L_i=1$ corresponds to the original cleaning beam. In settings with asymmetric competence, agent 0 is the \emph{talented cleaner}, with a wider cleaning beam but lower harvesting efficiency, while agent 1 is the \emph{talented harvester}, with a narrower cleaning beam but higher harvesting efficiency.

We also vary the agents' initial positions. Under random spawn, two distinct positions are sampled uniformly from the 19 candidate spawn locations marked by \texttt{P} on the map. Under fixed spawn, one agent is initialized near the river and the other near the apple region. The river-side position is sampled from
$\{(6,3),(6,8),(6,25)\}$,
and the apple-side position is sampled from
$\{(13,3),(14,10),(15,15)\}$.

Combining positional and competence asymmetries yields the four scenarios summarized in Table~\ref{tab:two_agent_scenarios}.
In the \emph{Mismatch} scenario, initial position and task competence favor opposite role assignments: the talented harvester spawns on the river side, whereas the talented cleaner spawns on the apple side.

\begin{table}[t]
\centering
\caption{Experimental scenarios for the controlled two-agent Cleanup study.}
\label{tab:two_agent_scenarios}
\begin{tabular}{lccc}
\toprule
Scenario & Spawn & $(\epsilon_0,\epsilon_1)$ & $(L_0,L_1)$ \\
\midrule
No gap
& Random
& $(1,1)$
& $(4,4)$ \\
Competence gap
& Random
& $(0.1,1)$ or $(0.25,1)$
& $(4,1)$ \\
Spawn gap
& Fixed
& $(1,1)$
& $(4,4)$ \\
Mismatch
& Fixed
& $(0.25,1)$
& $(4,1)$ \\
\bottomrule
\end{tabular}
\end{table}

\subsection{Training Protocol}
\label{append:two_agent_cleanup_training}

We use LASE~\citep{lase} as the base gifting method for all two-agent experiments. The policy architecture and PPO training configuration follow those used in the main Cleanup experiments. Each model is trained for $3\times10^8$ timesteps. For AGiR, we use $\alpha=0.01$, $H=50$, and $C=1$ throughout the two-agent experiments.

We evaluate both task behavior and resulting returns. In addition to extrinsic return, we report the number of apples harvested and cleaning actions per episode to characterize the roles learned by each agent. We also report total return, defined as the extrinsic return minus outgoing gifts plus all incoming gifts. Refused gifts are therefore excluded only from the recipient's policy-update signal and are not removed from the total return.

\subsection{Additional Quantitative Results}
\label{append:two_agent_cleanup_results}

Table~\ref{tab:two_agent_results} reports the final extrinsic return, number of cleaning actions per episode, and total return for each scenario. The results complement the behavioral analysis in Section~\ref{sec:specialization} by providing the corresponding agent-level outcomes.

\noindent\textbf{No gap.}
In this symmetric two-agent setting, LASE itself does not exhibit clear role lock-in, and adding AGiR does not systematically improve the outcome. This does not contradict the results in Appendix~\ref{append:symmetric}, where gift over-reliance can arise even in standard environments: the present experiment differs in the number of agents, uses the Cleanup-\emph{Easy} map, and gives both agents a wider cleaning beam. We therefore treat the absence of lock-in here as specific to this controlled setting.

\noindent\textbf{Competence gap.}
When competence alone is asymmetric, AGiR allows the talented cleaner to explore harvesting before settling on a role. At $0.1$ times the harvester's reward per apple, this exploration does not persist, and the final behavioral division remains dominated by cleaning. At $0.25$ times, however, the talented cleaner maintains substantially more harvesting while continuing to contribute to cleaning, consistent with the learning curves in Fig.~\ref{fig:competence_gap}. The talented harvester's return decreases under AGiR in these settings, so these results should be interpreted as a change in learned role allocation rather than a uniform improvement for both agents.

\noindent\textbf{Spawn gap.}
With equal competence but asymmetric initial positions, the river-side agent performs more cleaning and earns less under LASE. 
Despite performing more cleaning, the river-side agent's total return under LASE is only 246.7, barely above its extrinsic return of 245.7, indicating that gift transfers provide little net compensation for its cleaning burden.
Under AGiR, its extrinsic return increases from $245.7$ to $276.1$ while its cleaning actions decrease from $49.0$ to $36.8$ per episode. The apple-side agent's extrinsic return also increases slightly, from $291.5$ to $302.1$.

\noindent\textbf{Mismatch.}
When position and competence favor opposite roles, LASE largely assigns roles according to initial position, whereas AGiR shifts behavior toward the agents' relative competences. In particular, the talented harvester placed near the river increases its extrinsic return from $136.3$ to $246.0$ and reduces cleaning from $119.4$ to $8.4$ actions per episode. 
Its harvesting also increases from 68 to 123 apples per episode, illustrating the shift from position-driven to competence-driven role allocation.
Conversely, the talented cleaner takes on most of the cleaning. The corresponding total returns increase from $128.4$ to $176.5$ for the talented harvester and from $8.1$ to $81.3$ for the talented cleaner.

\begin{table}[t]
\centering
\caption{Final results in the controlled two-agent Cleanup experiments. Values are averaged across independent runs.}
\label{tab:two_agent_results}
\resizebox{\textwidth}{!}{
\begin{tabular}{llcrrr|rrr}
\toprule
& & & \multicolumn{3}{c|}{LASE} & \multicolumn{3}{c}{LASE + AGiR} \\
Scenario & Agent & $\epsilon$ &
Extrinsic & Cleaning & Total &
Extrinsic & Cleaning & Total \\
\midrule
No gap
& Agent 0 & 1
& 297.4 & 44.9 & 298.8
& 283.6 & 49.0 & 275.3 \\
& Agent 1 & 1
& 268.4 & 31.9 & 266.9
& 276.8 & 16.9 & 285.0 \\
\midrule
Competence gap
& Talented cleaner & 0.1
& 0.02 & 142.7 & 23.4
& 0.003 & 149.7 & 116.9 \\
& Talented harvester & 1
& 423.8 & 4.62 & 400.4
& 352.0 & 0.03 & 235.1 \\
\cmidrule{2-9}
& Talented cleaner & 0.25
& 0.48 & 169.1 & 28.0
& 41.0 & 138.6 & 136.5 \\
& Talented harvester & 1
& 464.3 & 0.48 & 436.8
& 386.5 & 0.01 & 291.1 \\
\midrule
Spawn gap
& Apple-side agent & 1
& 291.5 & 35.9 & 290.5
& 302.1 & 33.5 & 294.7 \\
& River-side agent & 1
& 245.7 & 49.0 & 246.7
& 276.1 & 36.8 & 283.5 \\
\midrule
Mismatch
& Talented cleaner & 0.25
& 0.3 & 137.8 & 8.1
& 11.8 & 143.8 & 81.3 \\
& Talented harvester & 1
& 136.3 & 119.4 & 128.4
& 246.0 & 8.4 & 176.5 \\
\bottomrule
\end{tabular}
}
\end{table}

\subsection{Additional Learning Curves}
\label{append:two_agent_cleanup_curves}

We provide learning curves for the \textit{Competence gap} and \textit{Mismatch} scenarios to illustrate how the learned roles emerge and evolve throughout training, beyond their final outcomes.

\paragraph{Competence gap.}
% Figure~\ref{fig:competence_gap_additional} shows the cleaning behavior of the talented harvester for both harvesting-efficiency settings. Together with Fig.~\ref{fig:competence_gap}, these results illustrate how the two agents' behaviors evolve as the talented cleaner either returns to the cleaning role or continues harvesting.

\begin{figure}[H]
    \centering
    \begin{subfigure}[t]{0.47\linewidth}
        \centering
        \includegraphics[width=0.49\linewidth,trim={0 0 12 0},clip]{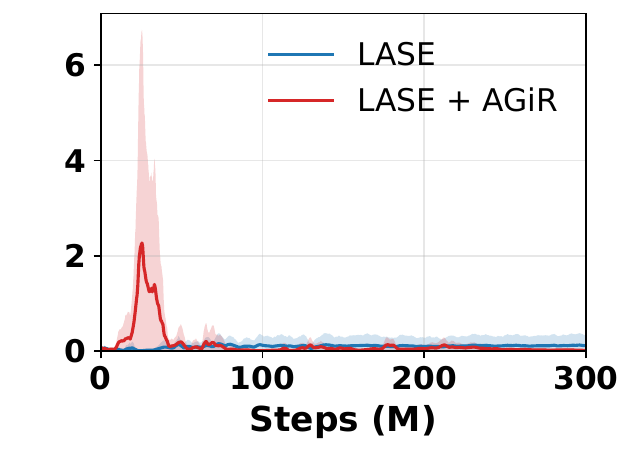}%
        \hfill
        \includegraphics[width=0.49\linewidth]{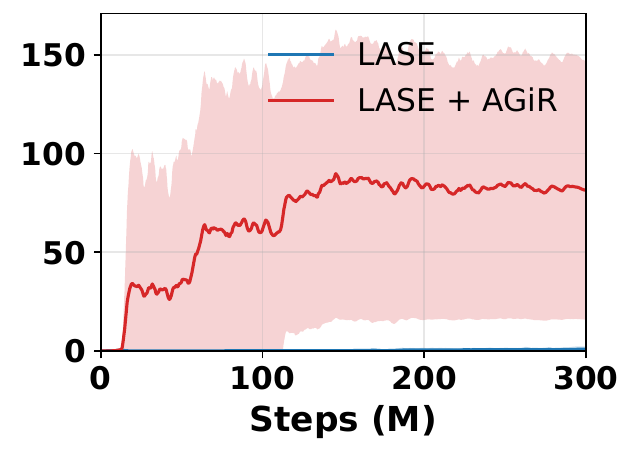}
        \caption{Apples harvested by the talented cleaner.}
        \label{fig:competence_gap}
    \end{subfigure}%
    \hfill
    \begin{subfigure}[t]{0.52\linewidth}
        \centering
        \includegraphics[width=0.49\linewidth]{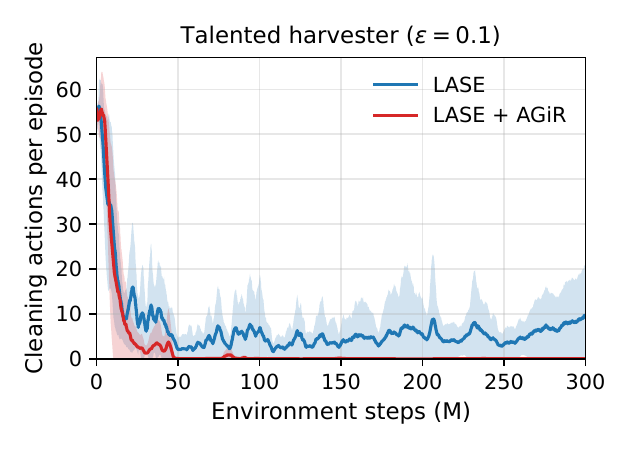}%
        \hfill
        \includegraphics[width=0.49\linewidth]{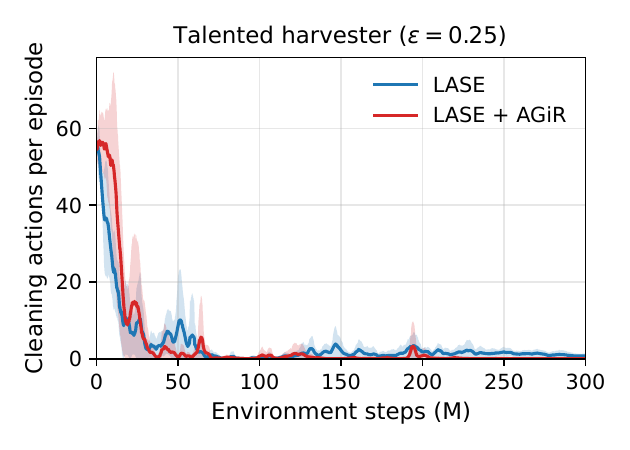}
        \caption{Cleaning actions by the talented harvester.}
        \label{fig:competence_gap_additional}
    \end{subfigure}
    \caption{Agent behaviors in the \emph{Competence gap} scenario. Within each pair, the left and right plots correspond to $0.1\times$ and $0.25\times$ reward per apple, respectively.}
\end{figure}

Figures~\ref{fig:competence_gap} and~\ref{fig:competence_gap_additional}
show the talented cleaner's harvesting behavior and the
talented harvester's cleaning behavior, respectively, under
both reward settings. Together, they illustrate how the
agents' roles evolve as the talented cleaner either returns
to cleaning or continues harvesting.

% \begin{figure}[H]
%     \centering
%     \begin{subfigure}[t]{0.48\linewidth}
%         \centering
%         \includegraphics[width=\linewidth]{image/competence_gap_eps0.1_harvester_cleaning.pdf}
%         \caption{$0.1\times$ reward per apple.}
%     \end{subfigure}
%     \hfill
%     \begin{subfigure}[t]{0.48\linewidth}
%         \centering
%         \includegraphics[width=\linewidth]{image/competence_gap_eps0.25_harvester_cleaning.pdf}
%         \caption{$0.25\times$ reward per apple.}
%     \end{subfigure}
%     \caption{Cleaning actions per episode by the talented harvester in the \emph{Competence gap} scenario.}
%     \label{fig:competence_gap_additional}
% \end{figure}

\begin{figure*}[t]
    \centering
    \begin{subfigure}[t]{0.32\textwidth}
        \centering
        \includegraphics[width=\linewidth]{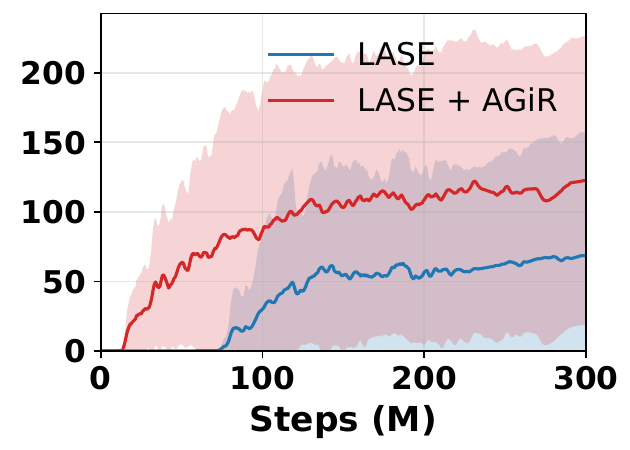}
        \caption{Harvester: apples}
    \end{subfigure}
    \hfill
    \begin{subfigure}[t]{0.32\textwidth}
        \centering
        \includegraphics[width=\linewidth]{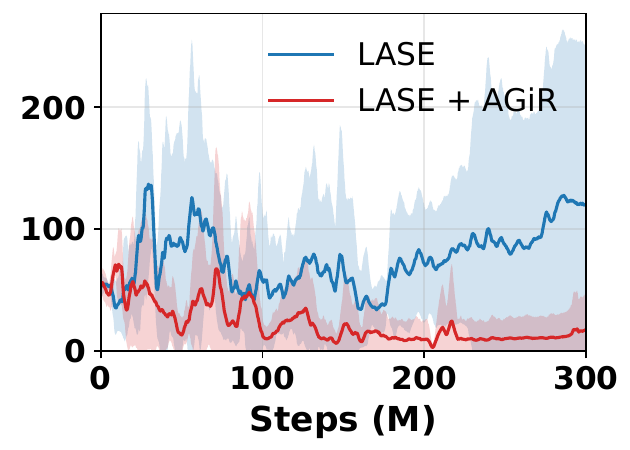}
        \caption{Harvester: cleaning}
    \end{subfigure}
    \hfill
    \begin{subfigure}[t]{0.32\textwidth}
        \centering
        \includegraphics[width=\linewidth]{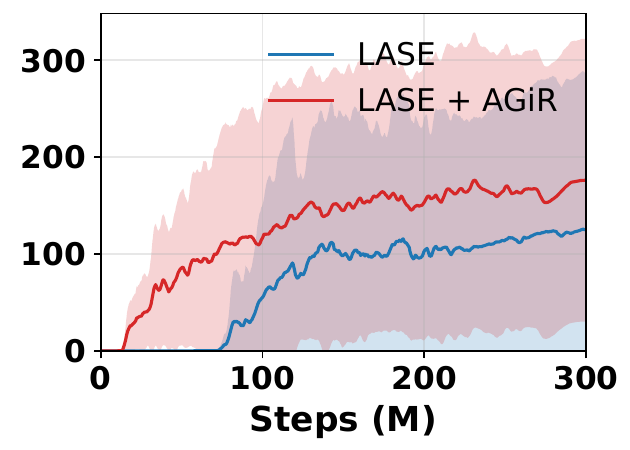}
        \caption{Harvester: total return}
    \end{subfigure}

    \vspace{0.5em}

    \begin{subfigure}[t]{0.32\textwidth}
        \centering
        \includegraphics[width=\linewidth]{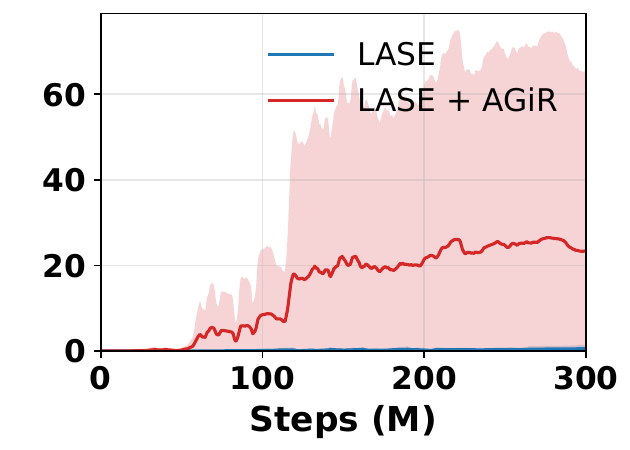}
        \caption{Cleaner: apples}
    \end{subfigure}
    \hfill
    \begin{subfigure}[t]{0.32\textwidth}
        \centering
        \includegraphics[width=\linewidth]{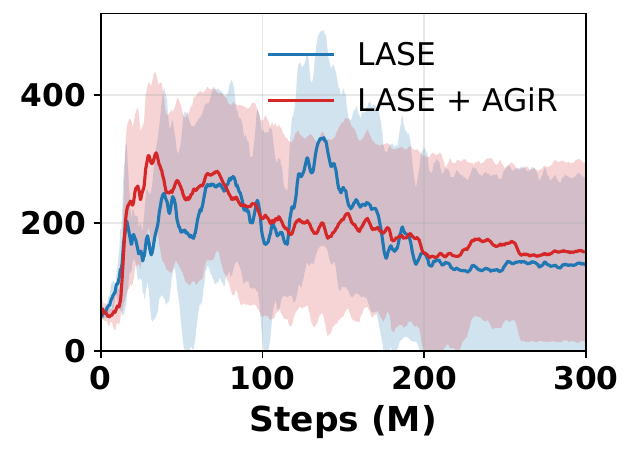}
        \caption{Cleaner: cleaning}
    \end{subfigure}
    \hfill
    \begin{subfigure}[t]{0.32\textwidth}
        \centering
        \includegraphics[width=\linewidth]{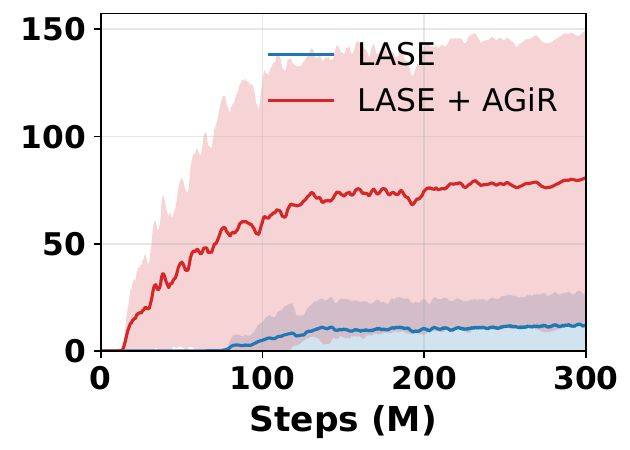}
        \caption{Cleaner: total return}
    \end{subfigure}

    \caption{Learning dynamics in the \emph{Mismatch} scenario for the talented harvester (top) and talented cleaner (bottom).}
    \label{fig:mismatch_additional}
\end{figure*}

\paragraph{Mismatch.}
Figure~\ref{fig:mismatch_additional} provides the corresponding learning curves for both agents in the \emph{Mismatch} scenario. The curves show the transition from position-driven roles under LASE toward behavior that is more consistent with the agents' relative competences under AGiR.

\FloatBarrier

\section{Ablation Studies}\label{append:ablation}
\subsection{Ablation Study on Gift Refunding} \label{append:ablation_refund}

In this section, we provide detailed experimental results comparing the gift handling policies: \textit{not accounted} (our default AGiR) and \textit{naive refunding}. For this ablation study, we used LASE~\cite{lase} as the base gifting method.

In the \textit{not accounted} setting, which corresponds to our proposed AGiR mechanism, the policy is updated using the shaped reward defined as $r^{i,\text{AGiR}} = r^{i,\text{ext}} - g_t^{i,-i} + w_t^i g_t^{-i,i}$. In the \textit{refunded} setting, we evaluate a naive approach where the policy is updated by incorporating returned gifts: $r^{i,\text{shaped}} = r^{i,\text{ext}} - g_t^{i,-i} + w_t^i g_t^{-i,i} + \text{(refunded gifts)}_t$. Note that in both cases, the acceptance weights $w_t^i$ are computed via the AGiR mechanism to ensure a fair comparison.

To provide a scale-invariant analysis across environments with different reward scales, we evaluate the performance using a normalized improvement score: $\frac{(\text{Gifting with refusal}) - (\text{Vanilla gifting})}{|(\text{Vanilla gifting})|}\times100\%$. Table~\ref{tab:ablation_refund_results} summarizes these normalized results for both gift handling policies. On average, the \textit{not accounted} approach demonstrates superior effectiveness compared to the naive refunding approach. Thus, we use the \textit{not accounted} scheme as our default because it yields stronger average improvements in GM and Min than \textit{naive refunding}.
% Thus, we establish the \textit{not accounted} scheme as our default for its reliable generalizability and sufficient fairness improvements over vanilla baselines. 

We hypothesize that without a sophisticated strategy to interpret the intent behind a rejection, the returned gifts might introduce noise into the agent's policy optimization. Developing advanced mechanisms for agents to reason about and utilize refunded gifts remains a subject for future work.

\begin{table}[h]
\centering
\caption{Ablation study on gift handling: Normalized improvement score (\%)}
\label{tab:ablation_refund_results}
\small
\begin{tabular}{llccc}
\toprule
Env. &  & AM Impr. (\%) $\uparrow$ & GM Impr. (\%) $\uparrow$ & Min Impr. (\%) $\uparrow$ \\
\midrule
\multirow{2}{*}{Cleanup-Easy} & Refund & -28.68 & -28.49 & -28.78 \\
 & Not accounted & +12.29 & +12.78 & +14.44 \\
\midrule
\multirow{2}{*}{Cleanup-Hard} & Refund & +4.28 & +144.58 & +128.65 \\
 & Not accounted & -8.20 & +115.37 & +96.88 \\
\midrule
\multirow{2}{*}{Harvest} & Refund & -3.37 & +44.65 & +73.68 \\
 & Not accounted & -3.15 & +44.76 & +65.79 \\
\midrule
\multirow{2}{*}{Coin Game} & Refund & -167.09 & -103.99 & -113.31 \\
 & Not accounted & +33.78 & +6.59 & +37.71 \\
\midrule
\hline
\multirow{2}{*}{Average} & Refund & -48.72 & +14.19 & +15.06 \\
 & Not accounted & \textbf{+8.68} & \textbf{+44.88} & \textbf{+53.71} \\
\bottomrule
\end{tabular}
\end{table}

\subsection{Ablation Study on Refusal-Aware Policy} \label{append:ablation_no_w_pi}

We ablate the refusal-aware policy proposed in Section~\ref{sec:main_method} by comparing the original policy $\pi(a \mid o, w)$ with a variant that does not condition on the refusal weight, namely $\pi(a \mid o)$. We use LASE as the base gifting method for this comparison. As shown in Table~\ref{tab:ablation_no_w_pi}, removing $w$ degrades performance, particularly GM and Min, across the evaluated environments. This result suggests that conditioning the policy on the current refusal state is important for AGiR to adapt its behavior to the refusal context.

\begin{table}[h]
\centering
\caption{Ablation study on refusal-aware policy}
\label{tab:ablation_no_w_pi}
\small
\begin{tabular}{llccc}
\toprule
Env. &  & AM $\uparrow$ & GM $\uparrow$ & Min $\uparrow$ \\
% \midrule
% \multirow{2}{*}{Cleanup-Easy} & w/ $w$ & 0 & 0 & 0 \\
%  & w/o $w$ & 0 & 0 & 0 \\
\midrule
\multirow{2}{*}{Cleanup-Hard} & w/ $w$ & 288 (41) & 283 (42) & 221 (46) \\
 & w/o $w$ & 289 (58) & 230 (117) & 169 (87) \\
\midrule
\multirow{2}{*}{Harvest} & w/ $w$ & 49 (0) & 41 (2) & 19 (5) \\
 & w/o $w$ & 49 (0) & 41 (1) & 17 (3) \\
\midrule
\multirow{2}{*}{Coin Game} & w/ $w$ & 20 (1) & -57 (3) & -40 (4) \\
 & w/o $w$ & 18 (25) & -59 (5) & -43 (7) \\
% \midrule
% \hline
% \multirow{2}{*}{Average} & Refund & -48.72 & +14.19 & +15.06 \\
%  & Not accounted & \textbf{+8.68} & \textbf{+44.88} & \textbf{+53.71} \\
\bottomrule
\end{tabular}
\end{table}

\subsection{Hyperparameter Sensitivity Analysis and Non-Adaptive Alternatives} \label{append:ablation_modulator}

\begin{figure}[h]
  \centering

    \includegraphics[width=0.9\linewidth]{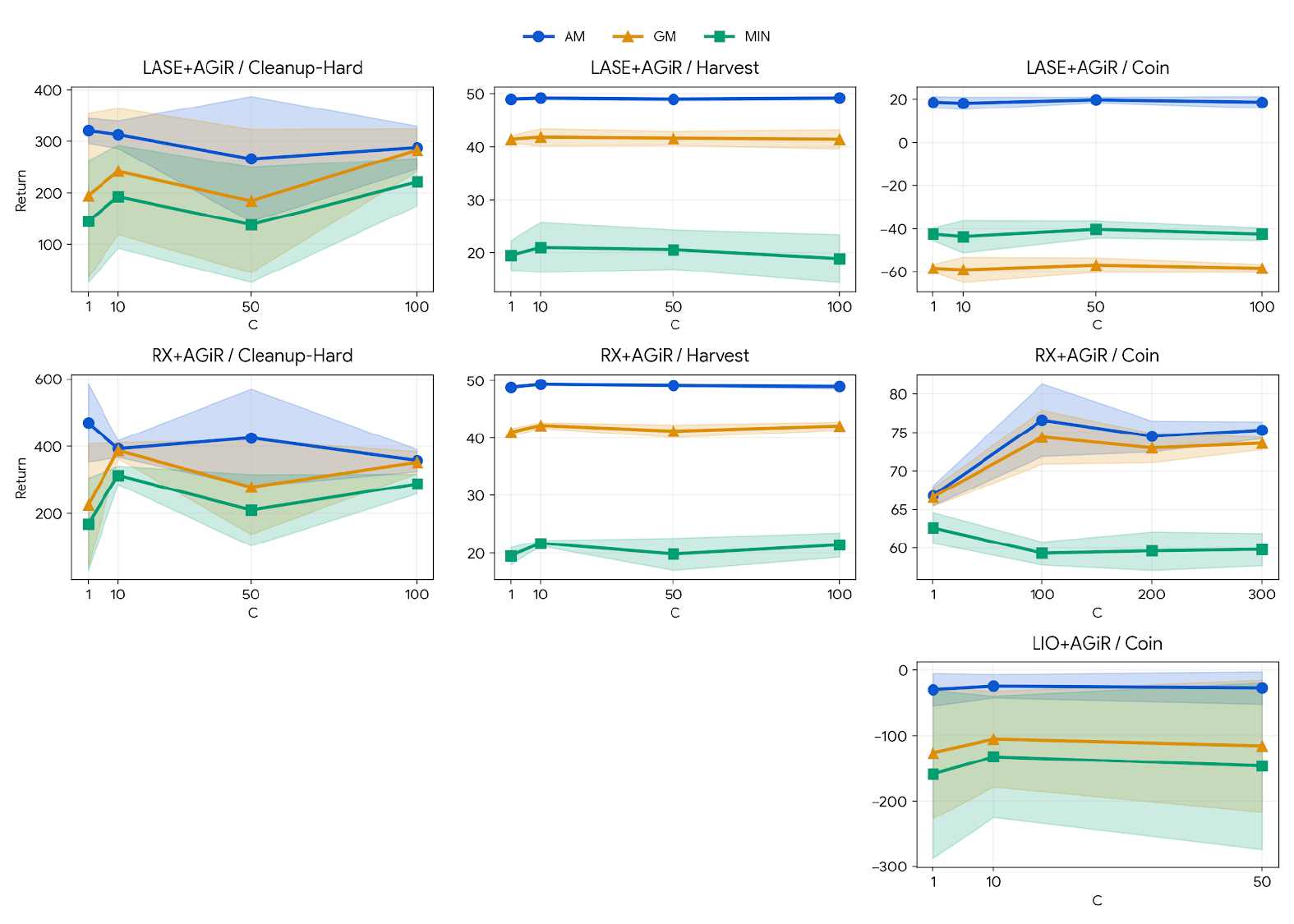}

  \caption{Ablation on modulator coefficient $C$ across environments and base gifting methods}
  \label{fig:ablation_c}
\end{figure}

\paragraph{Hyperparameter sensitivity analysis.}
The step-size modulator $S(R_t^i) := 2\sigma(\frac{-R_t^i}{C})$ is designed to adjust the learning rate based on the accumulated reward, where the coefficient $C$ determines the smoothness of this transition. A larger $C$ ensures that the modulator decreases more gradually as $R_t^i$ grows. Fig.~\ref{fig:ablation_c} presents the sensitivity analysis of the three $\alpha$-fairness metrics (AM, GM, and Min) with respect to varying values of $C$. 

In Fig.~\ref{fig:ablation_c}, each row corresponds to a specific base gifting method (LASE, RX, or LIO), and each column represents a different environment (Cleanup-Hard, Harvest, or Coin Game). We observe that when $C$ is set to an excessively small value, the sigmoid function decays too sharply, causing the step size to vanish prematurely. This leads to suboptimal performance in certain settings, as the agent fails to adapt sufficiently before the refusal-weight update becomes negligible. However, as $C$ increases, the performance becomes notably robust across all environments and algorithms. This stability justifies our choice of $C$ values, as it allows for a sufficiently gradual decay that supports stable learning.

\begin{figure}[H]
    \centering
    % 첫 번째 행 (A, B)
    \begin{subfigure}{0.4\textwidth}
        \centering
        \includegraphics[width=\linewidth]{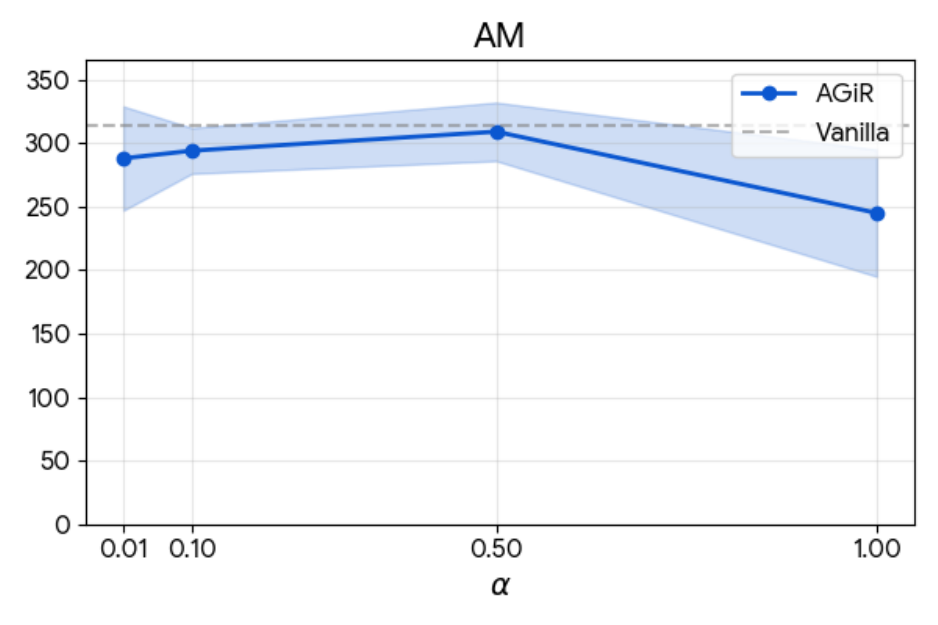}
        % \caption{Description A}
    \end{subfigure}
    % \hfill % 이미지 사이 간격 벌리기
    \begin{subfigure}{0.4\textwidth}
        \centering
        \includegraphics[width=\linewidth]{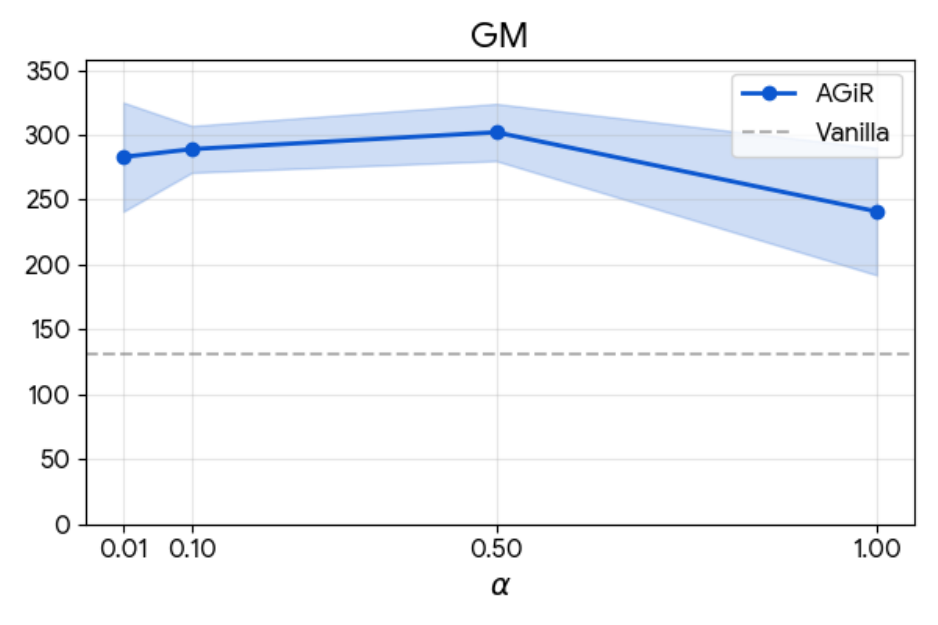}
        % \caption{Description B}
    \end{subfigure}

    \vspace{10pt} % 위아래 행 사이의 간격

    % 두 번째 행 (C, D)
    \begin{subfigure}{0.4\textwidth}
        \centering
        \includegraphics[width=\linewidth]{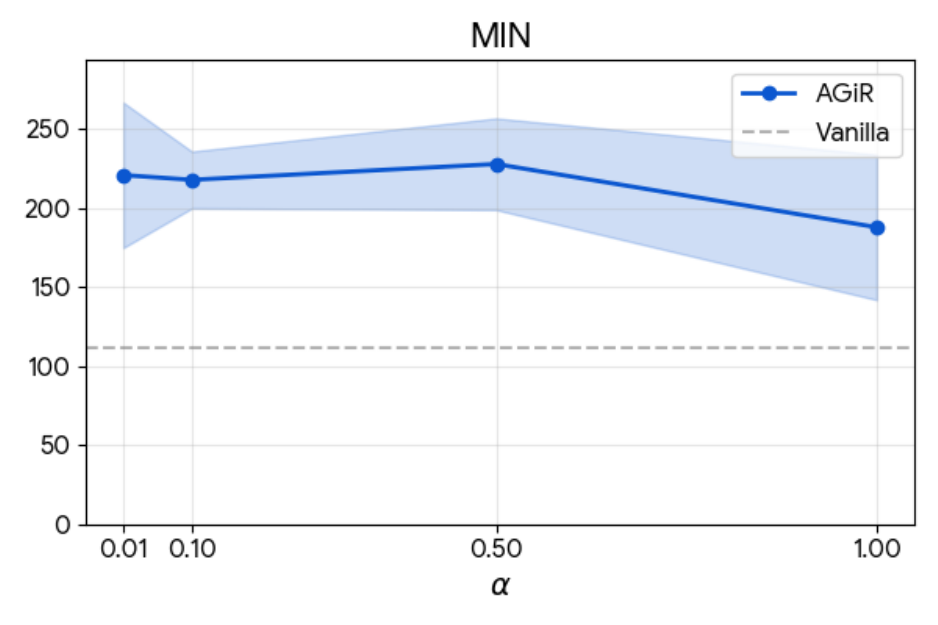}
        % \caption{Description C}
    \end{subfigure}
    % \hfill
    \begin{subfigure}{0.4\textwidth}
        \centering
        \includegraphics[width=\linewidth]{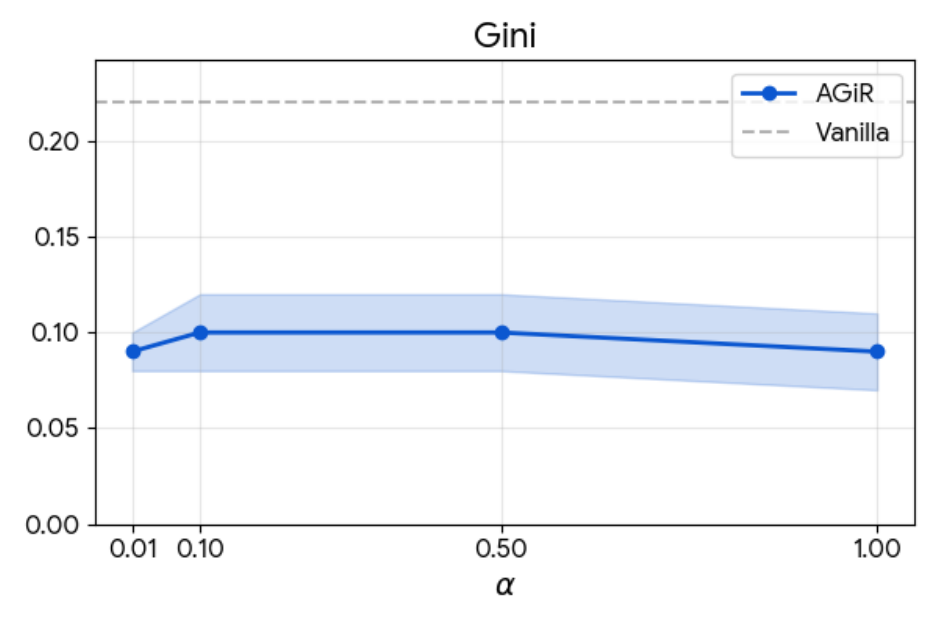}
        % \caption{Description D}
    \end{subfigure}
    \caption{Sensitivity analysis on the stepsize $\alpha$ in Cleanup-Hard (LASE)}
    \label{fig:ablation_alpha}
\end{figure}

\begin{figure}[H]
    \centering
    % 첫 번째 행 (A, B)
    \begin{subfigure}{0.4\textwidth}
        \centering
        \includegraphics[width=\linewidth]{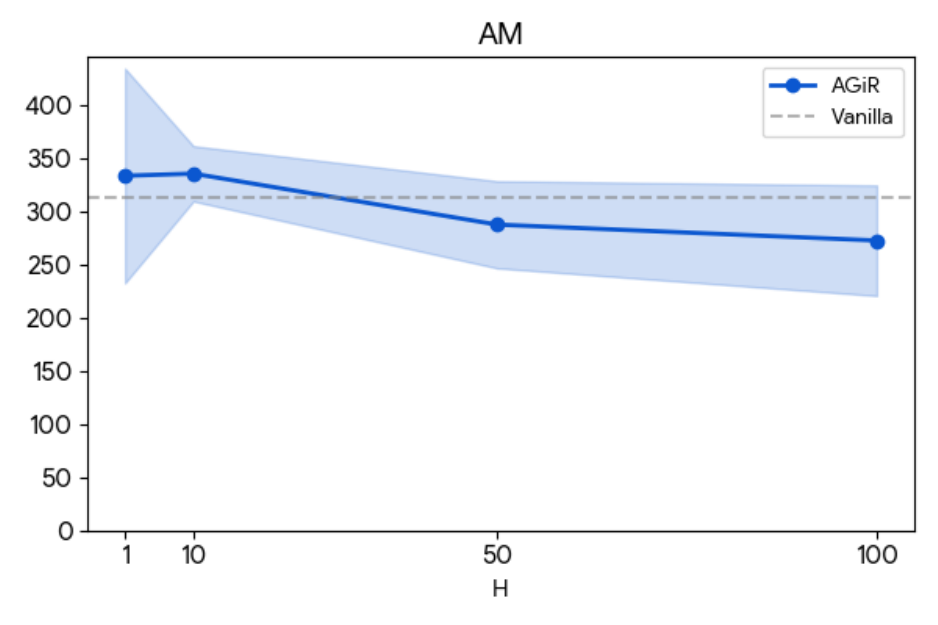}
        % \caption{Description A}
    \end{subfigure}
    % \hfill % 이미지 사이 간격 벌리기
    \begin{subfigure}{0.4\textwidth}
        \centering
        \includegraphics[width=\linewidth]{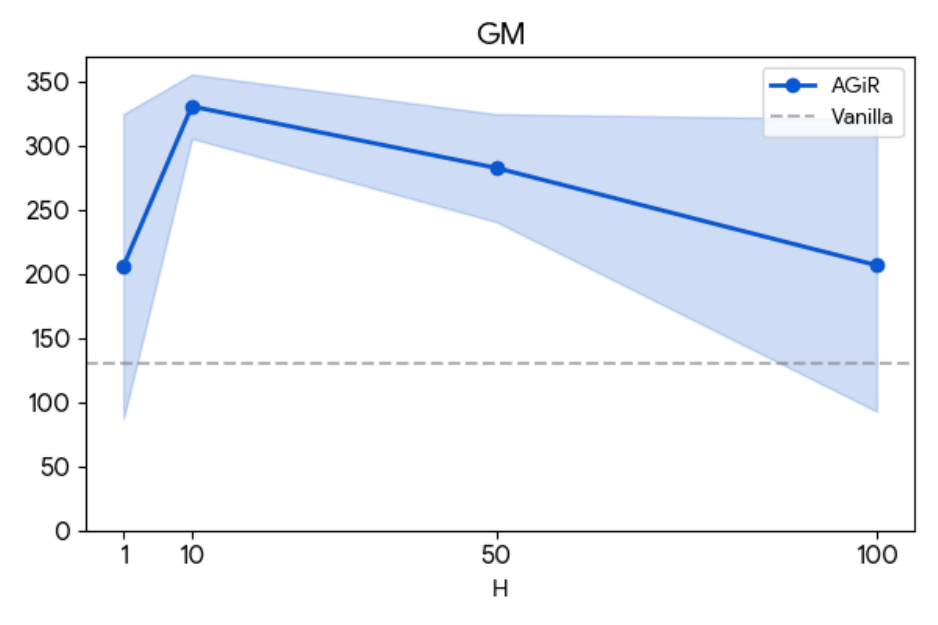}
        % \caption{Description B}
    \end{subfigure}

    \vspace{10pt} % 위아래 행 사이의 간격

    % 두 번째 행 (C, D)
    \begin{subfigure}{0.4\textwidth}
        \centering
        \includegraphics[width=\linewidth]{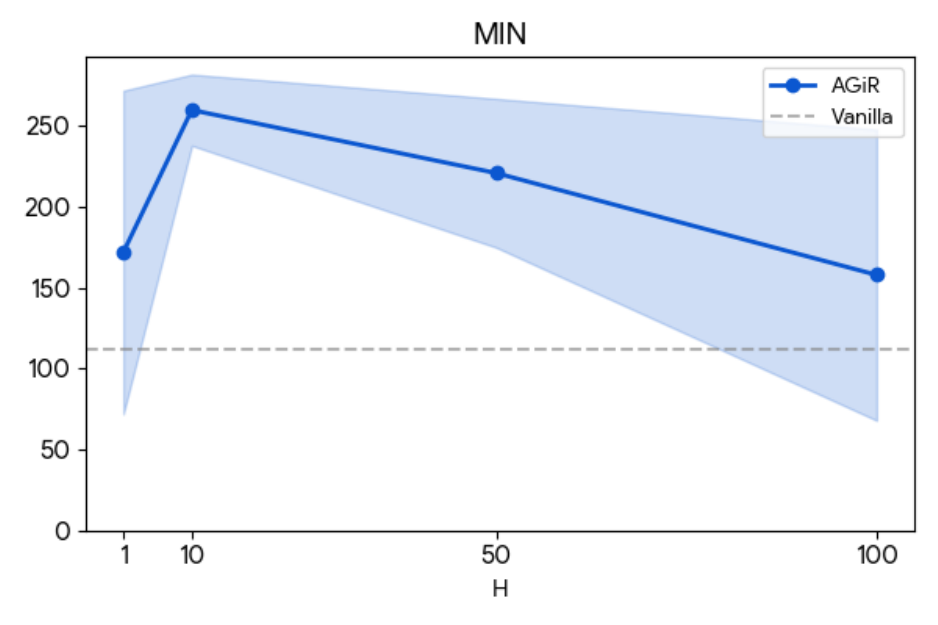}
        % \caption{Description C}
    \end{subfigure}
    % \hfill
    \begin{subfigure}{0.4\textwidth}
        \centering
        \includegraphics[width=\linewidth]{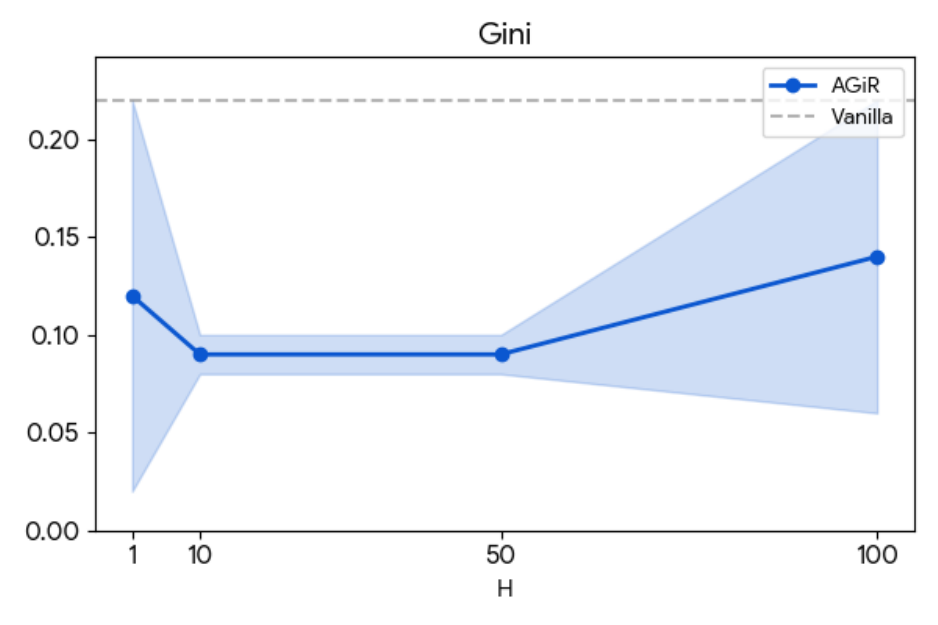}
        % \caption{Description D}
    \end{subfigure}
    \caption{Sensitivity analysis on the window size $H$ in Cleanup-Hard (LASE)}
    \label{fig:ablation_H}
\end{figure}

\begin{figure}[h]
    \centering
    % 첫 번째 행 (A, B)
    \begin{subfigure}{0.4\textwidth}
        \centering
        \includegraphics[width=\linewidth]{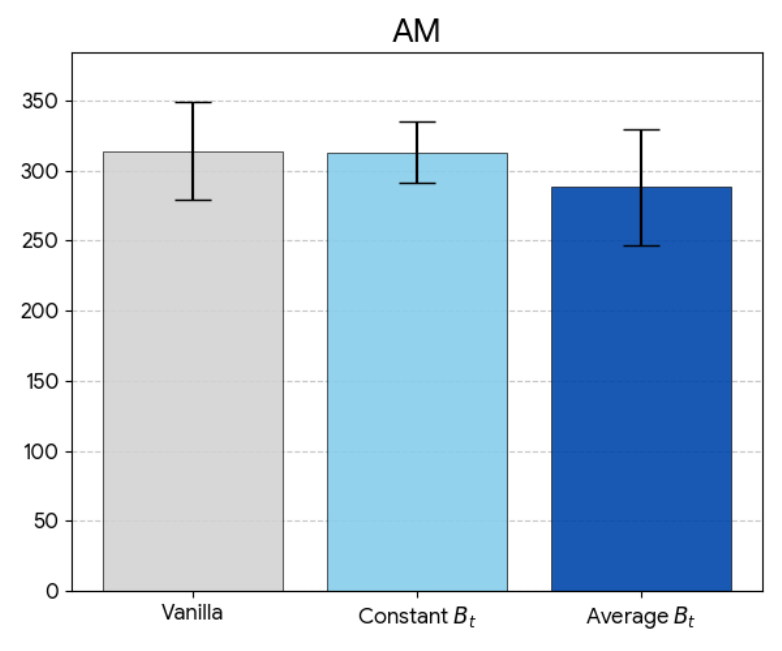}
        % \caption{Description A}
    \end{subfigure}
    % \hfill % 이미지 사이 간격 벌리기
    \begin{subfigure}{0.4\textwidth}
        \centering
        \includegraphics[width=\linewidth]{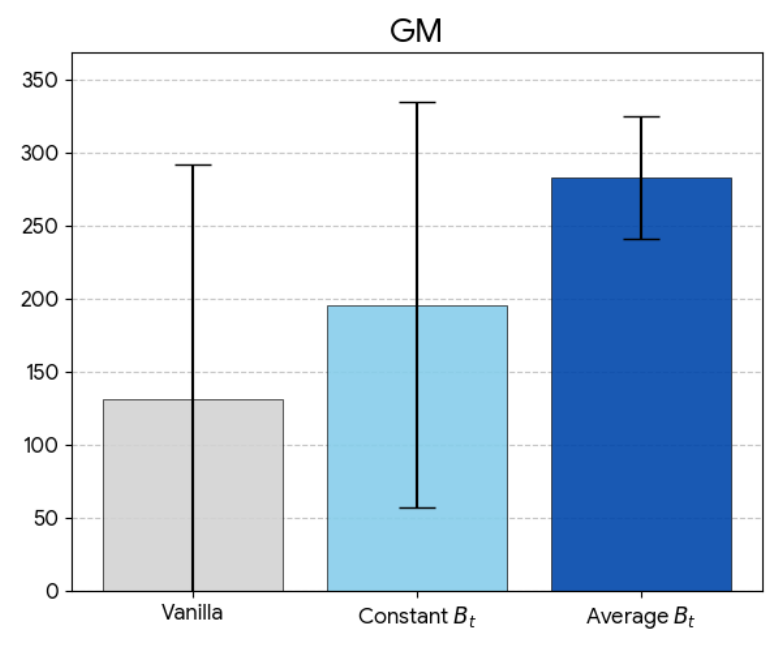}
        % \caption{Description B}
    \end{subfigure}

    \vspace{10pt} % 위아래 행 사이의 간격

    % 두 번째 행 (C, D)
    \begin{subfigure}{0.4\textwidth}
        \centering
        \includegraphics[width=\linewidth]{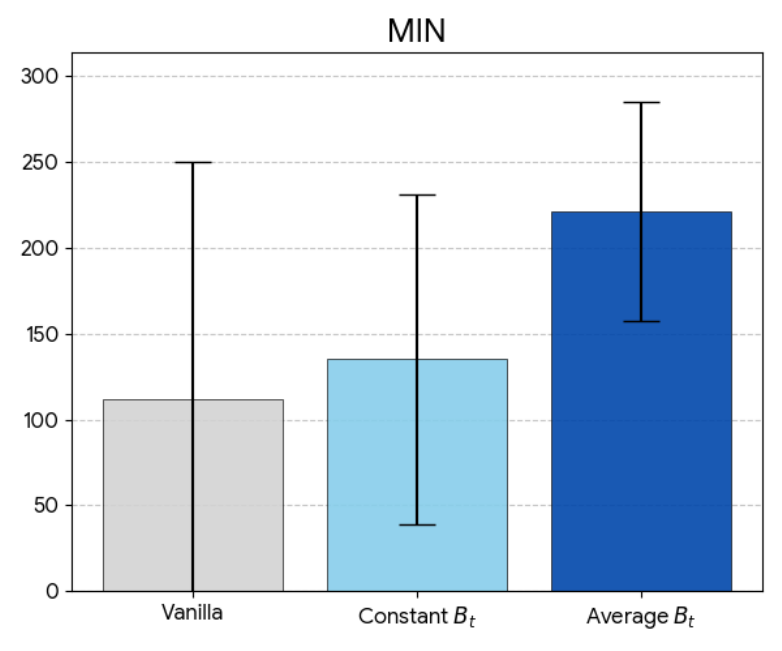}
        % \caption{Description C}
    \end{subfigure}
    % \hfill
    \begin{subfigure}{0.4\textwidth}
        \centering
        \includegraphics[width=\linewidth]{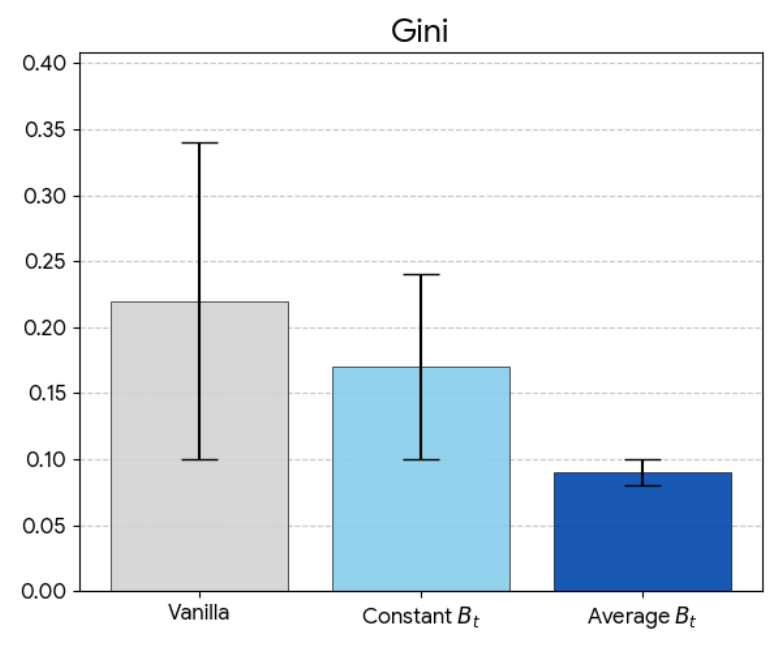}
        % \caption{Description D}
    \end{subfigure}
    \caption{Ablation on the baseline $B_t$ in Cleanup-Hard (LASE)}
    \label{fig:ablation_B}
\end{figure}

We further examine the sensitivity of AGiR to the step size $\alpha$, the window size $H$, and the baseline design $B_t$ using LASE in Cleanup-Hard. 
As shown in Fig.~\ref{fig:ablation_alpha}, AGiR consistently achieves higher GM and Min and lower Gini coefficient than vanilla gifting across all tested $\alpha$ values, while largely maintaining AM unless $\alpha$ is too large.
Fig.~\ref{fig:ablation_H} shows that AGiR consistently achieves higher GM and Min and lower Gini coefficient than vanilla gifting across a wide range of window sizes $H$, while largely maintaining AM.
Finally, Fig.~\ref{fig:ablation_B} shows that replacing the adaptive baseline with a fixed baseline ($B_t=\pi/4$) yields lower GM and Min than the adaptive design, while still outperforming vanilla gifting on these metrics.

\paragraph{Non-adaptive alternatives.}
We further compare AGiR with simple non-adaptive strategies for limiting incoming gifts in Unfair Cleanup-\emph{Hard}, using LASE as the base gifting method. We consider fixed acceptance weights, gift clipping, and linear annealing of the acceptance weight. As shown in Table~\ref{tab:nonadaptive_ablation}, all of these alternatives yield lower GM and Min than AGiR. In particular, linear annealing attains the same AM as AGiR but substantially lower GM and Min, indicating that simply reducing gift acceptance over training does not reproduce the behavior of adaptive refusal.

%%%%%% Ablation C %%%%%%

%%%%%% Ablation alpha %%%%%%

%%%%%% Ablation H %%%%%%

%%%%%% Ablation B_t %%%%%%

\begin{table}[H]
\centering
\caption{Comparison with non-adaptive gift-limiting strategies in Unfair Cleanup-\emph{Hard} with LASE as the base gifting method.}
\label{tab:nonadaptive_ablation}
\begin{tabular}{lcccc}
\toprule
Method & AM $\uparrow$ & GM $\uparrow$ & Min $\uparrow$ & Gini $\downarrow$ \\
\midrule
Fixed $w=0.25$ & 269 & 260 & 173 & 0.13 \\
Fixed $w=0.50$ & 263 & 259 & 193 & \textbf{0.09} \\
Fixed $w=0.75$ & 277 & 272 & 205 & \textbf{0.09} \\
Gift clipping ($0.5$) & 261 & 257 & 191 & 0.10 \\
Gift clipping ($0.7$) & 233 & 230 & 190 & \textbf{0.09} \\
Linear annealing & \textbf{288} & 168 & 98 & 0.22 \\
\midrule
AGiR & \textbf{288} & \textbf{283} & \textbf{221} & \textbf{0.09} \\
\bottomrule
\end{tabular}
\end{table}

\section{Limitations, Future Work, and Broader Societal Impacts}\label{append:limitation_futurework}
While this study establishes the theoretical and empirical foundations of AGiR, several avenues remain for future exploration.

% Limited theoretical analysis
First, while our rollout-axis analysis characterizes refusal-weight dynamics under fixed policies, understanding how simultaneous policy updates shape the joint value landscape in general MMGs remains an open problem. Moreover, the sufficient conditions for stability and uniqueness in Theorem~\ref{thm:uniqueness} have not been empirically verified. Investigating whether low refusal weights exhibit recovery and under which environments and gifting schemes these conditions hold is left for future work.

% Scalability and Large-scale Populations 
Second, due to computational resource constraints, our experiments were conducted with a limited number of players (up to four). Investigating how gift-refusal dynamics evolve in larger populations is a compelling direction for future research. Specifically, extending our framework to Mean-Field Markov Games could provide a mathematically principled way to analyze AGiR in systems with a massive number of agents, where individual interactions are abstracted into collective population dynamics.

% Diversity of Reinforcement Learning Backbones 
Third, our implementation utilized the PPO backbone provided by the SocialJax framework to ensure controlled and consistent experiments. Consequently, the performance of AGiR under value-based reinforcement learning algorithms remains unexplored. Future work could investigate the robustness of our adaptive mechanism across diverse algorithmic architectures to ensure its generalizability beyond policy-gradient methods.

% MO MMG

Fourth, extending AGiR to multi-objective mixed-motive games is a promising direction. Building on recent advances in multi-objective reinforcement learning~\citep{morl_0,morl_1,morl_2} and multi-objective Markov games and social dilemmas~\citep{momg_1,momg_2}, future work could investigate how multiple, potentially conflicting objectives affect gifting dynamics, gift reliance, and emergent agent behavior.

% Sophisticated Intent Inference in Refunded Settings 
Finally, while we conducted an ablation study on the "refunded setting" where refused gifts are returned to the giver, our ablation assumes a naive giver. In this setup, the giver does not perform any higher-level reasoning regarding why a gift was refused. A promising research direction would be to incorporate sophisticated intention inference, where givers adjust their strategies based on the perceived motives behind a refusal. Exploring such recursive social reasoning would not only enhance the strategic depth of the agents but also provide deeper insights into complex human social phenomena. % and the nuances of reciprocity.

% Broader Societal Impacts
This work contributes to the methodological understanding of adaptive incentive regulation in mixed-motive multi-agent systems. While our contributions are primarily algorithmic, they carry potential societal implications for the design of collaborative AI systems. On the positive side, our approach to mitigating "gift over-reliance" can inform the development of more robust and equitable resource-sharing protocols in automated socio-technical systems.

\newpage
\section*{NeurIPS Paper Checklist}

%%% BEGIN INSTRUCTIONS %%%

%%% END INSTRUCTIONS %%%

\begin{enumerate}

\item {\bf Claims}
    \item[] Question: Do the main claims made in the abstract and introduction accurately reflect the paper's contributions and scope?
    \item[] Answer: \answerYes{} % Replace by \answerYes{}, \answerNo{}, or \answerNA{}.
    \item[] Justification: {As specified in the abstract and introduction, we identify the "gift over-reliance" phenomenon and provide a comprehensive algorithm along with supporting theoretical and empirical analyses.}

\item {\bf Limitations}
    \item[] Question: Does the paper discuss the limitations of the work performed by the authors?
    \item[] Answer: \answerYes{} % Replace by \answerYes{}, \answerNo{}, or \answerNA{}.
    \item[] Justification: {We discuss the limitations in Appendix~\ref{append:limitation_futurework}.}

\item {\bf Theory assumptions and proofs}
    \item[] Question: For each theoretical result, does the paper provide the full set of assumptions and a complete (and correct) proof?
    \item[] Answer: \answerYes{} % Replace by \answerYes{}, \answerNo{}, or \answerNA{}.
    \item[] Justification: {We provide the full statement of each theorem, including all assumptions and complete proofs, in Appendix~\ref{append:proof}.}

    \item {\bf Experimental result reproducibility}
    \item[] Question: Does the paper fully disclose all the information needed to reproduce the main experimental results of the paper to the extent that it affects the main claims and/or conclusions of the paper (regardless of whether the code and data are provided or not)?
    \item[] Answer: \answerYes{} % Replace by \answerYes{}, \answerNo{}, or \answerNA{}.
    \item[] Justification: {We provide comprehensive details for reproducibility, including hyperparameters, environment configurations (e.g., maps and parameters), experimental setups, architecture, and baseline implementation details in Appendix~\ref{append:Implementation_Details} and~\ref{append:two_agent}.}

\item {\bf Open access to data and code}
    \item[] Question: Does the paper provide open access to the data and code, with sufficient instructions to faithfully reproduce the main experimental results, as described in supplemental material?
    \item[] Answer: \answerYes{} % Replace by \answerYes{}, \answerNo{}, or \answerNA{}.
    \item[] Justification: {Code is available at \url{https://github.com/whbyeon/AGiR}.}

\item {\bf Experimental setting/details}
    \item[] Question: Does the paper specify all the training and test details (e.g., data splits, hyperparameters, how they were chosen, type of optimizer) necessary to understand the results?
    \item[] Answer: \answerYes{} % Replace by \answerYes{}, \answerNo{}, or \answerNA{}.
    \item[] Justification: {All details are comprehensively detailed in Appendix~\ref{append:Implementation_Details} and~\ref{append:two_agent}.}

\item {\bf Experiment statistical significance}
    \item[] Question: Does the paper report error bars suitably and correctly defined or other appropriate information about the statistical significance of the experiments?
    \item[] Answer: \answerYes{} % Replace by \answerYes{}, \answerNo{}, or \answerNA{}.
    \item[] Justification: {We report the standard deviation and error bars across five independent runs (seeds) for all experimental results presented in Figures~\ref{fig:cleanup_agent_bar_hard},\ref{fig:harvest_agent_bar} as well as Tables~\ref{table_cleanup_hard},\ref{table_harvest},\ref{table_coingame_main}.}

\item {\bf Experiments compute resources}
    \item[] Question: For each experiment, does the paper provide sufficient information on the computer resources (type of compute workers, memory, time of execution) needed to reproduce the experiments?
    \item[] Answer: \answerYes{} % Replace by \answerYes{}, \answerNo{}, or \answerNA{}.
    \item[] Justification: {We specify the computing resources used for our experiments in Appendix~\ref{append:practical_algo}.}
    
\item {\bf Code of ethics}
    \item[] Question: Does the research conducted in the paper conform, in every respect, with the NeurIPS Code of Ethics \url{https://neurips.cc/public/EthicsGuidelines}?
    \item[] Answer: \answerYes{} % Replace by \answerYes{}, \answerNo{}, or \answerNA{}.
    \item[] Justification: {The research conducted in this paper fully conforms to the NeurIPS Code of Ethics.}

\item {\bf Broader impacts}
    \item[] Question: Does the paper discuss both potential positive societal impacts and negative societal impacts of the work performed?
    \item[] Answer: \answerYes{} % Replace by \answerYes{}, \answerNo{}, or \answerNA{}.
    \item[] Justification: {We discuss the potential societal impacts of our work in Appendix~\ref{append:limitation_futurework}.}
    
\item {\bf Safeguards}
    \item[] Question: Does the paper describe safeguards that have been put in place for responsible release of data or models that have a high risk for misuse (e.g., pre-trained language models, image generators, or scraped datasets)?
    \item[] Answer: \answerNA{} % Replace by \answerYes{}, \answerNo{}, or \answerNA{}.
    \item[] Justification: {This study presents no high-risk artifacts or data requiring specific safeguards against misuse.}

\item {\bf Licenses for existing assets}
    \item[] Question: Are the creators or original owners of assets (e.g., code, data, models), used in the paper, properly credited and are the license and terms of use explicitly mentioned and properly respected?
    \item[] Answer: \answerYes{} % Replace by \answerYes{}, \answerNo{}, or \answerNA{}.
    \item[] Justification: {We use SocialJax under the Apache-2.0 license and ensure it is properly cited.}

\item {\bf New assets}
    \item[] Question: Are new assets introduced in the paper well documented and is the documentation provided alongside the assets?
    \item[] Answer: \answerNA{} % Replace by \answerYes{}, \answerNo{}, or \answerNA{}.
    \item[] Justification: {The paper does not release new assets.}

\item {\bf Crowdsourcing and research with human subjects}
    \item[] Question: For crowdsourcing experiments and research with human subjects, does the paper include the full text of instructions given to participants and screenshots, if applicable, as well as details about compensation (if any)? 
    \item[] Answer: \answerNA{} % Replace by \answerYes{}, \answerNo{}, or \answerNA{}.
    \item[] Justification: {The paper does not include crowdsourcing experiments and research with human subjects.}

\item {\bf Institutional review board (IRB) approvals or equivalent for research with human subjects}
    \item[] Question: Does the paper describe potential risks incurred by study participants, whether such risks were disclosed to the subjects, and whether Institutional Review Board (IRB) approvals (or an equivalent approval/review based on the requirements of your country or institution) were obtained?
    \item[] Answer: \answerNA{} % Replace by \answerYes{}, \answerNo{}, or \answerNA{}.
    \item[] Justification: {The paper does not include research with human subjects and thus, does not require IRB approval.}

\item {\bf Declaration of LLM usage}
    \item[] Question: Does the paper describe the usage of LLMs if it is an important, original, or non-standard component of the core methods in this research? Note that if the LLM is used only for writing, editing, or formatting purposes and does \emph{not} impact the core methodology, scientific rigor, or originality of the research, declaration is not required.
    %this research? 
    \item[] Answer: \answerNA{} % Replace by \answerYes{}, \answerNo{}, or \answerNA{}.
    \item[] Justification: {LLMs were not used as a component of our core methodology.}

\end{enumerate}

\end{document}